\documentclass[fleqn,usenatbib]{mnras}
\usepackage{newtxtext,newtxmath}
\usepackage{orcidlink}
\usepackage{multirow}
\usepackage{hyperref}
\usepackage{longtable}
\usepackage{pdflscape}

\newcommand{\added}[1]{\textcolor{black}{#1}}

\usepackage[T1]{fontenc}

\DeclareRobustCommand{\VAN}[3]{#2}
\let\VANthebibliography\thebibliography
\def\thebibliography{\DeclareRobustCommand{\VAN}[3]{##3}\VANthebibliography}

\usepackage{graphicx}	
\usepackage{amsmath}	
\DeclareMathOperator{\sech}{sech}
\graphicspath{{figures/}}

\title[JWST detection of thin/thick discs at high z]{The emergence of galactic thin and thick discs across cosmic history}

\author[Takafumi Tsukui et al.]{
Takafumi Tsukui,$^{\orcidlink{0000-0002-1499-6377},1,2}$\thanks{E-mail: tsukuitk23@gmail.com (TT)}
Emily Wisnioski,$^{\orcidlink{0000-0003-1657-7878},1,2}$
Joss Bland-Hawthorn,$^{\orcidlink{0000-0001-7516-4016},2,3}$ 
Ken Freeman$^{\orcidlink{0000-0001-6280-1207},1,2}$
\\
$^{1}$Research School of Astronomy and Astrophysics, Australian National University, Cotter Road, Weston Creek, ACT 2611, Australia\\
$^{2}$ARC Centre of Excellence for All Sky Astrophysics in 3 Dimensions (ASTRO 3D), Australia\\
$^{3}$ Sydney Institute for Astronomy, School of Physics, A28, The University of Sydney, NSW 2006, Australia
}

\date{Accepted XXX. Received YYY; in original form ZZZ}

\pubyear{2024}

\begin{document}
\label{firstpage}
\pagerange{\pageref{firstpage}--\pageref{lastpage}}
\maketitle

\begin{abstract}
Present-day disc galaxies often exhibit distinct thin and thick discs. The formation mechanisms of the two discs and the timing of their onset remain open questions. To address these questions, we select edge-on galaxies from flagship JWST programs and investigate their disc structures in rest-frame, near-infrared bands. For the first time, we identify thick and thin discs at cosmological distances, dating back over 10 Gyr, and investigate their decomposed structural properties. We classify galaxies into those that require two (i.e. thin and thick) discs and those well fitted by a single disc. Disc radial sizes and vertical heights correlate strongly with the total galaxy mass and/or disc mass, independent of cosmic time. The structure of the thick disc resembles discs found in single-disc galaxies, suggesting that galaxies form a thick disc first, followed by the subsequent formation of an embedded thin disc. The transition from single to double discs occurred around 8 Gyr ago in high-mass galaxies ($10^{9.75} - 10^{11}M_\odot$), earlier than the transition which occurred 4 Gyr ago in low-mass galaxies ($10^{9.0} - 10^{9.75}M_\odot$), indicating sequential formation proceeds in a "downsizing" manner. Toomre $Q$-regulated disc formation explains the delayed thin disc formation in low-mass galaxies, leading to the observed anti-correlation between the thick-to-thin disc mass ratio and the total galaxy mass. Despite the dominant sequential formation, observations suggest that thick discs may continue to build up mass alongside their thin-disc counterparts.
\end{abstract}




\section{Introduction}
\label{sec:intro}
In the present universe, disc galaxies, including our Milky Way, commonly exhibit bimodal disc structures, i.e. geometrically thin and thick discs \citep{burstein_structure_1979, yoshii_density_1982, gilmore_new_1983, dalcanton_structural_2002, yoachim_kinematics_2008, comeron_thick_2011, comeron_evidence_2014}. These disc systems have vertical density profiles that are better described by two exponential (or $\sech^2$) functions rather than one. The stars in each disc are usually separable in some combination of stellar age, metal abundance properties, radial extent, and/or stellar kinematics \citep{hayden_chemical_2015}. Thick discs predominantly consist of old, metal-poor stars with enhanced [$\alpha$/Fe] abundance ratios, suggesting a rapid formation phase at early times. In contrast, thin discs mainly contain young, metal-rich stars with lower [$\alpha$/Fe] ratios, indicating later formation through prolonged star formation and effective metal accumulation in the interstellar medium (ISM). Furthermore, in line with the geometrical structure, the thick disc is kinematically hotter and exhibits lower mean rotation than the thin disc due to asymmetric drift \citep{lee_formation_2011}

Several mechanisms have been proposed to explain the disc dichotomy: 
\begin{enumerate}
    \item \textbf{The "born hot"} scenario \citep[e.g.,][]{burkert_collapse_1992, brook_emergence_2004, bournaud_thick_2009, bird_inside_2013, bird_inside_2021, leaman_unified_2017, grand_dual_2020, yu_bursty_2021, yu_born_2023} proposes that the thick disc forms first via intense star formation in a turbulent gas disc, followed by thin disc formation in a quiet gas disc. Observations show higher ionised gas turbulence in higher redshift galaxies \citep{forster_schreiber_sins_2009, genzel_sins_2011, wisnioski_kmos3d_2015, wisnioski_kmos3d_2019, ubler_evolution_2019}, presumably driven by higher gas accretion, gas fraction and star formation compared to today. A higher gas fraction in higher-redshift or lower-mass galaxies \citep[e.g.][]{mcgaugh_gas_1997, saintonge_cold_2022, tacconi_phibss_2013, tacconi_evolution_2020} can drive turbulence and inhibit thin disc formation \citep{elmegreen_star_2015, vandonkelaar_giant_2022}. In the turbulent phase of galaxy formation with high gas fractions ($f_{\rm gas} > 50\%$), typical in the high redshift universe \citep{carilli_cool_2013, tacconi_phibss_2013, tacconi_evolution_2020}, intense star formation expels gas from the disc and weakens the disc gravitational potential so that the stellar disc puffs up. This process is not fully reversed when some of the gas falls back to the disc \citep[][]{bland-hawthorn_turbulent_2024, Bland-Hawthorn_turbulent_2025}.
    \item The \textbf{"progressive thickening"} scenario suggests that stars form near the disc mid-plane and get heated up to form the thick disc by various density fluctuations or external perturbation \citep[giant molecular clouds: GMCs, spiral arms, giant clumps, galaxy interaction, e.g.,][]{wielen_diffusion_1977, lacey_influence_1984, villumsen_evolution_1985, quinn_heating_1993, Di_Matteo_formation_2011, inoue_properties_2014}. However, scattering from GMCs alone is shown to be insufficient to produce thick discs \citep{robin_constraining_2014, aumer_agevelocity_2016, leaman_unified_2017}, and is only effective for the thin-disc population near the disc midplane \citep{martig_dissecting_2014, mackereth_origin_2019}. The internal kinetic energy of star clusters may also contribute to the heating of these systems when they become unbound due to gas expulsion \citep{kroupa_thickening_2002}. Minor mergers can thicken discs and also induce disc flaring \citep{bournaud_thick_2009, comeron_thick_2011}.
    \item The \textbf{"ex situ"} scenario requires mergers or satellite accretion to form a thick disc \citep{abadi_simulations_2003, yoachim_structural_2006} or contribute to its growth \citep{pinna_fornax_2019, pinna_fornax_2019-1, martig_ngc_2021}. In this scenario, retrograde satellites produce counter-rotating stars in the thick disc, however the current observed fraction of thick discs with counter-rotating stars is  smaller than expected \citep{comeron_galactic_2015, comeron_kinematics_2019}.
\end{enumerate}

All mechanisms can explain observed qualitative signatures of the Milky Way's two discs, and they are not necessarily mutually exclusive \citep{pinna_stellar_2024}. A key challenge is to quantify their relative importance \citep{yu_bursty_2021, mccluskey_disc_2024, martig_dissecting_2014, agertz_vintergatan_2021} as a function of lookback time and galaxy properties (e.g., mass, environment, merger history), and addressing it requires measuring the properties of two discs for a large sample of galaxies spanning a wide range of lookback times and galaxy masses. Such measurements offer a time-machine perspective, as each galaxy has presumably been influenced by these processes to varying degrees. However, most insights so far come from studies at $z\sim0$, where thin and thick discs have been systematically decomposed using high resolution images. Leveraging JWST's imaging capability, this paper aims to systematically investigate thin and thick discs at higher redshifts, thereby exploring a much broader cosmic window.

Edge-on galaxies uniquely allow for direct study of their disc vertical structure and the decomposition of the thin and thick disc components \citep[e.g.,][]{van_der_kruit_surface_1981, yoachim_structural_2006, comeron_thick_2011, comeron_breaks_2012, comeron_evidence_2014}. Recently, the vertical height of high-redshift ($z\sim0.5-5$) disc galaxies has been studied using 1D profile fitting to the disc vertical profile. Using Hubble Space Telescope images at observed $VBI$ bands, \citealt{elmegreen_thick_2017} inferred the presence of thick and thin discs from an anti-correlation between height and disc intensity of the fitted profiles. \citet{hamilton-campos_physical_2023} measured disc heights at rest-frame 0.46–0.66$\micron$, suggesting that the population median height is similar to that of Milky Way's thick disc and does not evolve significantly over the explored epoch. In contrast, \citet{lian_thickness_2024} used JWST short bands (observing $\sim 1/(1+z)\micron$), showing a strong decreasing trend in disc height towards the present.

The limitation of previous investigations for high-redshift discs, which did not require two disc components, is the use of a relatively short rest-frame waveband that can be affected by the presence of star formation at the observed epoch. James Webb Space Telescope (JWST) observations, with long filter bands (F277W, F356W, F444W), provide the rest-frame $K_s$ and $H$ bands for a wide cosmic epoch ($z\lesssim 2$, $\sim 10$ Gyr ago) with unprecedented sensitivity and resolution. These passbands are optimal tracers for stellar mass distribution, much less sensitive to stellar age and minimally affected by dust extinction and emission. This offers the best contrast between thin and thick discs and a comprehensive picture of the stellar mass assembly of the two discs.

In this paper, we construct an edge-on galaxy sample with a careful visual inspection using ever-increasing JWST public images and investigate the thin and thick disc structures for the first time at cosmological redshifts up to redshift $z=2$, roughly 10 Gyr ago, leveraging the rest-frame, near-infrared bands with significantly improved imaging capability. The paper is organised as follows; in Section~\ref{sec:data} we introduce the sample and parameter measurements, in Section~\ref{sec:results} we show the resulting disc parameters including scale height and scale lengths with galaxy parameters (e.g. stellar mass and redshift), in Section~\ref{sec:discussion} we discuss the results with respect to the literature on disc formation, we summarise our results in Section~\ref{sec:summary}.  In this paper, we adopt a Chabrier initial mass function \citep{chabrier_galactic_2003} and flat $\Lambda$-cold dark matter ($\Lambda$CDM) dominated cosmology with a present-day Hubble constant of $H_0 = 70$ km s$^{-1}$ Mpc$^{-1}$, and a density parameter of pressureless matter $\Omega_{\mathrm{M}}$ = 0.3.

\section{Methods}
\label{sec:data}
\subsection{Data and sample selection}
We use the publicly released mosaic images from the DAWN JWST Archive (DJA) which have been homogeneously reduced using the \textsc{grizli} pipeline \citep{brammer_grizli_2023}. The data used primarily comes from flagship JWST observational programs including JADES \citep{rieke_jades_2023}, FRESCO \citep{oesch_jwst_2023}, CEERS \citep{bagley_ceers_2023}, COSMOS-Web \citep{casey_cosmos-web_2023}, PRIMER \citep{dunlop_primer_2021}, and NGDEEP \citep{bagley_next_2024}.  A summary of the mosaic fields, associated JWST programs, PI names, and the version of DJA reduction used in the paper is given in Table \ref{tab:taba1}.

A parent sample is selected based on the \textsc{SExtractor} \citep{bertin_sextractor_1996} source catalogues for the JWST mosaic images and then matched with the 3D HST catalogue \citep{brammer_3d-hst_2012, momcheva_3d-hst_2016}. The JWST source catalogues are publicly available through DJA, compiled by running \textsc{SExtractor} on the JWST detection image, which is produced by combining available long wavelength filters (F277W+F356W+F444W; \citealt{valentino_atlas_2023}). We extract galaxy parameters from the 3D HST catalogue including redshift, stellar mass, and star formation rate, while parameters from the DJA catalogues include the apparent axial ratio of the galaxy $q$.  Galaxies are selected to be edge-on with an axis ratio $q = a/b < 0.3$, a stellar mass $M_{*}>10^{8.5}M_{\odot}$, and well separated from nearby sources by more than $1.5"$. This results in 213 possible sources. 

Matching sources against the existing 3D HST catalogue effectively removes erroneous sources from the \textsc{SExtractor} catalog, including the Point Spread Function (PSF) wing of bright stars and parts of nearby spiral galaxies. Moreover, we visually inspect all galaxies, removing cases where galaxies show spiral features (indication of the disc being slightly face-on), significant curvature, and lopsidedness - potentially due to tidal tails and warping. We use multi-wavelength bands (F090W, F115W, F150W, F200W, F277W, F356W, F444W) for visual inspection. For some galaxies, the shortest band (F090W; 0.9\micron) reveals a straight line dust attenuated feature for edge-on galaxies and spiral features for slightly face-on galaxies, helping us to identify edge-on galaxies and non-edge-on galaxies. After visual inspection, the sample includes 132 galaxies.

For the disc structural analysis, we employ NIRCAM F277W: 2.7\micron, F356W: 3.6\micron, F444W: 4.4\micron~filter images for galaxies with redshift $z<0.46$, $0.46<z<0.82$, $0.46<z<1.45$, respectively, to maximise overlap with the rest-frame $K_s$ band. For galaxies at $1.45<z<2.24$ and $2.24<z<3$, we use the F444W band, which corresponds to the rest-frame $H$ and $J$ bands, respectively. Those near-infrared bands are minimally affected by dust extinction and trace the stellar mass distribution of galaxies as the mass-to-light ratio is insensitive to the stellar populations \citep{gavazzi_phenomenology_1996, bell_stellar_2001, courteau_galaxy_2014}. The systematic differences in the disc height measurements at those different near-infrared bands are small for nearby galaxies \citep[][8\% bigger for $H$ band and 16\% bigger for $J$ band compared to $K_s$ band]{bizyaev_structural_2009}. This is further expected to be small for our galaxies which have limited stellar populations born in the relatively narrow range of the age of the universe $\sim$ 2.2 Gyr (for the galaxies at $1.45<z<3$, corresponding $9.1-11.4$ Gyr in lookback time).

In these long wavelength bands, we do not see any notable dust lane signatures commonly seen in shorter bands. We further removed 21 galaxies from the sample due to the unavailability of an appropriate waveband as described above, leaving 111 galaxies in the final sample. A total of 12, 46, and 53 galaxies have spectroscopic, grism, and photometric redshifts from the 3D HST catalog, respectively. Although we did not impose any redshift cut, the sample only extends to a redshift of 3, with a visually well-defined disc.

\begin{figure*}
    \centering
    \includegraphics[width=\linewidth]{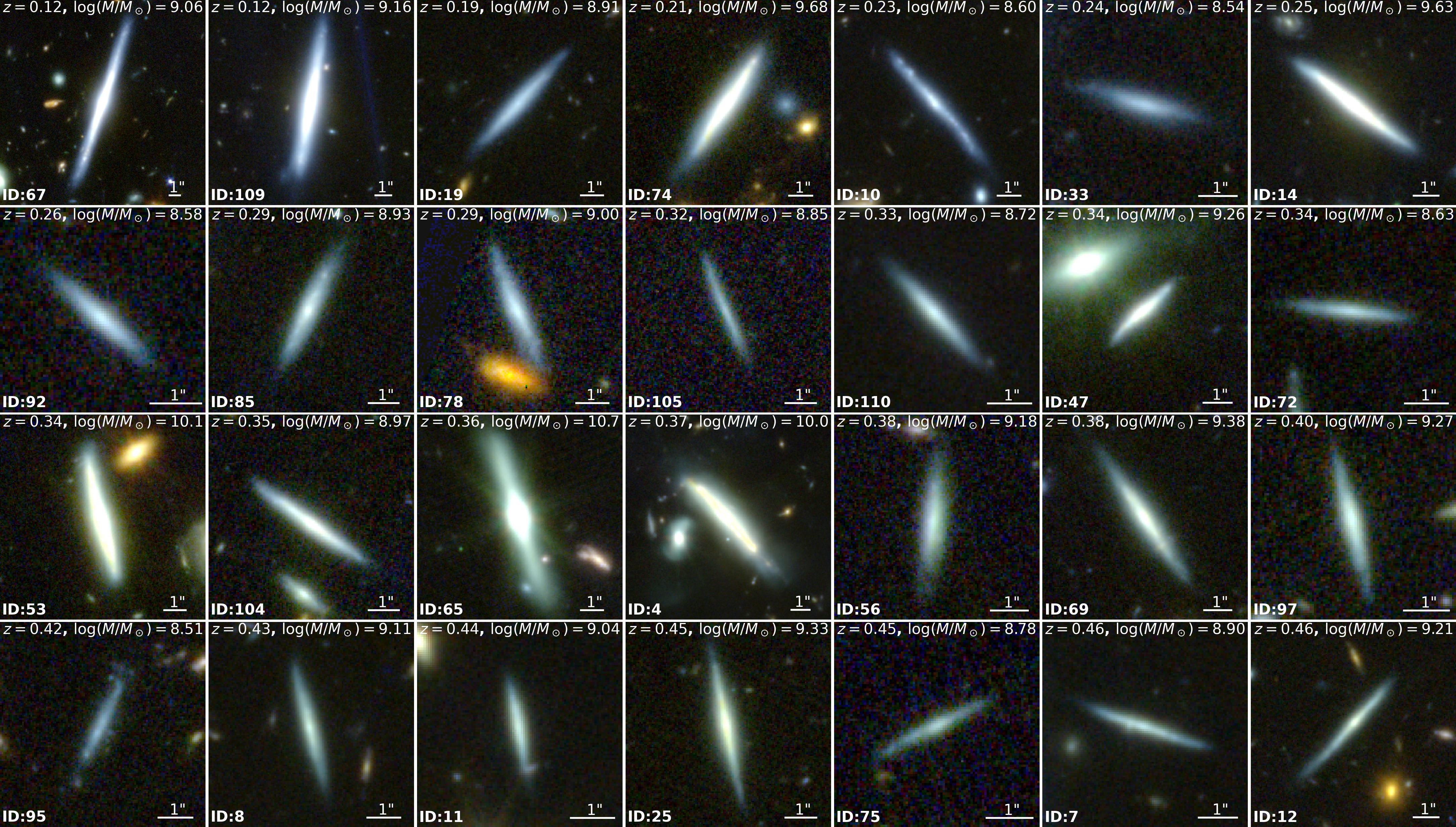}
    \caption{NIRCam F227W; F356W; F444W colour composite images of a quarter of our sample sorted by increasing redshift. The remainder of the sample is shown Fig.~\ref{fig:figb1}, \ref{fig:figb2} and \ref{fig:figb3}, in Appendix~\ref{sec:appendixb}. The white text on each image indicates the redshift and stellar mass from the 3D HST catalogue as well as the unique ID from this analysis in Table~\ref{tab:taba2}. 1" scale is denoted by a white bar in the lower right corner of each cutout.}
    \label{fig:fig001}
\end{figure*}

Figure \ref{fig:fig001} shows colour composite NIRCam images (F115W, F277W, F444W) for a quarter of our sample of the galaxies, where the rest of the galaxies are shown in Fig.~\ref{fig:figb1}, Fig.~\ref{fig:figb2}, and Fig.~\ref{fig:figb3} (Appendix~\ref{sec:appendixb}).

\begin{figure*}
    \centering
    \includegraphics[width=0.8\linewidth]{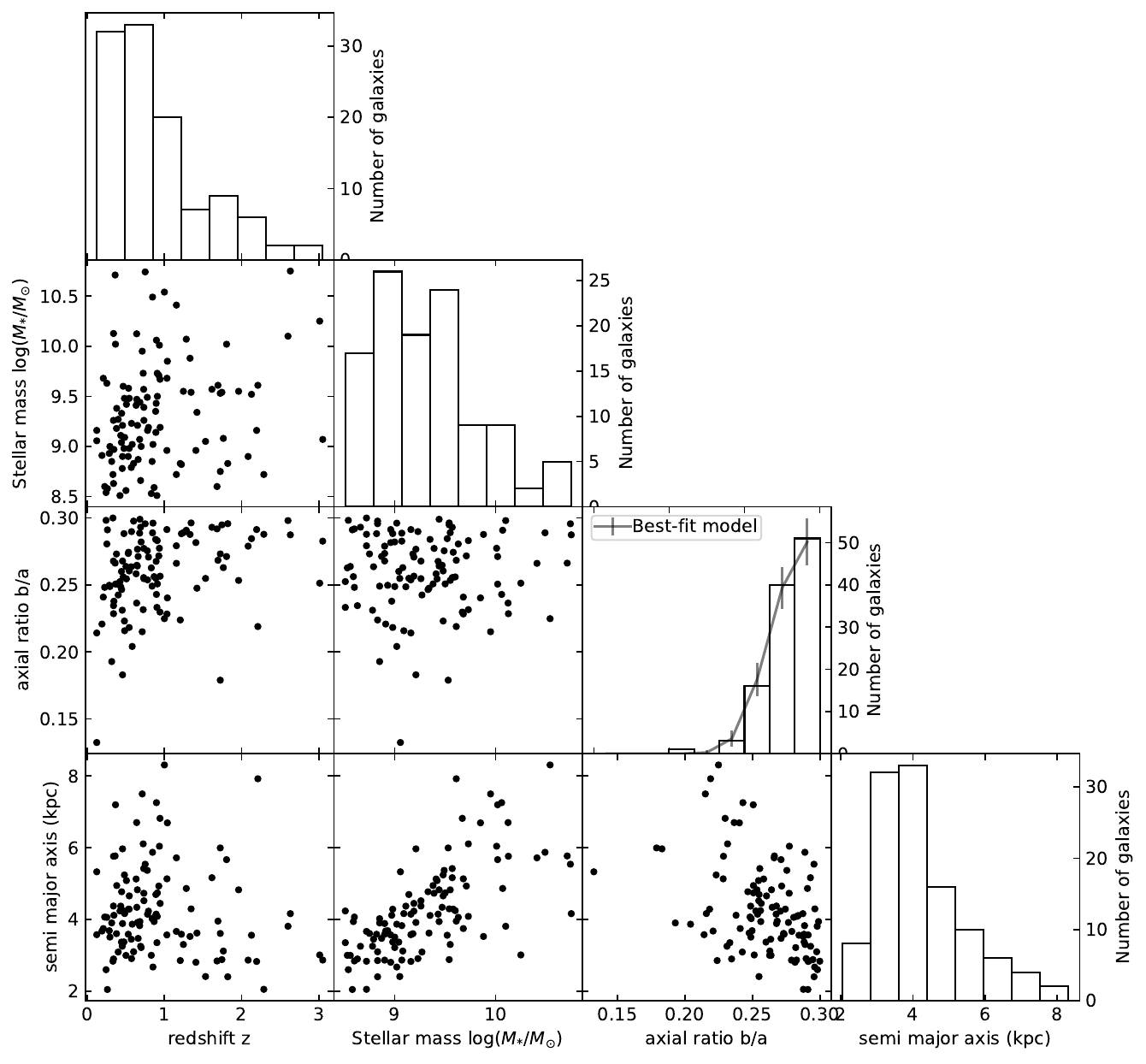}
    \caption{Summary of our 111 edge-on galaxies in redshift $z$, stellar mass $M_*$, apparent axial ratio $q=b/a$ and semimajor axis $a$. The scatter plots illustrate the relationships between each pair of parameters, while the histograms on the diagonal represent the number distribution of each parameter in our sample. The overlaid model line for the axial ratio $b/a$ is the best fit from our population model (Section~\ref{sec:model}).}
    \label{fig:fig002}
\end{figure*}

Figure \ref{fig:fig002} summarizes our galaxy sample distribution in stellar mass and redshift and morphological parameters. The stellar mass and redshift are extracted from the 3D HST catalogue while the apparent axial ratio $q=a/b$, semimajor axis $a$, semiminor axis $b$ is based on second order moment measurements of light distribution by \textsc{SExtractor} on the JWST detection images. The sample spans a redshift range of 0.1 to 3.0, corresponding to the lookback time of about 1.6 Gyr to 11.4 Gyr, $\sim70\%$ of the age of the universe, sampling a wide range of stellar masses at all redshifts.

\subsection{Inclination deviation from the edge-on orientation} \label{sec:model}

One can easily increase the sample size by allowing a higher axial ratio \citep[e.g., $q<0.4$:][]{hamilton-campos_physical_2023, lian_thickness_2024} than our study ($q < 0.3$ also adopted by \citealt[][]{elmegreen_thick_2017}). However, relaxing this limit may include many galaxies with large deviations from the perfectly edge-on case. 
To test this, we fit a probability distribution of the apparent axis ratios expected from the projection of a triaxial spheroid with principal axial length $A>B>C$ adopting random viewing angles \citep{binney_testing_1985}\footnote{The random viewing angles correspond to uniformly sampling $\cos(\theta)$ over $[-1, 1]$ and $\phi$ over $[0, 2\pi)$, where $\theta$ is the polar angle and $\phi$ is the azimuthal angle.}. We assume the galaxy's intrinsic axial ratio, $\gamma = C/A$, of the galaxy population follows a Gaussian distribution and $\epsilon = B/A$ follows a lognormal distribution \citep{ryden_ellipticity_2004}. Variables $\gamma$ and $\epsilon$ are separable for spiral galaxies: where $\gamma$ decides the shape of the distribution at small axial ratios $q$ and $\epsilon$ decides the shape at large axial ratios $q$. Our sample, with $q<0.3$, is only sensitive to $\gamma$. Therefore, we fix $\epsilon=0$ assuming the disc is circular (axis-symmetric) and obtain $\gamma=0.25\pm0.04$ (mean and standard deviation), which is consistent with the result obtained from the nearby disc galaxies \citep{ryden_ellipticity_2004}. Conversely, adopting typical lognormal $\epsilon$ distribution of disc galaxies \citep{ryden_ellipticity_2004} does not change the $\gamma$ value we obtained. 

\begin{figure}
    \centering
    \includegraphics[width=\linewidth]{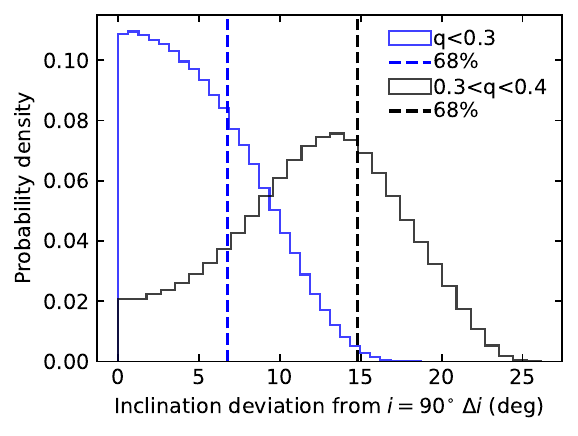}
    \caption{The probability distribution of inclination deviation $\Delta i$ from a perfect edge-on configuration ($i=90^\circ$) conditioned with the apparent axial ratio of the galaxies $q<0.3$ (blue; adopted in our sample), and $0.3<q<0.4$ (black; contribution if relaxing the axial ratio cut $q<0.4$, which outnumber galaxies with $q<0.3$). The distribution is based on the best-fit population model to the axial ratio $q=b/a$ number histogram in Fig.~\ref{fig:fig002}.}
    \label{fig:fig003}
\end{figure}

Based on the best-fit population model, Figure \ref{fig:fig003} shows the probability distribution of deviation from the perfect edge-on ($\Delta i$) for galaxies with apparent axial ratios of $q<0.3$ and $0.3<q<0.4$. Approximately 64 per cent of galaxies with $q<0.3$ have an inclination deviation of $\Delta i<7$ deg from the perfect edge-on orientation. In contrast, galaxies with $0.3<q<0.4$ have a median inclination deviation of $\Delta i\sim$ 13 deg. Additionally, galaxies with $0.3<q<0.4$ can outnumber those with $0.3<q$, comprising 70\% of total galaxies with $q<0.4$ in the model. This contamination of galaxies with larger deviations from edge-on poses significant challenges when measuring their structural parameters, such as vertical height, as these measurements become biased (see Section \ref{sec:3dmodel} and Appendix \ref{subsec:appendixc2}). To minimize potential biases in our structural measurements, we adopt a stricter axial ratio limit of $q<0.3$.

However, this choice may preferentially exclude bulge dominated galaxies and the thickest disc populations. Future work will address these population by incorporating higher-order information to better constrain inclination such as dust lanes, and radial changes in $q$, mitigating the bulge's effect on the global $q$ used in this study.

\subsection{Point spread function}
An accurate PSF is essential for studying the intrinsic light distribution of galaxies, particularly for identifying the faint, thick disc component, superimposed by the brighter thin disc component \citep{comeron_reports_2018}. We measured the effective PSF (ePSF) from each mosaic image, following star selection methods (\citealt{faisst_what_2022}, see also \citealt{ito_sizestellar_2024} and \citealt{zhuang_characterization_2024})\footnote{We used the \href{https://github.com/afaisst/JPP}{JWST PSF Pipeline} and its forked version \url{https://github.com/takafumi291/JPP}.}. For each field, we run \textsc{SExtractor} to identify candidate stars with source properties \texttt{class}\_\texttt{star} $>$ 0.8 and \texttt{elongation} $<$ 1.5. We select bright, unsaturated stars that lay on a horizontal locus in the magnitude-size plane, with a minimum signal-to-noise ratio (SNR) $>20$\footnote{This corresponds to AB magnitudes brighter than 22.5-23.5 depending on the filter and field.} and no brighter sources within 2". We then stacked the image cutouts of these selected stars using the Python package \textsc{Photutils} \citep{larry_bradley_2024_10967176, 2000PASP..112.1360A, anderson_empirical_2016}, maintaining the identical pixel sampling of the images to include the exact pixelization effect.

The measured ePSF per mosaic captures the point source response influenced by telescope jittering \citep{morishita_enhanced_2024}, the image drizzling process, and pixelization. This ePSF is expected to be broader than the simulated PSF by WebbPSF \citep{perrin_updated_2014}, which represents the intrinsic PSF properties without these effects. By fitting a Gaussian function to the ePSF, we measured the ePSF FWHM of 0.131", 0.146", and 0.166" for F277W, F356W, and F444W, respectively. The measured values do not change significantly across observations. Because the ePSF is only roughly approximated by a Gaussian, the measured ePSF widths depend on the pixel sampling; here, we used a pixel sampling size of 0.40". In physical scales, these FWHM corresponds to 0.3 kpc - 1.4 kpc with a median of 1.1 kpc in our galaxy sample, given the galaxy distance and band used. The limitation on the smallest measurable structural scale from the finite PSF width is evaluated in Sec.~\ref{sec:3dmodel} and in Appendix~\ref{sec:appendixc}.

\subsection{3D fitting and model selection}
\label{sec:3dmodel}

To measure the disc properties (thickness, size, etc) of our edge-on galaxy sample, we use a 3D disc model where the disc luminosity density, $\nu(R, z)$, follows an exponential profile radially and a $\sech^2$ profile vertically. In cylindrical coordinates ($R$, $z$), $\nu(R, z)$ is expressed as: 
\begin{equation}
    \nu(R, z)=\nu_0 \exp(-R/h_R) \sech^{2} (z/(2z_0))
\end{equation}
 where $\nu_0$ is the central luminosity density, and $h_R$ and $z_0$ represent the disc scale length and height, respectively. The $\sech^2$ profile, a solution for the self-gravitating isothermal sheet \citep{spitzer_dynamics_1942}, has been widely used to effectively describe and approximate the vertical thickness of galactic discs \citep{van_der_kruit_three-dimensional_1988, yoachim_kinematics_2008}.
The generalized function, $\sech^{2/N}$, proposed by \citet{van_der_kruit_surface_1982} describes non-isothermal disc profiles. The model varies from the standard $\sech^2$ profile only near the midplane \citep{van_der_kruit_surface_1982}, becoming either peaked or smoothed as $N$ approaches 1 or $\infty$, respectively. Despite these differences, all variants of the model asymptotically converge to the exponential profile $\exp(-z/z_0)$ at large distance, $z$, from the midplane. In the limiting case where $N\rightarrow\infty$, the model matches the exponential profile $\exp(-z/z_0)$ at all radii. 

For this study, we use the $\sech^2$ profile for consistency with existing literature. Detailed modelling of the midplane profile is beyond our scope, largely due to dust absorption and the limited resolution of JWST. We also assume an inclination of $90^{\circ}$. Our population model predicts that 64\% of our galaxies deviate from $90^{\circ}$ by no more than $7^{\circ}$ (see Section~\ref{sec:model}). Such a slight deviation introduces bias in the structural parameter measurements, which is known to be small \citep{de_grijs_z-structure_1997}. However, the impact depends on the intrinsic axial ratio of the disc and image quality (e.g., signal to noise ratio and instrumental resolution relative to the source extent). Section~\ref{uncertainty} discusses systematic uncertainties from inclination and other factors for structural measurements.

To model galaxy surface brightness distributions and measure the structural properties, we use the \textsc{imfit} package \citep{erwin_imfit_2015} throughout the paper. \textsc{imfit} projects the 3D luminosity along the line of sight onto 2D surface brightness in the sky, convolves with the user-supplied PSF, and then minimizes $\chi^2$ with the Levenberg–Marquardt algorithm. It takes the image, variance, and mask (to define the fitting region) as inputs and outputs the best-fit parameters and fit metrics such as the best fit $\chi^2$ and Bayesian Information Criterion (BIC). Additionally, \textsc{imfit} allows for the superposition of multiple components such as a second disc or 2D S\'ersic profile, which we employed in later stages. 

First, we fit a single 3D disc model to each cutout JWST image for the galaxy sample at the selected waveband for maximal overlap with rest-frame $K_s$ band (e.g., F277W; $z<0.46$, F356W; $0.46<z<0.82$, F444W; $0.82<z$). We set the lower boundary for the disc scale height to 0.2 pixels, based on our recovery experiment (see Appendix~\ref{subsec:appendixc1}). If the parameter reaches this boundary, we interpret it as a upper limit on the scale height. The best-fit single-disc model captures most of the galaxy's total flux in the data. However, some galaxies show systematic residual patterns in a model-subtracted data map. The most notable are disc-like excess light off the midplane and compact bulge components in the centre of galaxies. 

Secondly, to account for these structures and measure their structural properties, we consider two additional components: 1) a 2D S\'ersic profile to account for the centrally concentrated light and 2) a second thicker 3D disc component to account for the disc-like excess light. For the first component, we allow the S\'ersic index to range from 0.5 to 5, accommodating the range from classical bulges to disc-like bulges \citep{kormendy_secular_2004}. 

To find the best structural fit for each galaxy, we fit a series of increasingly complex models that include the second disc and central S\'ersic component using \textsc{imfit}. These include (i) the single disc model described above; (ii) two disc model (thin+thick); (iii)  a disc+bulge model; and (iv) a two disc (thin+thick) + bulge model. To avoid local minima in the $\chi^2$ landscape for the multiple component fits, we repeated Levenberg-Marquard optimization 15 times with initial starting points randomly drawn from a conservative wide range of parameters based on the single disc fits. To avoid erroneous solutions for the bulge being fit to the disc structure, and vice versa for the model including the bulge component, the $\chi^2$ minimum solution was bounded by a constraint: the bulge effective radius is smaller than both 2 kpc and the disc scale length. 

To ensure robust fits with well-behaved components motivated by data, we use the following procedures. We determine the necessity for additional components for each galaxy by comparing the BIC across models. We assigned the galaxies into 4 models, from the simplest to more complex models as described above. A more complex model is justified if the BIC improvement exceeds 15. We visually inspect the fit results using the model-subtracted residuals, checking each structural component fits the intended structure. 

Although the BIC classification is generally robust, we reclassify 6 galaxies from `two discs + bulge' to `a disc + bulge', and 1 from `two discs' to `single disc', based on the visual insignificance of the thick disc component.\footnote{In all 6 cases where classification changed based on visual inspection, the adopted classification corresponds to the second-best model when evaluating the BIC. Additionally, these 6 galaxies had some of the smallest differences in BIC between the best and second-best models among the full sample of galaxies.} One galaxy initially categorized as "single + bulge" was adjusted to `single disc', while three others were reclassified to `two discs' (2) and `two discs + bulge' (1) because the bulge component fits the visually significant disc feature. These 11 re-classifications are only a minor subset of the 111 galaxies $-$ eliminating them instead did not alter our conclusions of the paper. 

\begin{figure*}
    \centering
    \includegraphics[width=0.92\linewidth]{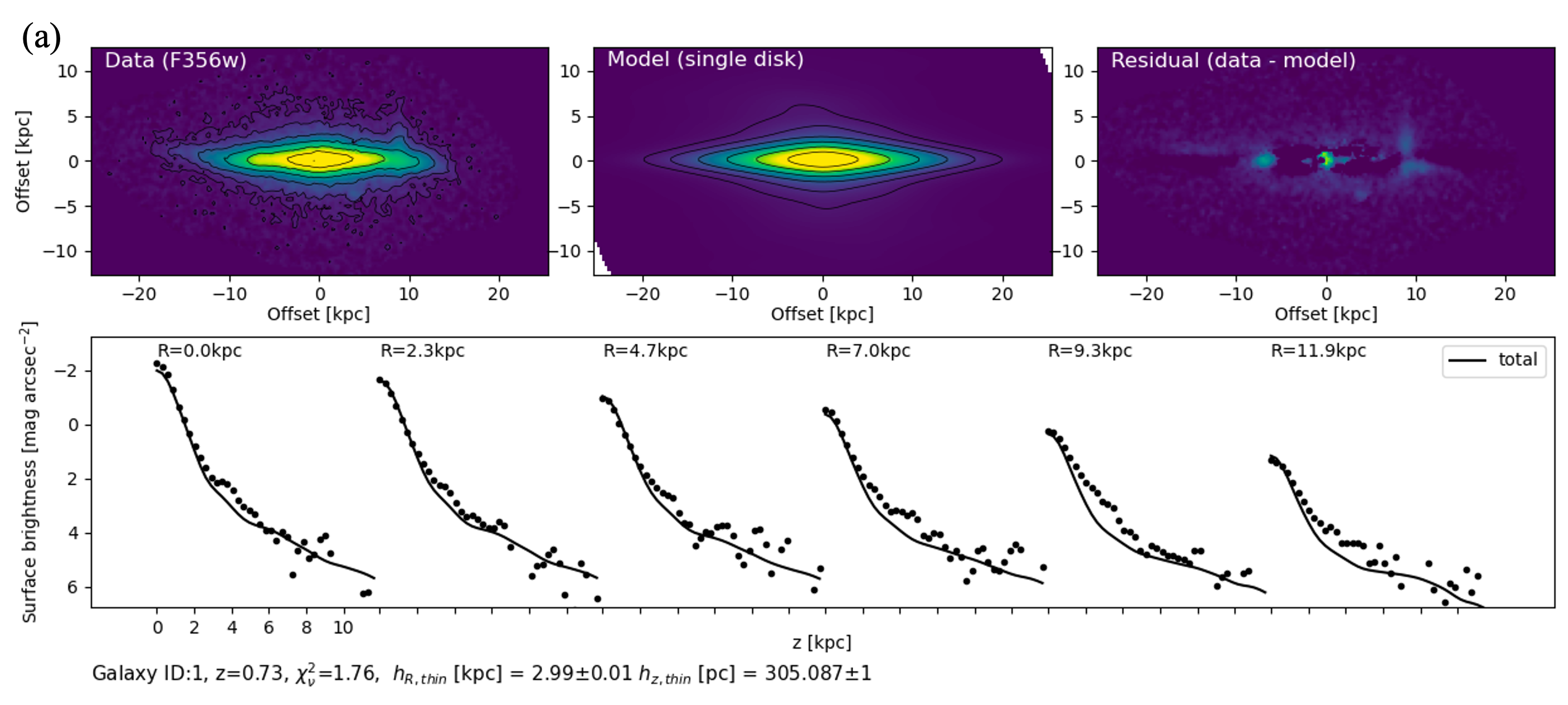}\vspace{-2.5mm}
    \includegraphics[width=0.92\linewidth]{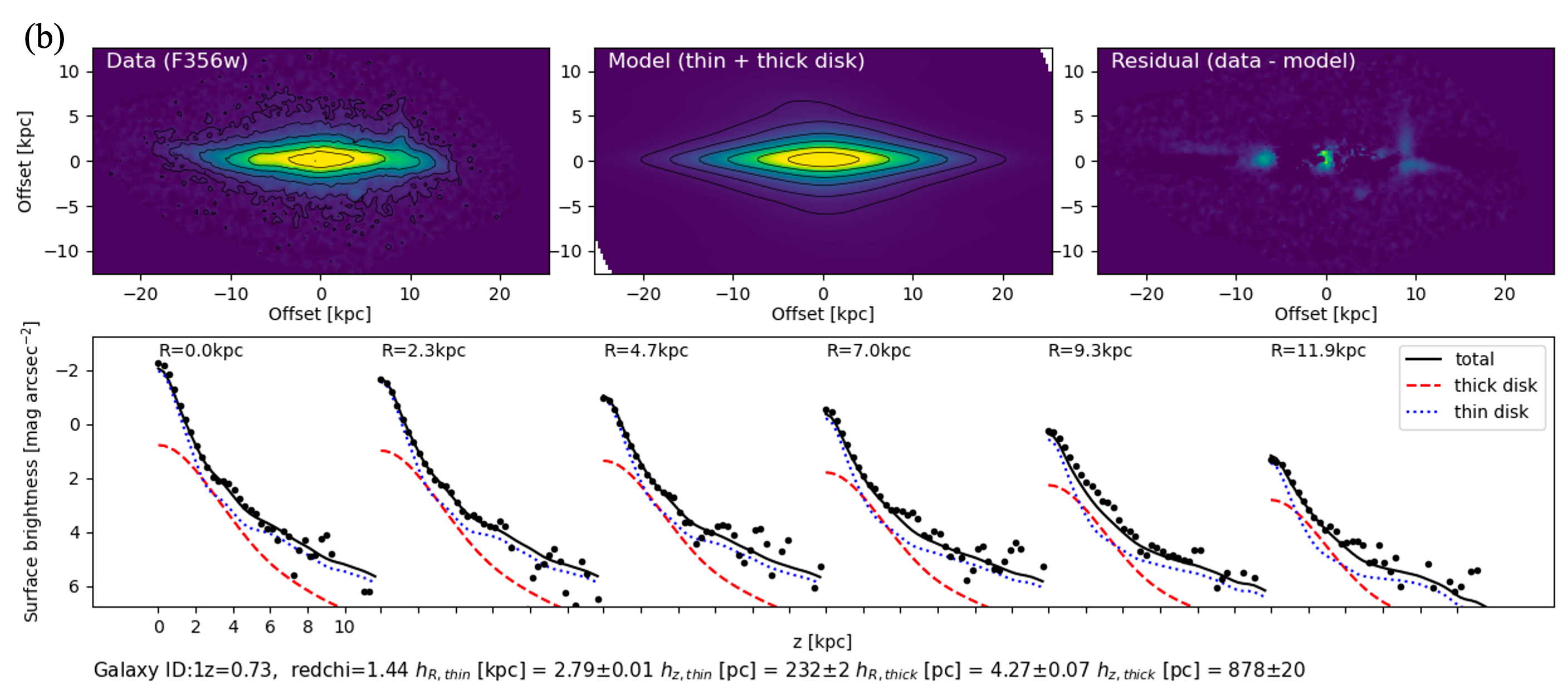}\vspace{-2.5mm}
    \includegraphics[width=0.92\linewidth]{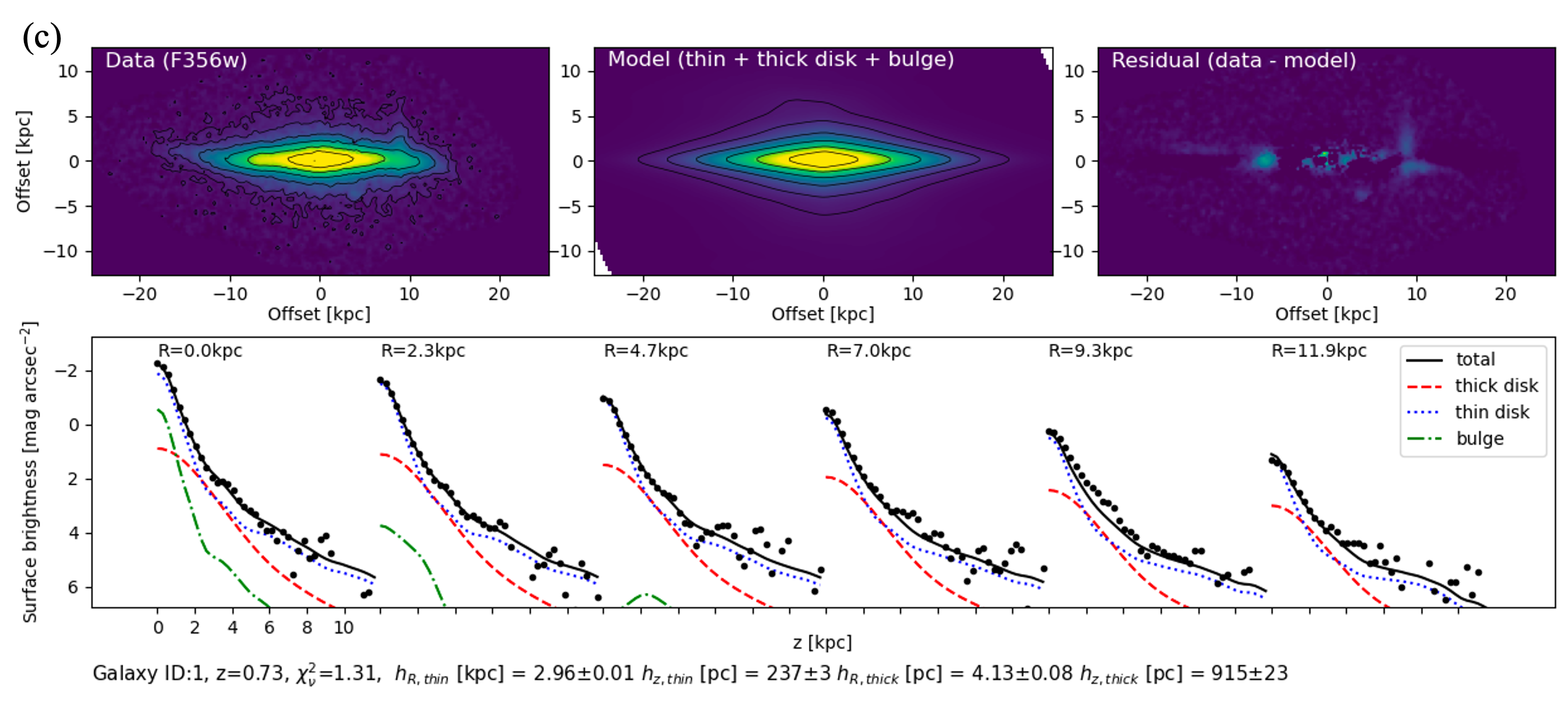}
    \caption{An example fit of a galaxy (ID=1) in our sample, with `two discs + bulge' model being the best model according to our criteria (Section~\ref{sec:3dmodel}). The three groups of panels from top to bottom display the best fit with (a) a single disc, (b) two discs, and (c) two discs + a bulge, respectively. Each group of panels show from left to right the data, model, and residual in the top row and the vertical surface brightness profiles at different radii in the bottom. \added{All images and profiles are PSF-convolved.} Each profile is extracted at the radius indicated above it. Assuming the two-fold symmetry of the disc galaxies, we show the average profile of four quadrants. The galaxy's redshift $z$, $\chi^2_\nu$ of each fit and derived disc parameters are also shown. Notably, there remains significant excess light above the disc midplane for the single disc model, which is eliminated by adding a second disc component. The central excess is accounted for by two discs + bulge models, but even ignoring it entirely, the derived parameters do not change by more than 20\%, consistent with the conservative systematic error of our sample (Appendix~\ref{sec:appendixc}).}
    \label{fig:fig004}
\end{figure*}

Figure~\ref{fig:fig004} shows an example galaxy (ID=1) best fitted with a `two discs + bulge' model alongside fits with a single disc (top), two discs (middle), and two discs + a bulge (bottom) models. For this galaxy, the `single disc + bulge' model provides a similar fit to two disc models as the bulge becomes the thin disc component. We omit it here to avoid redundancy. \added{Fitting a single disc model leaves systematic flux excesses above the disc midplane, clearly seen in both the residual image and vertical profile (Fig.~\ref{fig:fig004}a). Adding a second disc component successfully accounts for this excess light (Fig.~\ref{fig:fig004}b). The convolved profiles reveal diversity amongst the sample. Some galaxies are dominated by either thin or thick discs, while others transition between thin/thick dominance at a certain height. However, it is difficult to assess the relative dominance of the thin and thick discs in the PSF-convolved vertical profiles shown in Fig.~\ref{fig:fig004}. PSF wings can cause the thin disc light to extend to large heights. Characterising the PSF-deconvolved vertical profile of each component will be left for future work, which will require verifying the deconvolved profiles and assessing the uncertainties propagated from both the assumed model and statistical noise.
}

\subsection{Uncertainties for structural measurements}
\label{uncertainty}
JWST data allows us to measure the disc structural properties with high precision. We estimated statistical uncertainties using Levenberg-Marquardt optimization in IMFIT and verified them by bootstrap resampling, refitting the resampled data 500 times. We present our results with these statistical uncertainties throughout the paper. We also separately estimate systematic uncertainty arising from our assumptions and simplifications, as detailed in Appendix~\ref{sec:appendixc}, to help readers interpret our measurements and their limitations.

As the model is not always perfect, higher-order structures remain unmodelled, such as the outer truncation of the disc, and disc substructure (bar end or clump, see Fig.~\ref{fig:fig004} for residual at $-10$ kpc). Although dust extinction is not evident in our chosen bands, it may also bias our results. Such challenges are noted in previous works for nearby galaxies \citep{yoachim_structural_2006}. To test the robustness of our results, we have run additional tests that includes masking different regions of the galaxies (e.g., disc midplane, centre, outside typical truncation radius) and systematic effect (e.g., inclination, potential dust extinction). In summary, systematic uncertainty, mainly arising from deviations in the inclination from 90$^\circ$, results in median underestimation of 0.1\%, 12\%, 1\% for scale radii and median overestimations of 17\%, 27\%, 9\% for the scale heights for single, thin, thick disc respectively. Experiments masking the bulge region rather than modelling it confirm that the bulge structure does not influence the measured properties of the discs. The systematic effects are not significant compared to the reported trends in Section~\ref{sec:results} and as such do not affect our conclusions.

\section{Results and analysis}
\label{sec:results}
Of the 111 galaxies in our sample, we find that 28 galaxies are well fit by a single disc component, 39 are best fit by a single disc +  bulge, 19 are best fit by two disc components, and 25 are best fit with two disc components + bulge.  For the remainder of this work, we only focus on the disc components. We include galaxies that are well fitted by a single disc, with or without a bulge model, in our `single disc' category (67), which spans a redshift range $z=[0.2, 3]$. Similarly, we include galaxies with two discs, with or without a bulge, in our `two discs' category (44), which spans a redshift range $z=[0.1, 2]$. For galaxies with best fits including two disc components, we refer to the disc with the shorter scale height as a `thin disc' and the disc with the larger scale height as a `thick disc'.
We summarize structural parameters derived in Table \ref{tab:taba2} and show corner plots of all the measured parameters in Figs.~\ref{fig:fige1} and \ref{fig:fige2}.

When deriving the mass of individual disc components, we assume the same mass-to-light ratio (M/L) at rest-frame $K_s$/$H$ bands for all sub-components in the best-fit model including bulge, single disc, thin disc, and thick disc. With this assumption, each component's mass is derived using the luminosity ratio and the total stellar mass estimated in the 3D HST catalogue \citep{momcheva_3d-hst_2016}. The bulge component is not dominant in our galaxies, contributing only 2\% of the total luminosity in the median. Therefore the bulge's contribution is almost negligible but is accounted for as described above to derive the individual disc mass. 

This implies the relative mass-to-light ratio of thin and thin discs in the bands we used, $\Upsilon_\mathrm{thin}/\Upsilon_\mathrm{thick}$, is equal to 1. This allows for easy adjustments to the mass ratio in future studies when updated color or spectroscopic information constraining stellar population becomes available. However, in Sec. \ref{subsec:massratio}, we assume a value of 1.2 for comparison with previous studies, as suggested by the realistic star formation history of the Milky Way's thin and thick discs (see Appendix~\ref{sec:appendixd}). This slight adjustment does not affect the overall discussion.

\subsection{Scaling relations for disc scale length and scale height} 
\label{subsec:scalrel}
In this subsection, we investigate the dependence of the disc structural parameters (radial length, scale height, and the ratio of the two) on the total mass 
or the individual disc mass. 
\subsubsection{Existence of radial/vertical size - mass correlation}
\begin{figure*}
    \centering
    \includegraphics[width=\linewidth]{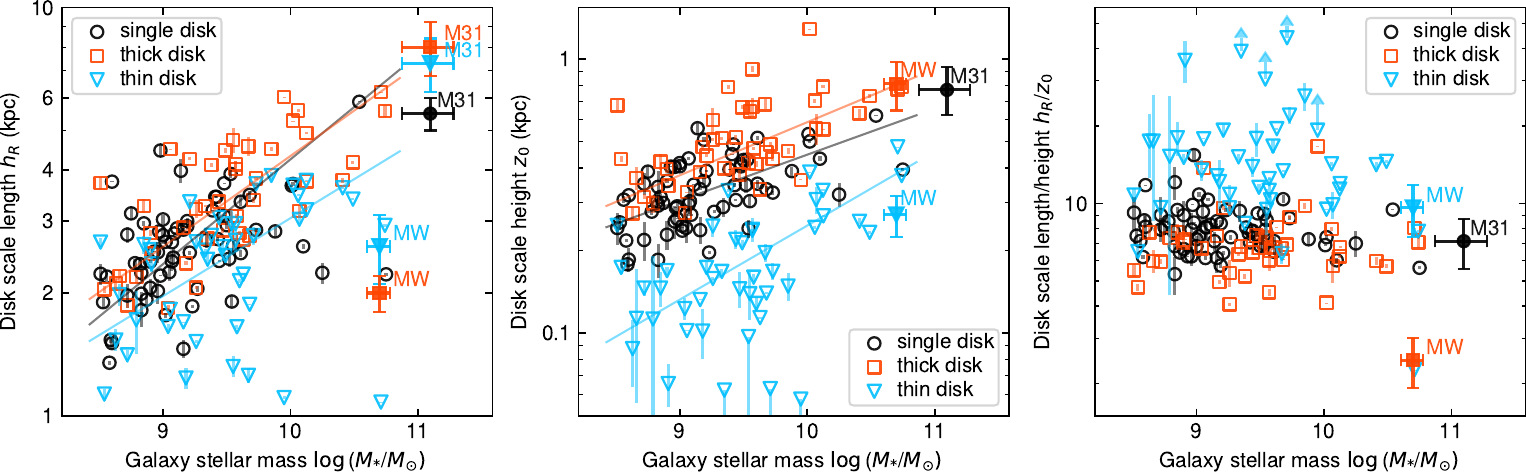}
    \caption{Disc geometrical parameters $h_R$ (left), $z_0$ (middle) and $h_R/z_0$ (right) are plotted against galaxy stellar mass $M_{*}$. The symbols used are: open black circles for single discs, open blue triangles for thin discs, and open red squares for thick discs in our galaxy sample, covering redshifts from $\sim0.1$ to 3. Similar filled symbols represent the thin and thick discs of the Milky Way \citep{bland-hawthorn_galaxy_2016} and M31 \citep{collins_kinematic_2011}. The measurements assuming a single disc component of M31 by \citet{dalcanton_panchromatic_2023} are also shown.}
    \label{fig:fig005}
\end{figure*}

Figure \ref{fig:fig005} shows the geometrical properties of discs, i.e., disc scale length ($h_R$), scale height ($z_0$), and the ratio ($h_R/z_0$) plotted against the total stellar mass of the galaxies ($M_{*}$). There is a positive correlation in the $M_* - h_R$ (left) and $M_* - z_0$ (middle) planes for all disc categories, suggesting, as expected, more massive galaxies have larger and thicker discs.  However, there is no evident correlation in the $M_{*} - h_R/z_0$ plane (right), where thin discs are separated from both single and thick discs to higher $h_R/z_0$. 

\begin{figure*}
    \centering
    \includegraphics[width=\linewidth]{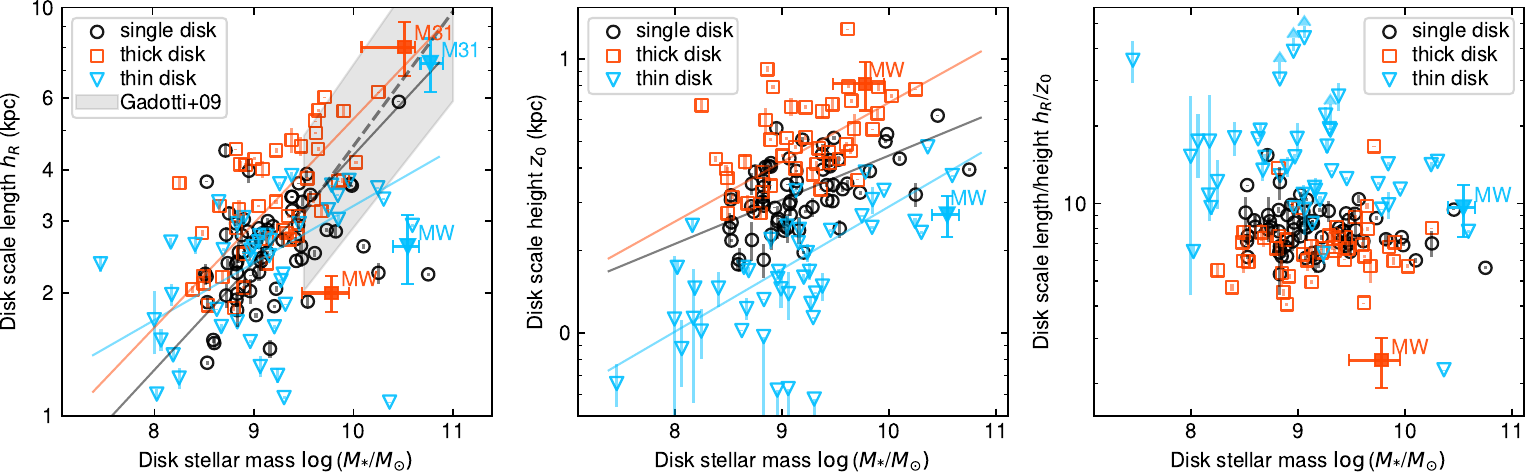}
    \caption{Same symbols as used in Fig.~\ref{fig:fig005}, but plotted against the stellar mass of each component rather than the galaxy's total stellar mass. In the left panel showing disc scale length $h_R$, the scaling relation for the $z=0$ SDSS sample derived by \citet{gadotti_structural_2009} is shown. We multiply the original relation by 2 to account for suggested systematics in \citet{boardman_are_2020} due to the galaxy's orientation (see text).}
    \label{fig:fig006}
\end{figure*}

When $h_R$ and $z_0$ are plotted against individual disc mass, $M_\mathrm{disc}$, as shown in Fig.~\ref{fig:fig006}, rather than the galaxy's total mass, $M_{*}$, the distributions of single, thin, and thick discs overlap more significantly, suggesting that the disc mass rather than the total galaxy mass is more fundamental to characterize the disc properties. 

In Figs~\ref{fig:fig005} and \ref{fig:fig006}, we show power-law fits for each population, where the length and height are proportional to the power of stellar masses, $M_*^\beta$. Table \ref{tab:tab001} summarizes the measured slope $\beta$ and intercept $\alpha$ with the Spearman's rank correlation coefficients and associated p-values against the null hypothesis that there is no correlation between the parameters. To find the best fits, robust to outliers, we used LTSLINEFIT \citep{cappellari_atlas3d_2013}, the implementation of the least trimmed squares (LTS) approach \citep{rousseeuw_robust_1987, rousseeuw_computing_2006}. The method effectively finds the global minima of the $\chi^2$ computed for all possible subsets of the data excluding potential outliers up to half of the sample. This method can also take into account the uncertainty for all coordinates. We used the statistical uncertainty for our disc length and height measurements and 0.2 dex uncertainties for stellar masses \citep{schreiber_constraints_2011}.

\begin{table}
    \centering
    \caption{Best fit linear relation to the disc geometrical parameters for a thick, thin, single disc with galaxy's total stellar mass $M_{*}$ or corresponding disc stellar mass $M_{\mathrm{disc}}$. For a geometrical parameter $X$ and either mass $M$, the linear relation $\log(X/\mathrm{kpc})= \alpha + \beta (\log (M/M_{\odot})-9.5)$ is fitted to find $\alpha$ and $\beta$. We chose the pivot mass of 9.5, close to the median disc mass, which does not affect the best-fit values and only weakly influences uncertainty. The 1$\sigma$ observed scatter around the best fit $\Delta$ is also shown. Additionally, for each parameter pair, the Spearman correlation coefficient $r$ is reported along with the p-value for the null hypothesis in parentheses.}
    \tabcolsep=0.15cm
    \begin{tabular}{|c|c|c|c|c|}
        \hline
        Parameters pair & $\alpha$ & $\beta$ & $\Delta$ & Spearman $r$\\
    	\hline
    	Single disc & & &\\
    	\hline
    	$h_R - M_{*}$ & $0.50 \pm 0.01$ & $0.26 \pm 0.03$ & $0.08$ & $0.44$ ($2\times10^{-4}$)\\
    	$z_0 - M_{*}$ & $-0.43 \pm 0.02$ & $0.17 \pm 0.02$ & $0.10$ &$0.53$ ($4\times10^{-6}$)\\
        $h_R/z_0 - M_{*}$ & - & -  & - & $-0.14$ ($0.26$)\\
    	$h_R - M_{\mathrm{disc}}$ & $0.51 \pm 0.02$ & $0.26 \pm 0.03$ & $0.08$ & $0.36$ ($2\times10^{-3}$)\\
    	$z_0 - M_{\mathrm{disc}}$ & $-0.43 \pm 0.02$ & $0.16 \pm 0.03$ & $0.10$ & $0.49$ ($3\times10^{-5}$)\\
        $h_R/z_0 - M_{\mathrm{disc}}$ & - & -  & - & $-0.15$ ($0.23$)\\
    	\hline
    	Thin disc & & &\\
    	\hline
    	$h_R - M_{*}$ & $0.37 \pm 0.02$ & $0.20 \pm 0.04$ & $0.13$ & $0.44$ ($3\times10^{-3}$)\\
    	$z_0 - M_{*}$ & $-0.74 \pm 0.02$ & $0.27 \pm 0.04$ & $0.14$ &
        $0.58$ ($8\times10^{-5}$)\\
        $h_R/z_0 - M_{*}$ & - & -  & - & $-0.20$ ($0.22$)\\
    	$h_R - M_{\mathrm{disc}}$ & $0.43 \pm 0.03$ & $0.18 \pm 0.04$ & $0.13$ &
        $0.44$ ($3\times10^{-3}$)\\
    	$z_0 - M_{\mathrm{disc}}$ & $-0.65 \pm 0.02$ & $0.23 \pm 0.03$ & $0.12$ &
        $0.65$ ($5\times10^{-6}$)\\
        $h_R/z_0 - M_{\mathrm{disc}}$ & - & -  & - & $-0.26$ ($0.11$)\\
    	\hline
    	Thick disc & & &\\
    	\hline
    	$h_R - M_{*}$ & $0.53 \pm 0.02$ & $0.22 \pm 0.03$ & $0.10$ & $0.66$ ($1\times10^{-6}$)\\
    	$z_0 - M_{*}$ & $-0.33 \pm 0.01$ & $0.19 \pm 0.03$ & $0.09$ & $0.58$ ($3\times10^{-5}$)\\
        $h_R/z_0 - M_{*}$ & - & -  & - & $0.09$ ($0.58$)\\
    	$h_R - M_{\mathrm{disc}}$ & $0.60 \pm 0.02$ & $0.26 \pm 0.04$ & $0.11$ & $0.62$ ($6\times10^{-6}$)\\
    	$z_0 - M_{\mathrm{disc}}$ & $-0.27 \pm 0.02$ & $0.22 \pm 0.04$ & $0.10$ & $0.49$ ($7\times10^{-4}$)\\
        $h_R/z_0 - M_{\mathrm{disc}}$ & - & -  & - & $0.17$ ($0.27$)\\
    	\hline
    \end{tabular}
    \label{tab:tab001}
\end{table}

The radial scaling relations, $h_R-M_*$ and $h_R-M_\mathrm{disc}$, show similar slopes, $\beta$, for thick and single discs, while thin discs have a shallower slope, suggesting a difference in their build-up process (i.e. a different total angular momentum history, \citealt{mo_formation_1998}). The slopes measured here for all disc and galaxy masses are similar to what was found in \citet{van_der_wel_3d-hstcandels_2014}. Fits to the vertical height scaling relations, $z_0-M_*$ and $z_0-M_\mathrm{disc}$, reveal similar slopes for all disc types suggesting that the disc scale height may be determined by local vertical equilibrium. In fitting $h_R/z_0 - M_{*}$ and $h_R/z_0 - M_{\mathrm{disc}}$, there is no statistically significant correlation (see p-values in Table~\ref{tab:tab001}) for any disc type, so we do not attempt a fit to the data. 

\subsubsection{Galaxies form a thick disc first, then a thin disc}\label{subsec:sequential}
Notably, the thick discs and single discs show significant overlap in all diagrams (Figs.~$\ref{fig:fig005}$ and $\ref{fig:fig006}$). This similarity implies that single discs correspond to the thick discs observed in galaxies with two disc components. This may suggest that most galaxies initially form a thick disc, which is observed as a single disc, followed by the later formation of a thin disc. This observational insight of the sequential formation is further explored by later analyses throughout this paper (Sec.~\ref{subsec:doubleevo}, \ref{subsec:subsec33}, \ref{subsec:massratio}, \ref{sec:discussion}). 

\subsubsection{Comparison with other studies (z$\sim$0, MW, high-z)}
The distribution of the thick and single disc populations in the $h_R-M_{\mathrm{disc}}$ plane are consistent with an extrapolation of measurements at $z\sim0$ for SDSS disc galaxies \citep{gadotti_structural_2009} after the following correction. Radii measured for edge-on galaxies are typically $\sim2$ times larger than those measured for face-on galaxies \citep{boardman_are_2020}, presumably due to unaccounted projection effects and the increased sensitivity of the edge-on configuration. Accordingly, we multiplied the relation originally derived in \citet{gadotti_structural_2009} for $z\sim0$ galaxies by 2. Some of the scatter in the $h_R-M_*$ relation may partly result from the mild size evolution across the large range of redshifts covered by this sample z $\sim$ 0.1 to 3 (e.g. \citealt{van_der_wel_3d-hstcandels_2014}), which is discussed in the next section.

The scaling relations $z_0-M_{*}$ defined above align well with the Milky Way’s values \citep{bland-hawthorn_galaxy_2016}\footnote{These values enclose the recent determination of thick and thin disc properties of MW using a rather similar approach in this study: decomposing the edge-on integrated light of MW \citep{mosenkov_structure_2021}.} and M31’s values \citep{dalcanton_panchromatic_2023}, as shown in Fig.~\ref{fig:fig005}. However, the scale length $h_R$ of the Milky Way thin and thick discs are comparatively shorter than in the scaling relation $z_0-M_{*}$, confirming the long-standing notion that the Milky Way's disc scale-length is shorter than the typical disc scale length expected for a galaxy of its stellar mass \citep{licquia_does_2016, boardman_are_2020}. However, we note that being inside of the Milky Way makes radial length measurement uncertain \citep{bland-hawthorn_galaxy_2016}. Recently, \citet{lian_size_2024} suggested that the Milky Way's disc exhibits a broken-exponential profile which could lead to an underestimated scale length by the steeply declining outer part, potentially reconciling the observed discrepancy with other galaxies.

The measured $z_0-M_{*}$ for single and thick discs in this sample are well aligned with the $z=0.5-3.5$ results from \citet{elmegreen_thick_2017} for edge-on galaxies using the $HST$ / F814W filter (0.8\micron) in two Frontier Field Parallels, assuming a single disc fit\footnote{Note the difference of the adopted definition of $H$ in \citet{elmegreen_thick_2017}, their values need to be halved to compare with our data, $z_0=H/2$ equivalent to an exponential scale height).}. However, \citet{lian_thickness_2024} reports no significant $z_0-M_{*}$ correlation for a sample of galaxies at $z=0.2-5$ using the F115W filter (1.2\micron) of $JWST$ / NIRCAM. Both studies measured disc thickness assuming a single disc component. The discrepancy may be attributed to differences in the selection criteria for edge-on galaxies. Our study and that of \citet{elmegreen_thick_2017} adopt stricter criteria, selecting galaxies with an axial ratio of $< 0.3$ and visually eliminating those with warping and tidal tail structures. In contrast, \citet{lian_thickness_2024} uses an axial ratio of $<0.4$, which may result in a sample dominated by the galaxies with axial ratio between $0.3-0.4$ (See Fig.~\ref{fig:fig003} and associated discussion in Section~\ref{sec:model} as well as Figure 3 in \citealt{hamilton-campos_physical_2023}) and such measurements may be affected by warping and tidal structure. 

\subsection{Evolution of disc scale length and scale height}
\subsubsection{Single discs}
\label{subsec:singleevo}
\begin{figure*}
    \centering
    \includegraphics[width=\linewidth]{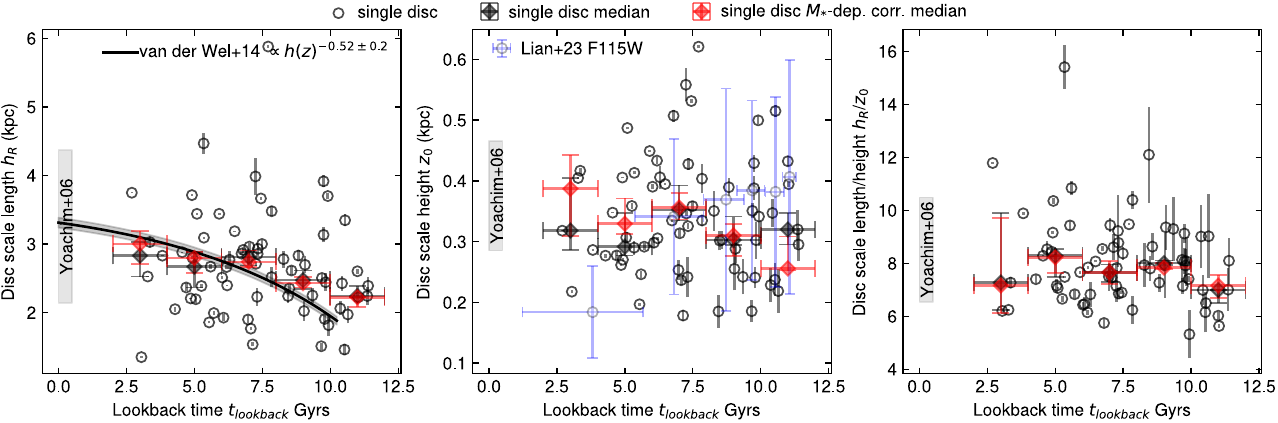}
    \caption{The properties of single discs, $h_R$ (left), $z_0$ (middle) and $h_R / z_0$ (right) are plotted against galaxies' lookback time $\tau_{\mathrm{lookback}}$. This figure includes only galaxies classified as having a single disc or single disc with a bulge. Black circles show individual measurements with associated $1\sigma$ statistical uncertainty. The large black diamonds indicate the median in each lookback time bin of 2 Gyr widths, with confidence intervals estimated via bootstrap resampling. The red diamonds show the same but linearly corrected for the relation seen in $\log M_{*}-\log h_R$ and $\log M_{*}- \log z_0$ to a median $\log (M_{*}[$M$_\odot])=9.2$ in each bin. The evolutionary trend from \citealt{van_der_wel_3d-hstcandels_2014} for galaxies with stellar mass $\log (M_{*}[$M$_\odot])=9-9.5$  is shown by a black line with an associated 1$\sigma$ confidence interval for the median. The grey shading near $\tau_{\mathrm{lookback}}\sim0$ shows the range of values derived by \citealt{yoachim_structural_2006} for z$\sim0$ sample. The lookback time evolution of $z_0$ from \citealt{lian_thickness_2024} is shown in blue.}
    \label{fig:fig007}
\end{figure*}

Figure \ref{fig:fig007} shows the disc scale length, $h_R$, scale height, $z_0$, and the ratio $h_R/z_0$ as a function of lookback time for single disc galaxies. This subsample provides a good reference to see how the size and thickness of single discs evolve and to be compared with the literature fitting single components, without potential systematics from the thin/thick disc decomposition. We include individual measurements (black circles) as well as the median in each lookback time bin of 2 Gyr widths (black diamonds). The median of each bin along with 1$\sigma$ statistical uncertainties\footnote{We use the \textsc{bootstrap} function from Python \textsc{scipy} package with 10$^4$ samples to derive the uncertainty on median values. For the upper bound of the first and final bin, 68th confidence interval provides the same value as the median, so we instead use the upper bound of 95th confidence interval\label{bootstrap}} are overlaid in Fig.~\ref{fig:fig007}. We also include for comparison in each panel a $z=0$ reference sample from \citealt{yoachim_structural_2006}.

To account for the different numbers of galaxies with a range of masses in each bin, we correct the size and thickness dependency on the galaxy's stellar mass, as seen in Fig.~\ref{fig:fig005}. For example, for $h_R$, we first derive the covariance Cov$(\log h_R, \log M_{*})$ and variance Var$(\log M_{*})$ and find the slope $\beta =\, $Cov$(\log h_R, \log M_{*})$/Var$(\log M_{*})$ of the linear relationship between $\log h_R$ and $\log M_{*}$. Then, we subtract the mass dependency $\log h_{R,\mathrm{corr}} = \log h_{R} - \beta (\log M_{*}-\log \widetilde{M_{*}})$ for the single disc sample with a median stellar mass $\log \widetilde{M_{*}}=9.2$. After making a correction for the $M_{*}$ dependency with $z_0$ and $h_R$ (red diamonds), both $h_R$ (left) and $z_0$ (middle) show a mild increasing trend from 12 Gyr to present. 

The mild rising trend in the median values of $h_R$ towards the present are consistent with the result of \citet{van_der_wel_3d-hstcandels_2014}. We compare our measurements with their median evolutionary trends, rather than with the absolute values (which are typically about 1.7 times smaller than ours on average), for the following reasons.
The evolutionary trend and absolute values by \citet{van_der_wel_3d-hstcandels_2014} shown in Fig.~\ref{fig:fig007} correspond to effective radii derived from fitting a sigle 2D S\'ersic profile to galaxies with a wide range of inclinations. If we convert these values to disc scale radii by dividing by 1.678, appropriate for an exponential profile, their measurements fall significantly below those presented here. As discussed in the previous section (Sec.~\ref{subsec:scalrel}), radii measured for face-on galaxies tend to be $\sim$2 times shorter than those measured for edge-on galaxies \citep{boardman_are_2020}. Additionally, the combination of a disc and a young compact bulge fitted with a single S\'ersic can result in a shorter measured radius, whereas in this study, we use multiple components to mitigate the effects of the bulge in edge-on galaxies.

The increasing trend in $z_0$ toward the present is consistent with similar studies using local galaxies \citep{yoachim_structural_2006}, but contrasts with the declining trend reported by \citet{lian_thickness_2024}. This discrepancy can be attributed to the sequential formation of the thick disc followed by the thin disc and the use of shorter wavebands in \citet{lian_thickness_2024}, which are sensitive to younger stellar populations. Consequently, their measurements capture the thickness associated with the ongoing formation of thick discs at earlier times and the subsequent formation of thin discs at later times. This is further clarified in the next subsection (Sec.~\ref{subsec:doubleevo}), which examines the evolution of thin and thick discs.

No median evolution is seen for the ratio of scale length to scale height ($h_R/z_0$, Fig.~\ref{fig:fig007} right). This suggests that the discs at all of our explored epochs, $\tau_{\mathrm{lookback}}\sim 1.6-11.4$ Gyrs, have already developed geometrically similar structures to present-day galaxies (as denoted by the grey band at $z=0$).

\begin{figure*}
    \centering
    \includegraphics[width=\linewidth]{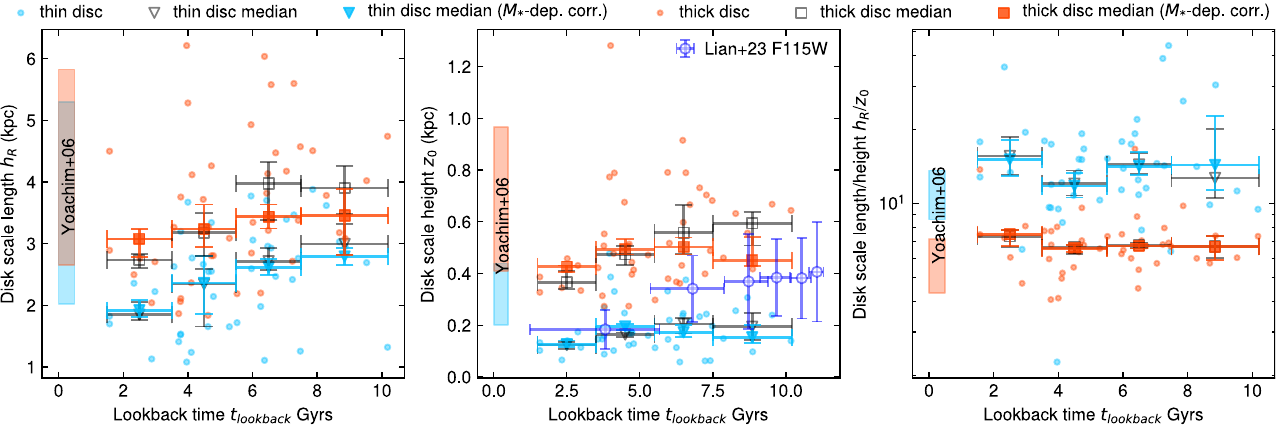}
    \caption{The properties of the thick and thin discs plotted against a galaxy's lookback time. The individual measurements for thin and thick discs are shown in blue and red points, respectively. The black and blue triangles are the median trends of thin disc without and with $M_{*}$ dependency correction to the stellar mass $\log(M_{*}/M_{\odot})=9.5$. The black and red squares are the same for thick discs. There is no significant evolution in all properties.}
    \label{fig:fig008}
\end{figure*}

\subsubsection{Thick and thin discs}
\label{subsec:doubleevo}
Fig.~\ref{fig:fig008} shows equivalent plots to those presented in Fig.~\ref{fig:fig007}, i.e. disc properties as a function of lookback time, but for the double disc subsample, showing both thin and thick discs. The measured scale lengths and heights for discs are almost constant as a function of lookback time after correcting the $M_{*}$ dependency. This indicates that there is no significant structural evolution in thin and thick discs. 

One notable exception is a mild decrease in scale lengths for the thin disc components towards late times. This trend is of interest as thin discs may be preferentially affected by scattering processes that reduce their scale radius by decreasing angular momentum, or because we observe the effect of reduced pressure support in the thin disc, causing it to shrink (see more discussion in Sec.~\ref{subsec:discussion1}). However, measuring the scale length of thinner discs is subject to larger uncertainties (see Appendix \ref{sec:appendixc}), and low number statistics at the bin at the latest lookback time. 

\citet{lian_thickness_2024}, fitting a single disc component to the F115W (rest-frame $\sim$1.2/$(1+z)\micron$) images, show a strong decrease in scale height towards the present, which contrasts with the mild increase shown in Fig.~\ref{fig:fig007} (middle). Their thickness measurements are, however, aligned with the thick disc at early times and then aligned with the thin disc at later times towards the present, Fig.~\ref{fig:fig008} (middle). The shorter wavelength bands used in \citet{lian_thickness_2024} trace young stellar populations that likely outshine the main mass component of the galaxy. At early times, $\sim6-12$ Gyr ago, their thickness measurements in short bands aligned well with our measurements for single disc galaxies, indicating that early young thick discs shone brightly across wave bands ($\lambda\sim0.3\mu$m and $\lambda\sim1.1\mu$m at $z=3$ for \citet{lian_thickness_2024} and this work respectively). In the recent $\sim6$ Gyr, due to the increasing prevalence of thin discs which outshine older thick discs, their measurements in short bands effectively measure the thin disc scale height ($\lambda\sim 0.8\mu$m and $\lambda\sim 2.4\mu$m at $z=0.5$ for \citet{lian_thickness_2024} and this work respectively). Those align with our measurements for thin discs decomposed from thick discs in our rest-frame IR band. The rest-frame IR band used in this study traces stellar mass, less sensitive to the age of the stars, thus providing better contrast to see the fainter thick disc remaining from early formation against the thin discs which will dominate in the shorter bands. 

\subsubsection{Interpreting evolutionary trends in disc formation}
Single discs show a mild increase in both scale length ($h_R$) and height ($z_0$) toward the present, whereas the thin and thick discs in two-disc galaxies show no significant evolution.
It is important to note that the median evolution of disc structural parameters \textit{at a given galaxy mass} does not reflect the specific evolutionary tracks of individual galaxies -- galaxies in different time bins are not the same progenitor populations. With the mass dependency already removed, the evolution in radius and height is primarily determined by angular momentum and vertical energy, respectively.

In the context of cosmic downsizing, for galaxies of the same mass, those observed at early times formed rapidly \citep{behroozi_average_2013} and completed their thick disc formation earlier \citep{comeron_prediction_2021} compared to galaxies observed at later times. The mild increase in $h_R$ at given galaxy mass can be attributed to halo evolution in the $\Lambda$CDM universe, where halo virial radii grow larger toward the present. Gas accreting from more extended regions carries larger specific angular momentum, so discs that form later tend to be larger \citep{mo_formation_1998}. Meanwhile, lower halo concentrations at earlier epochs \citep{Bullock_profiles_2001} partially offsets this effect, resulting in a modest overall size increase \citep{somerville_explanation_2008}. Additional crucial processes, such as outflows removing low–angular momentum material \citep{brook_hierarchical_2011, guedes_forming_2011}, radial migration that move stars to outer radii on average \citep{Minchev_evolution_2012}, and mergers also affect disc sizes \citep{governato_forming_2009}. The increase in disc thickness $z_0$ is expected because stars do not cool and are continuously thickened by satellite perturbations; consequently, discs observed at later times have more chance to be heated by such events. Moreover, for a given velocity dispersion, a disc with a larger scale length at the same mass has a lower surface mass density, leading to a larger scale height in gravitational equilibrium. The absence of an increase in $z_0$ for the thick disc in two-disc galaxies \added{(middle panel of Fig. \ref{fig:fig008}) could} be due to the presence of the thin disc. \added{Simulations have shown that gas in a thin disc can reduce heating from minor mergers and thin disc growth can lead to the adiabatic vertical contraction of the thick disc \citep{moster_can_2010}.} In our disc sample, the mechanisms responsible for radial and vertical height growth seem to maintain the geometrical proportion of the discs (see constant $h_R/z_0$ in Figs. \ref{fig:fig007} and \ref{fig:fig008}).

For evolutionary tracks of individual galaxies, it is more informative to examine their size and mass relations. As galaxies and discs grow in mass, both their radii ($h_R$) and heights ($z_0$) tend to increase on average, aligning with the scaling relations in Figs. \ref{fig:fig005} and \ref{fig:fig006}. An increase of about two orders of magnitude in mass can roughly triple these sizes. In contrast, the redshift evolution affects the sizes by a factor of at most 1.5, contributing to the observed scatter in the scaling relation. The relatively constant or mildly evolving mass-corrected medians shown in Figs. \ref{fig:fig007} and \ref{fig:fig008} indicate that the fundamental scaling relations remain largely unchanged over time. 

\subsection{Emerging thin discs at later epochs}\label{subsec:subsec33}
\begin{figure}
    \centering
    \makebox[\linewidth][c]{\includegraphics[width=1.02\linewidth]{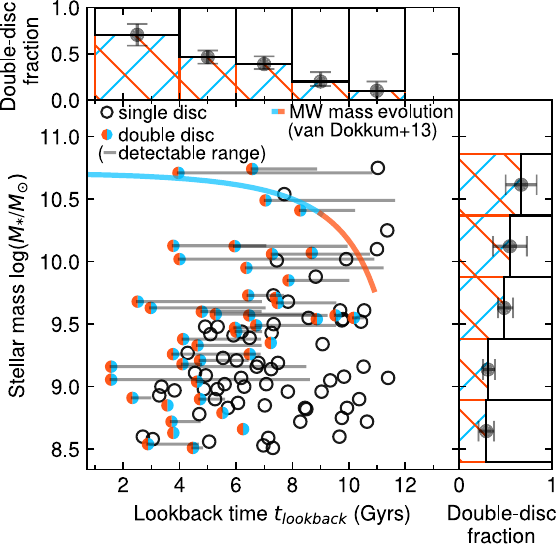}}
    \caption{Distribution of two-disc (red-blue points) and single-disc galaxies (black circles) in lookback time and stellar mass. The top and right panels display the fraction of the two disc galaxies as a function of lookback time and stellar mass, with associated 1$\sigma$ statistical uncertainties$^{\ref{bootstrap}}$. For two-disc galaxies, gray lines indicate the range of lookback times at which galaxies can be identified as having two discs when artificially redshifted. The averaged evolutionary track of Milky Way mass galaxies \citep{van_dokkum_assembly_2013} is shown by a blue-red line, where the red part marks the thick disc formation period, and the blue part marks the thin disc formation, with a transition estimated to be around 9 Gyr ago based on stellar age measurements.}
    \label{fig:fig009}
\end{figure}

Figure~\ref{fig:fig009} shows the distribution of two-disc and single-disc galaxies in stellar mass and lookback time, along with fractions of two-disc galaxies as a function of each variable. Two-disc galaxies are prevalently found at high stellar masses at earlier times, extending back to $\sim$10 Gyr ago ($z\sim2$). At later times (i.e., shorter lookback times), they are increasingly found at lower stellar masses, resulting in a increasing fraction of two-disc galaxies with cosmic time and stellar mass.
The distribution includes observational biases: identifying a two-disc structure becomes more difficult for more distant galaxies or lower mass galaxies. To dissect observational effects from the onset of thin disc formation, Fig.~\ref{fig:fig009} shows the range of lookback times at which two discs can be identified when a galaxy (of similar properties) with two discs is artificially redshifted (gray line for each data point).

To find the maximum lookback time, we calculate redshifted surface brightness, disc radius ($h_R$), and disc height ($z_0$) by redshift increments ($dz=0.02$) using galaxies' best-fit model parameters for galaxies categorised as having two discs (Sec.~\ref{sec:results}). We consider the surface brightness evolution including not only the surface brightness dimming ($\propto (1+z)^{-4}$) but also k-correction and intrinsic evolution (see Fig.~\ref{fig:figd2} in Appendix~\ref{sec:appendixd}). We then generate simulated images with noise added and the identical PSF, and refit the images using both a two-disc model and one-disc model. The noise incorporates variance from sky noise, read noise, and Poisson variance due to photon counting from the object itself. Bulge components are subject to redshifting but fixed during the refit. We consider the two-discs to be detectable if the BIC difference between the two-disc fit and one-disc fit is conservatively more than 100\footnote{Compared to a difference of 15 used in Sec.~\ref{sec:3dmodel}. In this simulation, we fit a perfect model (two discs) to a noisy version of itself, so the model fit is already near optimal, resulting in a high likelihood contrast with the case of fitting an imperfect model (single disc). Consequently, we need to adopt a higher BIC threshold compared to fitting models to real data, where both models are imperfect representations of reality. In the real case, unaccounted structures in the data contribute to the $\chi^2$, reducing the likelihood contrast between the models.}. We visually inspect the fitting results to confirm that the two-discs are identifiable with clarity comparable to the visual inspection process described in Sec.~\ref{sec:3dmodel}.

By examining the lookback time ranges where two-disc structures are detectable, we find that identifying two-disc galaxies becomes increasingly difficult for those with masses below $10^9 M_\odot$, making it challenging to assess their fractional changes over time. In contrast, galaxies with masses above $10^9 M_\odot$ extend well over some single disc galaxies, which allowed us to see the onset of two disc galaxies. For the most massive galaxies above $10^{9.75} M_{\odot}$, there is evidence of a transition from single discs to two discs around 8 Gyr ago. Lower mass galaxies with masses between $10^9 M_{\odot}$ and $10^{9.75} M_{\odot}$, on the other hand, begin forming thin discs approximately 4 Gyr ago. This shift in the onset of two disc formation suggests that thin disc formation proceeds in a "downsizing" manner, where more massive galaxies develop thin discs at earlier epochs. 

The averaged evolutionary track of Milky Way mass galaxies \citep{van_dokkum_assembly_2013} crosses the detectable lookback time ranges of two-disc galaxies, demonstrating the power of JWST to directly constrain the thin disc onset of MW-sized galaxies. The thin disc formation for high mass galaxies, $\sim$ 8 Gyr, aligns with the thin disc formation period, starting $\sim9$ Gyr, for Milky Way \citep{mukremin_age_2017, conroy_birth_2022, yaqian_timing_2023, ciuca_chasing_2024}.

\subsection{The mass ratio of thin and thick discs}\label{subsec:massratio}
\begin{figure}
    \centering
    \includegraphics[width=\linewidth]{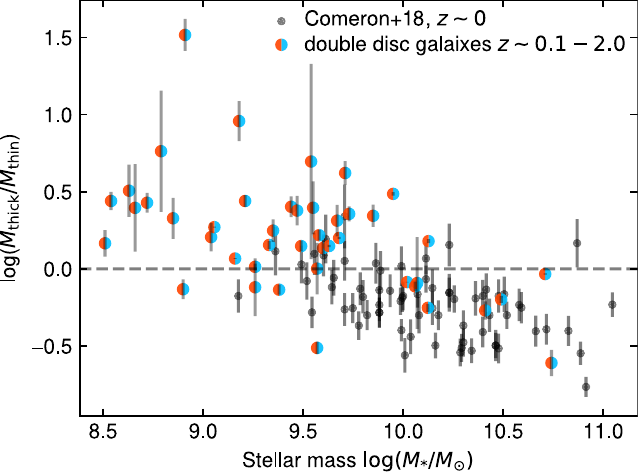}
    \caption{The mass ratio of thick to thin discs plotted against the total stellar mass for two disc galaxies. The red-blue points represent measurements of this work, spanning a redshift range of $z\sim0.1$ to 2.0.  These measurements are taken in bands approximately overlapping rest $K_s$ band at $z<1.45$ and $H$ band at $1.45<z<2.25$. Black points denote measurements from \citet{comeron_reports_2018} for galaxies at $z\approx0$, obtained using \textit{Spitzer} 3.5$\micron$ and 4.5$\micron$ bands. Error bars indicate the uncertainties associated with each measurements. For visual reference, the dashed line marks the ratio $M_\mathrm{thick}/M_\mathrm{thin} = 1$.}
    \label{fig:fig010}
\end{figure}

In this section, we explore the mass fraction of thick and thin discs in galaxies that host two disc structures. The mass ratio is calculated multiplying the luminosity ratio with the ratio of mass-to-light ratios for thick and thin discs, $M_{\mathrm{thick}}/M_{\mathrm{thin}} = \frac{\Upsilon_\mathrm{thick}}{\Upsilon_\mathrm{thin}}\frac{L_{\mathrm{thick}}}{L_{\mathrm{thin}}}$. We adopt the fiducial ratio of thin and thick mass-to-light ratio, $\Upsilon_\mathrm{thick}/\Upsilon_\mathrm{thin} = 1.2$, derived by \citet{comeron_thick_2011} assuming the star formation history (SFH) of the Milky Way's thin and thick discs at the F356W band \citep[][but see additional assumptions made in \citealt{comeron_thick_2011}]{pilyugin_chemical_1996, nykytyuk_galactic_2006}.  Figure \ref{fig:fig010} shows the mass ratio of thick to thin discs for our sample, as a function of the total stellar mass, compared with $z=0$ galaxies from \citet{comeron_evidence_2014} using Spitzer $3.5\micron$ band (equivalent to F356W used in this study).

The value $\Upsilon_\mathrm{thick}/\Upsilon_\mathrm{thin} = 1.2$ is applied to galaxies with a assumed specific SFH (observed at $z=0$ with the F356W band). Using the same SFH, we consider the spectral energy distribution (SED) evolution of galaxies and rest-frame band shifting with redshift and find that $\Upsilon_\mathrm{thick}/\Upsilon_\mathrm{thin}$ does not significantly change across redshifts of the sample and the bands used in this study (F227W, F356W, F444W: see Appendix \ref{sec:appendixd}). Therefore, we can safely assume $\Upsilon_\mathrm{thick}/\Upsilon_\mathrm{thin} = 1.2$ for comparison between our measurement with the $z=0$ sample \citep{comeron_thick_2011}. \citet{comeron_thick_2011} also varied the assumption on the SFHs of thin and thick discs and found values ranging from 1.2 to 2.4. We tested the constancy across redshift, $z=0-3$, for the same set of SFHs. While a delayed formation of thin disc relative to thick disc provides higher values, all cases show a constant $\Upsilon_\mathrm{thick}/\Upsilon_\mathrm{thin}$ across redshifts and observed bands, with differences among the three bands being small, less than 0.25.

The thick and thin disc mass ratios $M_{\mathrm{thick}}/M_{\mathrm{thin}}$ for our two disc galaxies at $z\sim0.1-2$ show a decreasing trend as a function of stellar mass. This result is well aligned with massive galaxies at $z = 0$ \citep{yoachim_structural_2006, comeron_thick_2011, comeron_breaks_2012, comeron_evidence_2014}. \citet{yoachim_structural_2006} also derived the mass ratio for their  $z=0$ sample using the $R$ band, where the mass-to-light ratios are strongly influenced by galaxies' SFH. Therefore, we do not make a comparison, but a similar decreasing trend is observed.

\begin{figure}
    \centering
    \includegraphics[width=\linewidth]{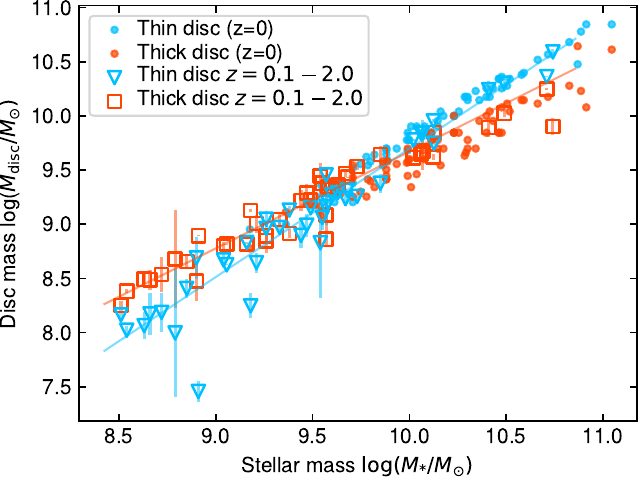}
    \caption{The decomposed thin and thick disc masses of two disc galaxies are plotted against their total stellar masses. Blue triangles represent thin disc measurements, and red squares represent thick disc measurements from this study spanning the redshift range of 0.1 to 2. For comparison, blue and red points indicate thin and thick disc measurements at redshift $z\approx0$ from \citet{comeron_reports_2018}, respectively.}
    \label{fig:fig011}
\end{figure}

To further explore what drives the decreasing trend of thick-to-thin disc luminosity with stellar mass, Fig.~\ref{fig:fig011} shows the individual disc mass of thin and thick discs plotted against total stellar mass. The best-fit slopes $M_{*}-M_{\mathrm{disc}}$ for thin and thick discs are shown. Fits are derived from the two best-fit relations $M_{*}-z_0$ and $M_{\mathrm{disc}}-z_0$ (Table  \ref{tab:tab001}), demonstrating that a single power-law can describe the thin and thick disc sequences separately. The thin and thick disc sequences are consistent with the local results \citep{comeron_reports_2018}, showing two clear distinct sequences with a shallow slope for $M_{\mathrm{thick}}-M_{*}$ and a steep slope for $M_{\mathrm{thin}}-M_{*}$ that cross at a $\log(M_*[M_\odot])\sim10$. The different slopes of the thin and thick disc sequences are responsible for the decreasing disc mass ratio as a function of galaxy masses seen in Fig.~\ref{fig:fig010}. 

Figs.~\ref{fig:fig010} and \ref{fig:fig011} demonstrate that higher mass galaxies tend to have more massive thin discs compared to their thick discs. This aligns with the downsizing formation trend shown in the previous section (Sec. \ref{subsec:subsec33}), where thin discs begin to form earlier in more massive galaxies. Supporting this, numerical simulations of Milky Way-sized galaxies suggest that an earlier transition from a bursty thick disc phase to a steady thin disc formation phase results in higher thin disc-fractions \citep{yu_bursty_2021}.

The different slopes may also reflect the efficiency with which the thin and thick discs gain mass along with total galaxy mass growth. For instance, it is expected that Milky Way-sized galaxies have increased its mass by approximately 0.6 dex on average from redshift $z=2$ to $z=0.1$, which spans the redshift range of our two-disc galaxy sample \citep{van_dokkum_assembly_2013}. While this 0.6 dex growth is relatively modest compared to the mass range we explored, the slopes for both the thick and thin discs appear to align those at redshift $z=0$. For this relationship to remain unchanged as galaxies grow, both the thick and thin discs must continue to gain mass. Therefore, even after a galaxy possesses both a thick and a thin disc, the thick disc continues to grow, albeit less efficiently than the thin disc. This ongoing growth of the thick disc may be driven by the variable gravitational stability of gas discs modulating between thin and thick disc regimes due to episodic events such as merging or accretion, which is discussed in next Section~\ref{subsec:toomreq}. Additionally, satellite accretion or gradual scattering and heating from thin disc stars could contribute to the growth of thick discs. This overlapping formation could explain the continuous age populations seen in the Milky Way \citep{ciuca_unveiling_2021, beraldo_e_silva_co-formation_2021, bovy_milky_2012}. This contrasts with a purely sequential scenario where the two discs would form in entirely separate epochs.

Disentangling the initial ratio of the two discs when they first become observationally distinguishable from the later build-up process requires additional information, such as star formation histories and stellar kinematics of these galaxies. 

\section{Discussion}
\label{sec:discussion}
We confirm the presence of single discs up to $z\sim~3$ and two discs up to $z\sim2$, which already exhibit radial and vertical size-mass relations (Sec.\ref{subsec:scalrel}). The edge-on confirmation of stellar discs complements the identification of spiral structures in face-on stellar discs \citep{kuhn_jwst_2024} at redshift up to $z\sim3$. Well developed stellar discs are internally unstable or dynamically responsive to external perturbations, forming spiral patterns \citep{byrd_tidal_1992, law_high_2012, pettitt_gas_2016, bland-hawthorn_rapid_2023,tsukui_detecting_2024}. Spiral-inducing mechanisms driven by external perturbations are presumably more significant at higher redshifts, where merger rates are elevated \citep{rodriguez-gomez_merger_2015}.

In this Section, building on the main observational findings in preceding sections we discuss the evolution from gaseous to stellar discs and the emergence of thick and thin discs across cosmic history, linking gaseous disc measurements with the structural measurements of stellar discs in hand.

\subsection{Toomre \textit{Q} self-regulated disc formation}
\label{subsec:toomreq}
\begin{figure}
    \centering
    \includegraphics[width=1\linewidth]{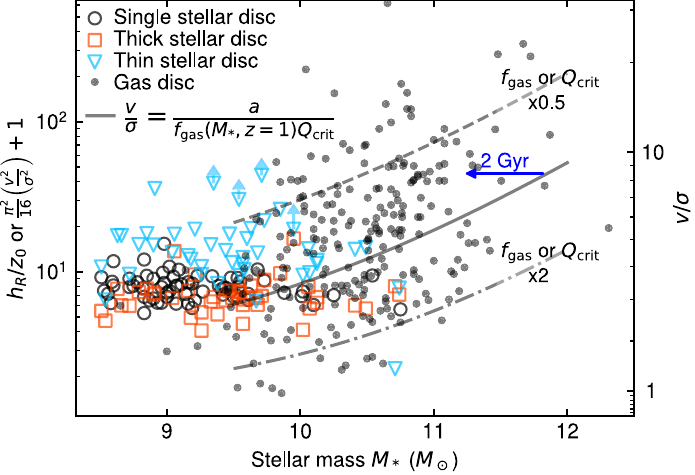}
    \caption{The measured axial ratios for the single disc (black circle), thin discs (blue triangles), and thick discs (red square). The black dots represent the $v/\sigma$ values for gas discs \citep[][and others$^{\ref{gasdiscsource}}$]{ubler_evolution_2019}, or the expected axial ratios for stellar discs that form from the gas discs in the simplified scenario where the gas disc is entirely converted into a stellar disc while conserving $v/\sigma$. The black line indicates the predicted curve for a Toomre $Q$-regulated gas disc, with $Q_{\mathrm{crit}}=1$ and $f_{\mathrm{gas}}(M_{*})$ for a main-sequence galaxy at $z=1$ \citep{tacconi_evolution_2020}. A 2 Gyr time evolution is illustrated by the blue arrow, which horizontally shifts the curve to the left, allowing more low-mass galaxies to enter the thin-disc formation regime. Variations in the predicted curve, resulting from doubling or halving $Q_{\mathrm{crit}}$, $f_{\mathrm{gas}}$, or the product $Q_{\mathrm{crit}} f_{\mathrm{gas}}$, are shown by dashed and dash-dotted lines.}
    \label{fig:fig012}
\end{figure}

The kinematics of gaseous discs have been systematically characterized up to redshift $z\sim2.7$ \citep[e.g.,][]{wisnioski_kmos3d_2015, ubler_evolution_2019}, with more recent studies extending this exploration to redshifts beyond $z\sim4$ \citep[e.g.,][]{neeleman_cold_2020, rizzo_dynamically_2020, lelli_massive_2021, tsukui_spiral_2021}. The commonly measured kinematic parameter, $v$/$\sigma$, serves as a proxy for the dynamical support of gas discs and provides insights into the geometric proportion of the resulting stellar disc formed from star formation in the gas disc. 
Fig.~\ref{fig:fig012} shows the intrinsic vertical to radial `axial ratio' of the discs, $h_R/z_0$, for all disc categories as a function of total stellar mass, as shown in Fig.~\ref{fig:fig005} (right). We compare these measured geometrical proportions with $v/\sigma$ values from the literature for galaxies at similar redshifts $z\sim0.1-2.7$, mainly measured using optical emission lines \citep{ubler_evolution_2019}, and for galaxies at earlier redshifts ($>2$) primarily using far-infrared (FIR) emission lines (e.g., [C\textsc{ii}], [C\textsc{i}]; sources in the footnote\footnote{\citealt{neeleman_cold_2020, rizzo_dynamically_2020, rizzo_dynamical_2021, lelli_massive_2021, tsukui_spiral_2021, fraternali_fast_2021, rizzo_alma-alpaka_2023, roman-oliveira_regular_2023, parlanti_alma_2023, amvrosiadis_kinematics_2023, rowland_rebels-25_2024, fujimoto_primordial_2024}\label{gasdiscsource}}).
Although measurements by optical emission lines outnumber those using FIR emission measurements, both distributions appear similar when plotted separately in the diagram. 

Assuming stellar components inherit the same $v/\sigma$ of the gas disc they form out of, 
a simple application of the tensor virial theorem \citep{binney_rotation_2005, binney_galactic_2008} predicts a relation between $h_R/z_0$ and $v/\sigma$ of self-gravitating axisymmetric stellar structure. Under the assumption of an isotropic velocity dispersion, the approximate relation is given by \citet{kormendy_observations_1982} as,
\begin{equation}
    h_R/z_0 = \frac{\pi^2}{16}\left(\frac{v}{\sigma}\right)^2+1.
\end{equation}
We assume an isotropic velocity dispersion as we do not have evidence to the contrary at high redshift
\citep{genzel_strongly_2017}. 

Gas discs exhibit a weak correlation between $v/\sigma$ (or the expected $h_R/z_0$) and stellar mass, with more massive galaxies typically located in the thin disc regime and less massive galaxies in the thick disc regime or below. This trend is consistent with results from simulations \citep{pillepich_first_2019, kohandel_dynamically_2024} and can be explained by self-regulated star formation of gas discs, supporting the downsizing thin disc formation discussed earlier. The Toomre stability criterion for gas-dominated discs \citep{toomre_gravitational_1964} is:
\begin{equation}\label{eq:toomreQ}
    Q_{\mathrm{gas}} = \frac{\sigma}{v}\frac{a}{f_\mathrm{gas}}
\end{equation}
where $a$ is a constant depends on the shape of rotation curve ($a=\sqrt{2}$ for a constant rotation curve), and $f_\mathrm{gas}$ is the gas mass fraction within the disc (see \citealt{genzel_sins_2011}). In models where disc star formation is self-regulated by gravitational instability and star-formation feedback, Toomre $Q_\mathrm{gas}$ is maintained around a critical value ($Q_{\mathrm{crit}} \approx 1$), keeping the disc marginally unstable. Thus, $v/\sigma \approx a/f_{\text{gas}}/Q_{\text{crit}}$, which links $v/\sigma$ to the gas mass fraction $f_\mathrm{gas}$. 

More massive galaxies, which have lower gas fractions \citep{mcgaugh_gas_1997, tacconi_phibss_2013, tacconi_evolution_2020} due to efficient star formation \citep{behroozi_average_2013}, achieve high $v/\sigma$ values earlier, leading to the earlier onset of thin discs in these galaxies. Conversely, less massive galaxies, with higher gas fractions, exhibit lower $v/\sigma$ and cannot form thin discs until their stellar component develops sufficiently to reduce the gas fraction and support higher $v/\sigma$. To illustrate this, in Fig.~\ref{fig:fig012}, we overplot the $v/\sigma$ expected for gas discs from Eq.~\ref{eq:toomreQ}, using the averaged gas fraction $f_\mathrm{gas}(M_{*})$ for main-sequence galaxies at $z=1$ \citep{tacconi_evolution_2020} and assuming $Q_{\mathrm{crit}}=1$. This predicted curve shows that high-mass galaxies ($>10^{10}M_\odot$) lie in the thin disc formation regime, while low-mass galaxies lie in thick disc regime. The blue arrow indicates the 2 Gyr evolution of the curve from $z=1$, which shifts the curve horizontally, allowing less massive galaxies to enter the thin disc formation regime as their gas fraction decrease over time. This explains the earlier thin disc formation in massive galaxies (downsizing thin disc formation) hinted by Fig.~\ref{fig:fig009}, explaining the dominance of thin discs in high-mass galaxies and thick discs in low-mass galaxies (Fig.~\ref{fig:fig011}). 

The expected $h_R/z_0$ ratio for gas discs shows a broader range than observed in stellar discs, indicating that gas discs are subject to significant temporal variation of the Toomre $Q_\mathrm{gas}$ parameter. These fluctuations in $Q_\mathrm{gas}$ are driven by episodic events such as gas accretion and mergers, which increase the gas fraction ($f_\mathrm{gas}$), and subsequent starburst that enhance turbulent energy ($\sigma$), and its dissipation \citep[see e.g.,][]{tacchella_confinement_2016}. As a result, the $Q_\mathrm{gas}$ parameter for gas discs can vary widely, oscillating around the marginal stable value ($Q_\mathrm{crit}$). Also, $Q_\mathrm{crit}$ itself is variable; it can decrease to about $0.7$ when the finite scale height stabilizes a disc \citep[][as opposed to an infinitesimally thin disc]{kim_three-dimensional_2002, bacchini_3d_2024}, but can increase to $2-3$ under a condition that gas turbulence efficiently dissipates (\citealt{elmegreen_gravitational_2011}). In Fig.~\ref{fig:fig012}, we show the variations in $v/\sigma$ expected from the averaged population by assuming either a 2x higher or lower value for $f_{\mathrm{gas}}$ or $Q_{\mathrm{crit}}$, or the product $f_{\mathrm{gas}}Q_{\mathrm{crit}}$, which encompasses the range of gas disc measurements.

In contrast, stellar distributions are shaped by the cumulative effect of star formation, occurring under the variable conditions of the gas disc, presumably leading to a convergence toward a narrower range of axial ratios over time. As the stellar component develops in gas discs, the total disc Toomre parameter, $Q_{\mathrm{tot}}^{-1}=Q_{\mathrm{*}}^{-1}+Q_{\mathrm{gas}}^{-1}$, is subject to the disc stability criterion. This is valid if all components have similar velocity dispersions \citep{romeo_effective_2011}. Considering a simple scenario, where the gas distribution and stellar distribution are coupled with the same velocity structure and distribution and $f_{\mathrm{gas}}+f_{\mathrm{star}}=1$, the marginally unstable disc would have $h_R/z_0 \sim (v/\sigma)^2 = (a/Q_\mathrm{crit})^2$, which does not depend on galaxy stellar mass, consistent with no correlation seen in $h_R/z$ for stellar discs in Fig.~\ref{fig:fig012}. Note also that during or after their formation from gaseous discs, stellar discs undergo distinct processes from gaseous discs such as heating by disc substructure or mergers. Unlike gaseous discs, once heated, the stellar components do not cool. 

Finally, note that stellar masses of galaxies with available optical line kinematics, derived using the same methods (3D-HST; \citealt{ubler_evolution_2019}), are larger than those in our sample. This reflects the current sensitivity limitations of spectroscopic observations and the challenge of obtaining kinematics for low-mass galaxies.

\subsection{Vertical equilibrium of the disc}
\begin{figure}
   \centering
   \includegraphics[width=\linewidth]{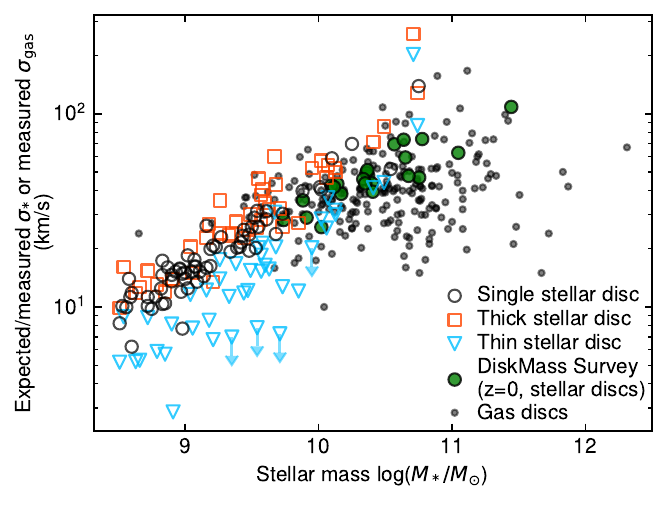}
   \caption{Expected stellar velocity dispersion from vertical equilibrium for single, thin and thick discs, with black circles, blue triangles, and red squares, respectively. The arrows indicate upper limit measurements. Green circles indicate stellar velocity dispersion measurements from the DiskMass survey for face-on galaxies at $z=0$ \citep{martinsson_diskmass_2013, martinsson_diskmassvi_2013}. These stellar velocity dispersions are compared with those of gas discs (black dots).}
   \label{fig:fig013}
\end{figure}

In the previous section (Sec.\ref{subsec:toomreq}), we demonstrated that a gas disc with widely varying gas fractions $f_\mathrm{gas}$, can form stellar thin and thick discs with similar geometric proportions $-$ the relative height to the radial length $h_R/z_0$. However, this does not necessarily mean that turbulent gaseous discs are sufficiently thick or thin to directly produce thick or thin stellar discs. In this section, we address this question by examining the vertical equilibrium of these system. Figure~\ref{fig:fig013} shows the expected velocity dispersion of the stellar disc assuming vertical equilibrium. The scale height $z_0$ of a self-gravitating isothermal sheet is given by $z_0 = \sigma_{*}^2/(2\pi G \Sigma{*})$ \citep{binney_galactic_2008}, where $\sigma_{*}$ is the stellar velocity dispersion, and $\Sigma{*}$ is the surface density of the stellar disc. Using measurements of the scale height $z_0$ and the surface density $\Sigma{*} = M_{*}/(2\pi h_R^2)$, we compute the expected velocity dispersion for a single disc in vertical equilibrium as: 
\begin{equation}\label{eq:vertical}
  \sigma = \sqrt{2\pi G \Sigma_{*} z_0}.
\end{equation}
For galaxies with two disc components, we approximate the surface density as $\Sigma_{*}=\Sigma_\mathrm{thin}+\Sigma_\mathrm{thick}$ (refer to \citealt{aniyan_resolving_2018} for isothermal sheet solution with the presence of an additional thin disc) to derive the expected velocity dispersion of thick disc:
\begin{equation}\label{eq:vertical2}
  \sigma_{\mathrm{thick}} = \sqrt{2\pi G (\Sigma_{\mathrm{thin}}+\Sigma_{\mathrm{thick}}) z_{0,\mathrm{thick}}}.
\end{equation}
For thin disc, we use an approximate solution by \citet{forbes_evolving_2012}, which accounts for the gravitational influence of the thick disc: 
\begin{equation}\label{eq:vertical3}
  \sigma_{\mathrm{thin}} = \sqrt{2\pi G (\Sigma_{\mathrm{thin}}+\Sigma_{\mathrm{thick}}\times z_{0,\mathrm{thin}}/z_{0,\mathrm{thick}}) z_{0,\mathrm{thin}}}.
\end{equation}

For simplicity, we exclude contributions from dark matter, as baryons dominate in the disc mid-plane \citep{narayan_vertical_2002}, and neglect small bulges, which contribute a median of 2\% to the total disc luminosity in our sample. Their inclusion would marginally increase the estimated $\sigma$. The expected velocity dispersion of thin discs aligns with the measurements of vertical stellar velocity dispersion by the DiskMass Survey for local galaxies \citep{martinsson_diskmass_2013, martinsson_diskmassvi_2013}. The light weighted dispersion of the literature sample traces the thin disc stellar dispersion. This alignment suggests that discs achieve vertical equilibrium, and their measured thickness reflects the virialized state. The vertical equilibrium can be reached rapidly: for instance, the crossing time of a stellar particle within a typical vertical scale height of 0.5 kpc with a velocity dispersion of 50 km s$^{-1}$ is approximately 10 Myr. 

Gas velocity dispersions, from the high-redshift samples, span a range similar to those expected for thin and thick stellar discs. If galaxies of similar stellar mass have comparable disc density structures (Fig.~\ref{fig:fig005}) and have reached vertical equilibrium, then velocity dispersions serve as a good proxy for scale height and vice versa. The roughly consistent velocity dispersions between stars and gas suggest that gaseous discs are effectively forming corresponding structures with similar scale heights. Together, Fig.~\ref{fig:fig012} and Fig.~\ref{fig:fig013} confirm that early gaseous discs are able to form both thick and thin stellar discs. 

\subsection{On why thin discs are smaller than thick discs}
\label{subsec:discussion1}

The Milky Way's thin disc is more radially extended than its thick disc. This contrasts with our measurements, showing the thick disc is often larger than the thin disc, which is also shown in local galaxy samples structurally \citep{yoachim_structural_2006, comeron_breaks_2012} and chemically \citep{sattler_vertical_2023,Sattler_relatively_2024}. 

The larger size of thick discs in both radius and height than that of thin discs suggests compaction and expansion processes preferentially at work on the discs. One mechanism contributing to larger thick discs could be the inward radial flow of gas particles as the proto-gas disc dissipates turbulent energy, conserving the angular momentum. The less turbulent thin gas disc ends up having a shorter radius to increase the centrifugal force against gravity with less pressure support \citep{yoachim_structural_2006}. 

There is another mechanism that may selectively expand thick discs (or single thick discs before thin disc formation) in radius and vertical height. \citet{bland-hawthorn_turbulent_2024} demonstrate the gas-rich turbulent starburst phase involves significant mass ejection which weakens disc potentials. Following the weakening of the disc potential due to the mass loss with the axial ratio remaining unchanged, the velocity dispersion ratio of the ensemble disc stars $\sigma_z/\sigma_R$ is conserved and thus existing stars subsequently adiabatically expand vertically and radially. The episodic or continuous mass ejection makes the thick disc longer and thicker. When the thinner disc dominantly forms later with low gas fraction and less turbulence, the mass loading of outflow decreases \citep{hayward_how_2017}, making this process inefficient. 

In addition to the mechanisms active during the proto-gaseous disc phase, the thin disc is preferentially scattered by the density fluctuations in the disc midplane, including GMCs, clumps, spiral, bars, resulting in further compaction of the thin disc relative to the thick disc \citep{bournaud_thick_2009}. This may explain why the disc scale length has a shallower slope with mass relative to the thick discs and single discs (see left figures of Fig.~\ref{fig:fig005} and Fig.~\ref{fig:fig006}). The radial sizes of both thin and thick discs increase as the discs acquire stellar mass, but the thin discs are more prone to compaction due to heating, leading to smaller disc radial growth per unit mass increase compared to the thick disc.

\begin{figure}
    \centering
    \includegraphics[width=1\linewidth]{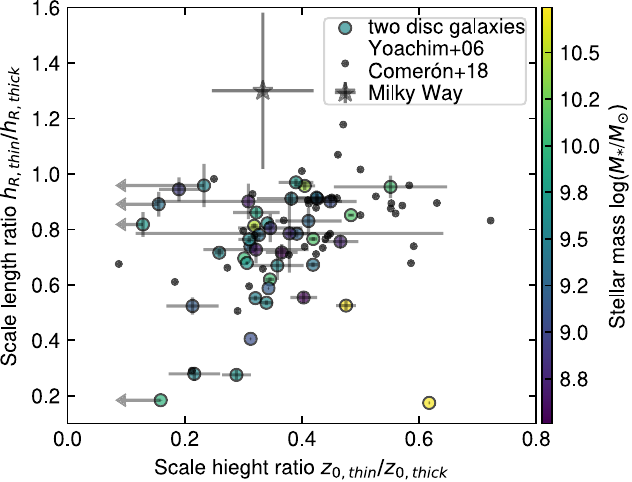}
    \caption{The relative thickness and radial sizes of thin and thick discs in individual two-disc galaxies. The ratio $h_\mathrm{thin}/h_\mathrm{thick}$ is plotted against $z_\mathrm{thin}/z_\mathrm{thick}$ for each galaxy, represented by black circles filled with color to indicate galaxy mass. Galaxies at z=0 \citep{yoachim_structural_2006, comeron_reports_2018} and the Milky Way \citep{bland-hawthorn_galaxy_2016} are overplotted with black dots and a black star, respectively.}
    \label{fig:fig014}
\end{figure}

Fig.~\ref{fig:fig014} compares the relative thickness and radial sizes of thin and thick discs in galaxies, showing that thin discs are shorter in both radius and vertical height compared to thick discs. Despite some outliers, galaxies appear to form a positive correlation, indicating that thinner thin discs, relative to thick discs, within the same galaxies are also radially shorter. However, the correlation of our measurements alone is not statistically significant with a spearman rank coefficient of $r=0.2$ and $p\mathrm{-value}$ of 0.25. To confirm this correlation, we include $z=0$ results from \citet{yoachim_structural_2006} and \citet{comeron_reports_2018}, restricting the latter to galaxies without disc truncations. This redshift $z=0$ sample shows positive correlation ($r=0.40$, $p\mathrm{-value}=0.01$) and when combined with our data provide higher statistical significance ($r=0.36$, $p\mathrm{-value}=0.001$).

The values ($z_\mathrm{thin}/z_\mathrm{thick}$, $h_\mathrm{thin}/h_\mathrm{thick}$) are expected to be noisier compared to ($h_\mathrm{thin}/z_\mathrm{thin}$, $h_\mathrm{thick}/z_\mathrm{thick}$) due to potential degeneracies and correlations between the denominator and numerator in the fitting process. This correlation may qualitatively support the disc evolution scenario discussed above e.g. the selective compaction or expansion of thin/thick discs. Galaxies may also move within this diagram over time. As thick disc growth slows while the thin disc continues to gain mass and expand in both height and radius \citep[inside-out disc evolution;][]{Minchev_chemodynamical_2013}, data points may shift from the lower left to the upper right along the sequence. In the downsizing thin disc formation, more massive galaxies might have had more time for their thin discs to grow; however, such trends are not clearly evident in the diagram. Constraining stellar ages could further help test these scenarios by revealing how galaxies evolve within the diagram. 

The formation of a smaller thin disc within an existing thick disc suggests an outside-in transformation when the thin disc emerges in pre-existing thick disc. \citealt{tadaki_structural_2020} found that in massive star-forming galaxies at $z \sim 2 $ ($ M_* > 10^{11} M_\odot $), the star-forming region is about half the size of the pre-existing stellar disc, possibly marking the onset of thin disc formation. According to the downsizing scenario, such massive galaxies may have formed thin discs earlier than the most massive galaxies in our sample ($\sim 8$ Gyr ago; see Fig.~\ref{fig:fig009}).

\subsection{Origin of thick and thin discs}
In this section, we discuss which formation scenarios of thin and thick discs are supported by our findings. By directly observing galaxies in the past, we confirm sequential stellar disc formation: galaxies first form a thick disc and later evolve into two-disc composite galaxies by forming a subsequent thin disc from within. The accretion of \textbf{ex situ stars from small satellites} cannot be the primary mechanism for the thick disc formation. 
Bringing satellites to the galactic disc scale via dynamical friction, in the absence of a pre-existing disc, would take longer \citep{Penarrubia_satellite_2002, villalobos_simulations_2008} than the observed early appearance of dissipative gas discs at z$\sim$4-7 (as early as $700$ Myr, \citealt{neeleman_cold_2020, rizzo_dynamically_2020, lelli_massive_2021, tsukui_spiral_2021, rowland_rebels-25_2024}). These early discs likely form through gas-rich mergers and/or cold gas accretion. Similarly, slow \textbf{"progressive thickening"} is inconsistent with our observations, as galaxies seem to have thick discs as early as $z\sim3$. However, it may still be viable if the process occurs on a much shorter timescale than our observations can capture. 

Over the long term, these mechanisms may contribute to the growth of the thick disc. Both thin and thick disc masses increase with galaxy mass, following different power-low slopes (Fig.\ref{fig:fig011}). If we interpret this as galaxies evolving along the tracks, they continue increasing their thick disc mass (albeit less efficiently than their thin discs) after forming a thin disc. Satellite accretion and heating may be a viable mechanism for thick disc growth at later stages \citep{pinna_fornax_2019, pinna_fornax_2019-1, martig_ngc_2021}, driving the continuous growth of thick discs as total galaxy mass increases.

We demonstrate that sequential formation, thick then thin, proceeds in a downsizing manner, where more massive galaxies form thin discs earlier. By linking to high-$z$ gas kinematics, we highlight the role of ISM turbulence in determining the timing of thin disc formation (Sec. \ref{subsec:toomreq}). This is consistent with the scenario in which thick discs rapidly form in chaotic gas-rich turbulent disc and thin discs subsequently form from quiescent low gas-fraction discs \citep{beraldo_e_silva_co-formation_2021, yu_born_2023, Bland-Hawthorn_turbulent_2025}\footnote{In gas-rich turbulent discs, \citet{yu_born_2023} suggest that thick disc stars form in hot orbits (\textbf{"born-hot"}), whereas \citet{beraldo_e_silva_co-formation_2021, Bland-Hawthorn_turbulent_2025} propose that most stars form near the disc mid-plane and are quickly heated (clumps and sloshing, \textbf{"instant thickening"}). The structural analysis in this paper cannot distinguish them, but color gradients or rest-frame ultra violet (UV) observations for a single (thick) disc in high-z galaxies may help differentiate them.}. Under gravitational stability-regulated disc formation, the gas turbulence is related to gas fraction $f_\mathrm{gas}$ by $v/\sigma \propto 1/f_{\mathrm{gas}}$ \citep{genzel_sins_2011}, suggesting that higher $f_\mathrm{gas}$ leads to higher turbulence (lower $v/\sigma$). High turbulence pressure may prohibit the thin disc formation \citep{vandonkelaar_giant_2022}. In cosmic downsizing, more massive galaxies convert gas into stars more efficiently \citep{behroozi_average_2013}, forming thick discs earlier \citep{comeron_prediction_2021}. As they achieve lower $f_\mathrm{gas}$ and lower turbulence, they transition to form thin discs earlier.

Archaeological studies of nearby edge-on galaxies also support the downsizing picture. More massive galaxies tend to have older thick discs with higher [$\alpha$/Fe] than low-mass galaxies, indicating rapid and early thick disc formation in massive galaxies \citep{Sattler_relatively_2024, sattler_vertical_2023, pinna_fornax_2019, pinna_fornax_2019-1, martig_ngc_2021}. Ground-based observations support this view showing that disc turbulence and $f_\mathrm{gas}$ increases at higher redshifts \citep{forster_schreiber_sins_2009, genzel_sins_2011, wisnioski_kmos3d_2015, ubler_evolution_2019, tacconi_evolution_2020, rizzo_alma-alpaka_2024}. Although recent ALMA observations have revealed surprisingly low relative turbulence, with $v/\sigma$ values as high as $\sim10$ at z$\sim$4, this remain consistent with this downsizing framework \citep{rizzo_dynamically_2020, lelli_massive_2021}. The available spatially resolved kinematic measurements are generally biased towards massive systems (Fig. \ref{fig:fig012}), which may already host substantial thick discs and thus exhibit lower $f_\mathrm{gas}$. Interestingly, some ALMA detected lower mass galaxies \citep[e.g.,][]{neeleman_cold_2020, tsukui_spiral_2021, parlanti_alma_2023} show enhanced turbulence (low $v/\sigma$) and high gas fractions, suggesting ongoing thick disc formation. For example, BRI 1335-0417 has a high gas fraction of $\sim$70\% and turbulence $v/\sigma\sim2.5\pm0.5$, corresponding to an axial ratio $q\sim5$, clearly placing it in the thick disc formation regime (Fig. \ref{fig:fig012}). Additionally, it uniquely shows spiral and bar structures \citep{tsukui_spiral_2021, tsukui_detecting_2024}.

Understanding the role of gas-rich turbulent discs in thick disc formation - and the puzzling presence of spiral and bar in such environments - has advanced significantly through numerical simulations. Recent simulations of gas-dominated discs show that high gas-fraction discs can rapidly develop spirals and bars, while young stellar bars form through disc shear flows \citep{bland-hawthorn_turbulent_2024}. These simulations also reveal an intriguing mechanism: stochastic star formation within complex gas substructures induces bulk motion (sloshing) of the gas disc relative to the halo potential, dispersing stars. The energy from this bulk motion is transferred to the stars, contributing to thick disc formation \citep{Bland-Hawthorn_turbulent_2025}. In these early epochs, clumps may have also contributed to stellar scattering \citep{beraldo_e_silva_geometric_2020, beraldo_e_silva_co-formation_2021}.

A potential contradiction to our findings "thick disc first, thin disc later" is the presence of an old, metal-poor thin disc in the Milky Way \citep{nepal_discovery_2024}. However, this low-mass component may have formed through later satellite accretion events, where dynamical friction drags satellites into the disc plane, preferentially from prograde satellites \citep{walker_quantifying_1996} , or it may consist of stars that survived heating in the gas-rich disc  \citep{beraldo_e_silva_co-formation_2021, yu_born_2023, Bland-Hawthorn_turbulent_2025}.

The relative importance of different growth mechanisms likely depends on a galaxy's properties (e.g., total mass at a given epoch) and formation history \citep[e.g.,][]{pinna_stellar_2024, yu_bursty_2021}. This study demonstrates the JWST potential to understand the Milky Way's formation history by directly examining Milky Way-sized progenitors at earlier epochs (Fig.~\ref{fig:fig009}) and determine if the Milky Way has a distinct formation history compared to others \citep[e.g.,][]{rey_vintergatan-gm_2023}.

\section{Summary}
\label{sec:summary}
We present the first systematic thin/thick disc decomposition of high-redshift galaxies using a sample of 111 edge-on galaxies from the flagship JWST imaging fields from JADES, FRESCO, CEERS, COSMOS-Web, PRIMER, and NGDEEP. To create a robust sample, we cross-match JWST detections with the 3D HST catalogue for reliable redshifts and galaxy parameters. The sample covers a wide redshift range, $0.1<z<3.0$, encompassing $\sim70\%$ of cosmic history up to a lookback time of 11.4 Gyr.

We fit a 3D disc model, where the 3D luminosity density follows a radially exponential and vertically $\sech^2$ function, to the JWST galaxy images, corresponding to the rest-frame $K_s$ band ($z<1.45$) or $H$ band ($z>1.45$). Most galaxies are well fit with a single disc model, while 44 galaxies show systematic excess light above the disc midplane, necessitating a second disc component. With Bayesian Information Criteria (BIC) and visual inspection, we classify galaxies into categories of (1) well-fitted by a single disc and (2) requiring two discs (thin and thick discs). We also assessed the need for a bulge component. We find 44 `two disc' galaxies (25 with bulge and 19 without bulge) and 67 `single disc' galaxies (39 with bulge and 28 without bulge). The most distant two disc galaxy we identify is at redshift up to $z=1.96$. 

We identify well-defined correlations between some measured disc parameters across our sample, despite the wide baseline across cosmic time. 
The radial length, $h_R$, and vertical height, $z_0$, of all disc categories (single disc, thin and thick discs) correlate strongly with total stellar mass and disc mass, independent of the cosmic time. However, the ratio of radial length and vertical height ($h_R/z_0$) does not correlate with host galaxy mass or individual disc mass. Single discs occupy similar regions to thick discs rather than thin discs, suggesting the sequential formation that thick discs dominantly form first before galaxies develop a second, detectable thin disc (Figs.~\ref{fig:fig005} and \ref{fig:fig006}). 

The transition from single to double discs occurred around $\sim$ 8 Gyr ago in high-mass galaxies ($10^{9.75} - 10^{11}M_\odot$ ), somewhat earlier than the transition around $\sim$ 4 Gyr ago in low-mass galaxies ($10^{9.0} - 10^{9.75}M_\odot$). The shift of the onset indicates the sequential thick then thin disc formation proceeds in a "downsizing" manner, where higher mass galaxies tend to form thin discs earlier and lower mass galaxies increasingly form thin discs at later time (Fig.~\ref{fig:fig009}).

Lower mass galaxies have higher thick-to-thin disc mass ratios (Fig.~\ref{fig:fig010}), consistent with the delayed formation of thin disc in low mass galaxies and aligning with the results for $z=0$ galaxies \citep{yoachim_structural_2006, comeron_evidence_2014}. Both thin and thick disc masses increase with total stellar mass, roughly described by single slopes across a wide range of masses (2.5-3 dex, Fig.~\ref{fig:fig011}). The slope for thin discs is steeper than for thick discs, crossing at a $\log(M_*[M_\odot])\sim10$, creating the anti-correlation between the thick-to-thin disc mass ratios and galaxy stellar masses. 

Despite the dominant sequential picture of thick to thin disc formation revealed in this study, Fig.~\ref{fig:fig011} indicates the co-evolution of the two discs, with the thick disc continuously growing as the galaxy grows (although less efficient than thin disc growth). This is in contrast to a simple sequential scenario where two discs form in \textit{entirely separate epochs}. 

We propose that the Toomre-$Q$ self-regulated star formation coherently explains the above findings (Sec.~\ref{subsec:toomreq}), linking our structural measurements for stellar discs with available gas kinematics of gas discs from recent ALMA and ground-based IFU surveys (Fig.~\ref{fig:fig012}). High-mass galaxies achieve lower gas fractions early on, enabling them to host less turbulent gas discs and form thin discs earlier in time. The declining gas fraction over time allows more lower-mass galaxies to form thin discs at later epochs.

\section*{Acknowledgements}
We are grateful to the anonymous reviewer for their constructive feedback, which has significantly improved the clarity and quality of this paper. TT is grateful to the conference organizers and financial support of the ELT Science in Light of JWST meeting, held in Miyagi, Japan in June 2024. Attending the conference helped to focus this research and the eventual paper. TT thanks Bruce Elmegreen, Eric Emsellem, Mark Krumholz, Andreas Burkert, Masashi Chiba, Trevor Mendel, Lucas Kimmig, Lucas Valenzuela and Federico Lelli for insightful discussions and Mahsa Kohandel and Hannah Übler for sharing their data \citep{kohandel_dynamically_2024, ubler_evolution_2019}. This research was supported by the Australian
Research Council Centre of Excellence for All Sky Astrophysics in 3 Dimensions (ASTRO 3D), through project
number CE170100013. This work was supported in part by Japan Foundation for Promotion of Astronomy. This work is based on observations made with the NASA/ESA/CSA James Webb Space Telescope. The data were obtained from the Mikulski Archive for Space Telescopes at the Space Telescope Science Institute, which is operated by the Association of Universities for Research in Astronomy, Inc., under NASA contract NAS 5-03127 for JWST. These observations are associated with program \# 1180, 1210, 1895, 2079, 2514, 1181, 3577, 1345, 2750, 2279, 1727, 1837, 1283, 2198. The data products presented herein were retrieved from the Dawn JWST Archive (DJA). DJA is an initiative of the Cosmic Dawn Center, which is funded by the Danish National Research Foundation under grant No. 140. This work made use of the following software packages: \texttt{astropy} \citep{astropy:2013, astropy:2018, astropy:2022}, \texttt{Jupyter} \citep{2007CSE.....9c..21P, kluyver2016jupyter}, \texttt{matplotlib} \citep{Hunter:2007}, \texttt{numpy} \citep{numpy}, \texttt{pandas} \citep{mckinney-proc-scipy-2010, pandas_13819579}, \texttt{python} \citep{python}, \texttt{scipy} \citep{2020SciPy-NMeth, scipy_14880408}, \texttt{Cython}, and \texttt{scikit-image} \citep{scikit-image}. This research has made use of NASA's Astrophysics Data System. This research made use of Photutils, an Astropy package for detection and photometry of astronomical sources \citep{larry_bradley_2024_10967176} and \textsc{Mathematica} \citep{Mathematica}. Software citation information aggregated using \texttt{\href{https://www.tomwagg.com/software-citation-station/}{The Software Citation Station}} \citep{software-citation-station-paper, software-citation-station-zenodo}.


\section*{Data Availability}

The data products are available at DJA The DAWN JWST Archive \url{https://dawn-cph.github.io/dja/index.html}. The specific observations included in the mosaic images used in the paper can be accessed via DOI: \href{http://dx.doi.org/10.17909/ehkk-th30}{10.17909/ehkk-th30}



\bibliographystyle{mnras}
\bibliography{mnras_template} 




\appendix
\section{Our used datasets and observational programs}\label{sec:appendixa}
Table \ref{tab:taba1} summarize the observational programs used in this studies.
\begin{table*}
    \centering
    \caption{Summaries of the JWST programs included in the DJA mosaic images used in the paper. The table lists: the mosaic field name defined by DJA, the version of the reduced mosaic images used in the paper, JWST program IDs used to produce the mosaic images, and PI names of the observation program.}
    \begin{tabular}{|c|c|c|c|}
        \hline
        \textbf{Mosaic field name} & \textbf{DJA Version} & \textbf{JWST Program IDs} & \textbf{PI Names} \\ 
        \hline
        \multirow{4}{*}{GOODS-South (-SW)} & \multirow{4}{*}{7.1 (7.0)} & \#1180/1210 & D.Eisenstein/N.Luetzgendorf \\
         & & \#1895 & P.Oesch \\
         & & \#2079 & S.Finkelstein \\   
         & & \#2514 & C.Williams \\
        \hline
        \multirow{4}{*}{GOODS-North} & \multirow{4}{*}{7.3} & \#1181 & Eisenstein \\
         & & \#1895 & P.Oesch \\
         & & \#2514 & C.Williams \\
         & & \#3577 & E.Egami \\
        \hline
        \multirow{4}{*}{CEERS-full} & \multirow{4}{*}{7.2} & \#1345 & S.Finkelstein \\
        & & \#2750 & P.Arrabal Haro \\
        & & \#2514 & C.Williams \\
        & & \#2279 & R.Naidu \\
        \hline
        \multirow{2}{*}{\parbox{3cm}{\centering PRIMER-COSMOS-East \\ PRIMER-COSMOS-West}} & \multirow{2}{*}{7.0} & \#1727 & J.Kartaltepe \\
        & & \#1837 & J.Dunlop \\
        \hline
        \multirow{2}{*}{\parbox{3cm}{\centering PRIMER-UDS-South\\PRIMER-UDS-North}} & \multirow{2}{*}{7.0} & \multirow{2}{*}{\#1837} & \multirow{2}{*}{J.Dunlop} \\
        & & & \\
        \hline
        \multirow{3}{*}{\centering NGDEEP} & \multirow{3}{*}{7.0} & \#2079 & S.Finkelstein \\ 
        & & \#1283 & G.Oestlin \\ 
        & & \#2198 & L.Barrufet \\ 
        \hline
    \end{tabular}
    \label{tab:taba1}
\end{table*}

\section{Composite color images of galaxies}\label{sec:appendixb}
Figures~\ref{fig:figb1}, \ref{fig:figb2} and \ref{fig:figb3} show the thee colour composite JWST NIRCam (F115W/F277W/F444W) images of our sample galaxies continued from Fig.~\ref{fig:fig001}. For galaxies without those filter bands available, we instead show other filter JWST NIRCam images which are indicated in lower right corner. If fewer than three filters are available, we show a grayscale image. 

\begin{figure*}
    \centering
    \includegraphics[width=\linewidth]{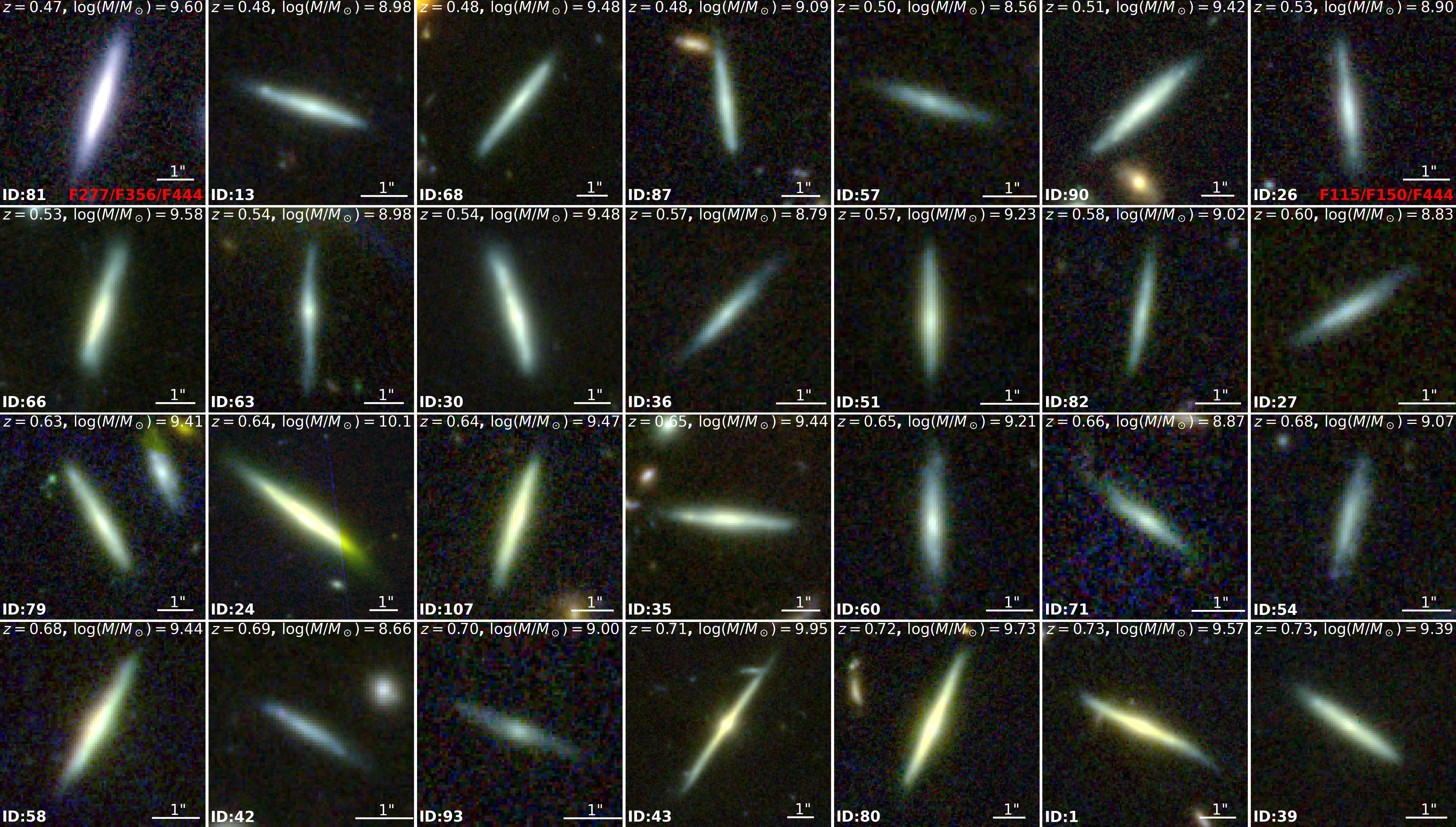}
    \caption{Continued from the Fig.\ref{fig:fig001}. NIRCam F227W; F356W; F444W colour composite images unless noted by red texts which denote the filters used to make the composite.}
    \label{fig:figb1}
\end{figure*}

\begin{figure*}
    \centering
    \includegraphics[width=\linewidth]{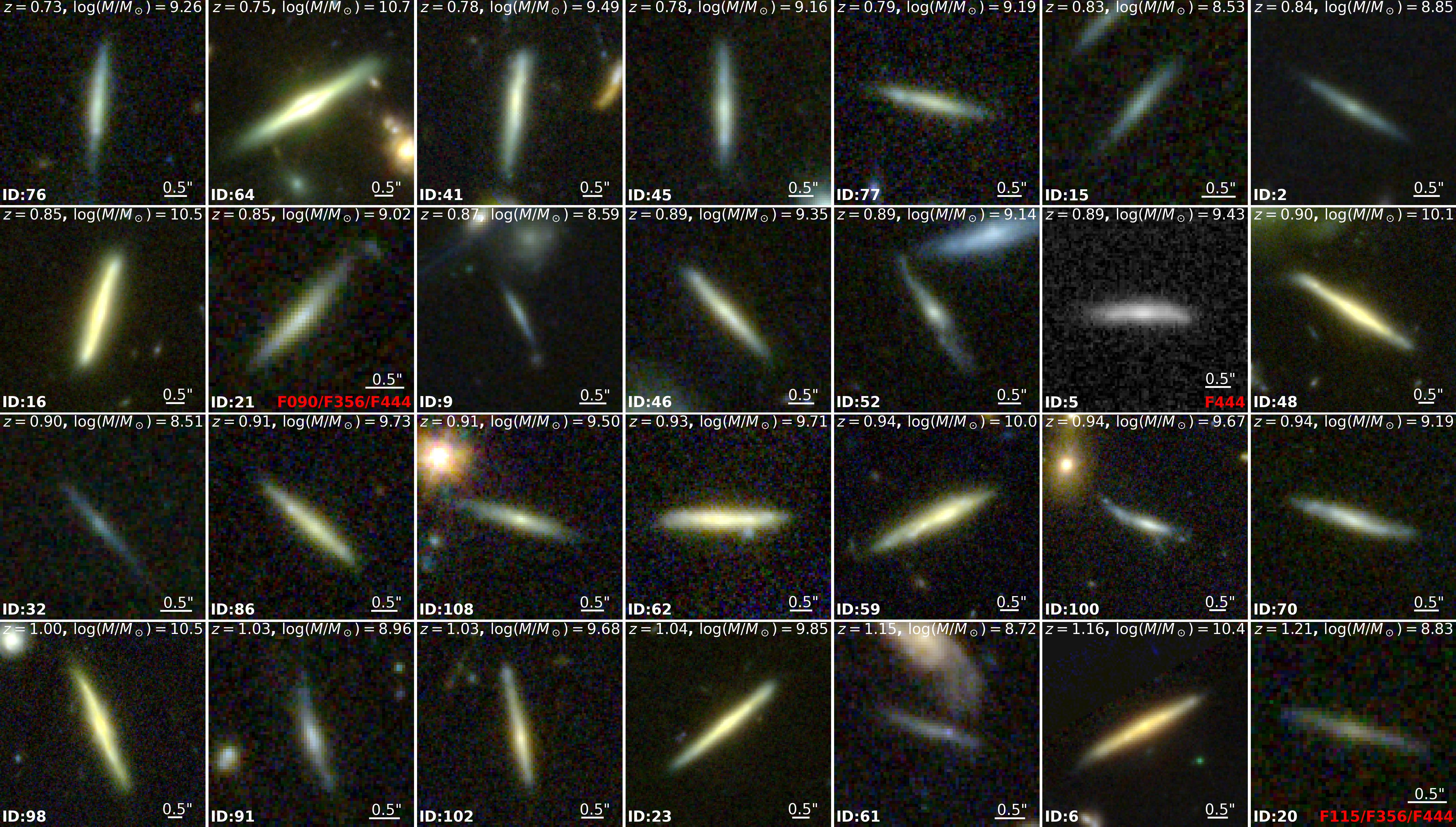}
    \caption{Continued from the Fig.\ref{fig:fig001}. 0.5" scale bar is shown instead of 1". }
    \label{fig:figb2}
\end{figure*}

\begin{figure*}
    \centering
    \includegraphics[width=\linewidth]{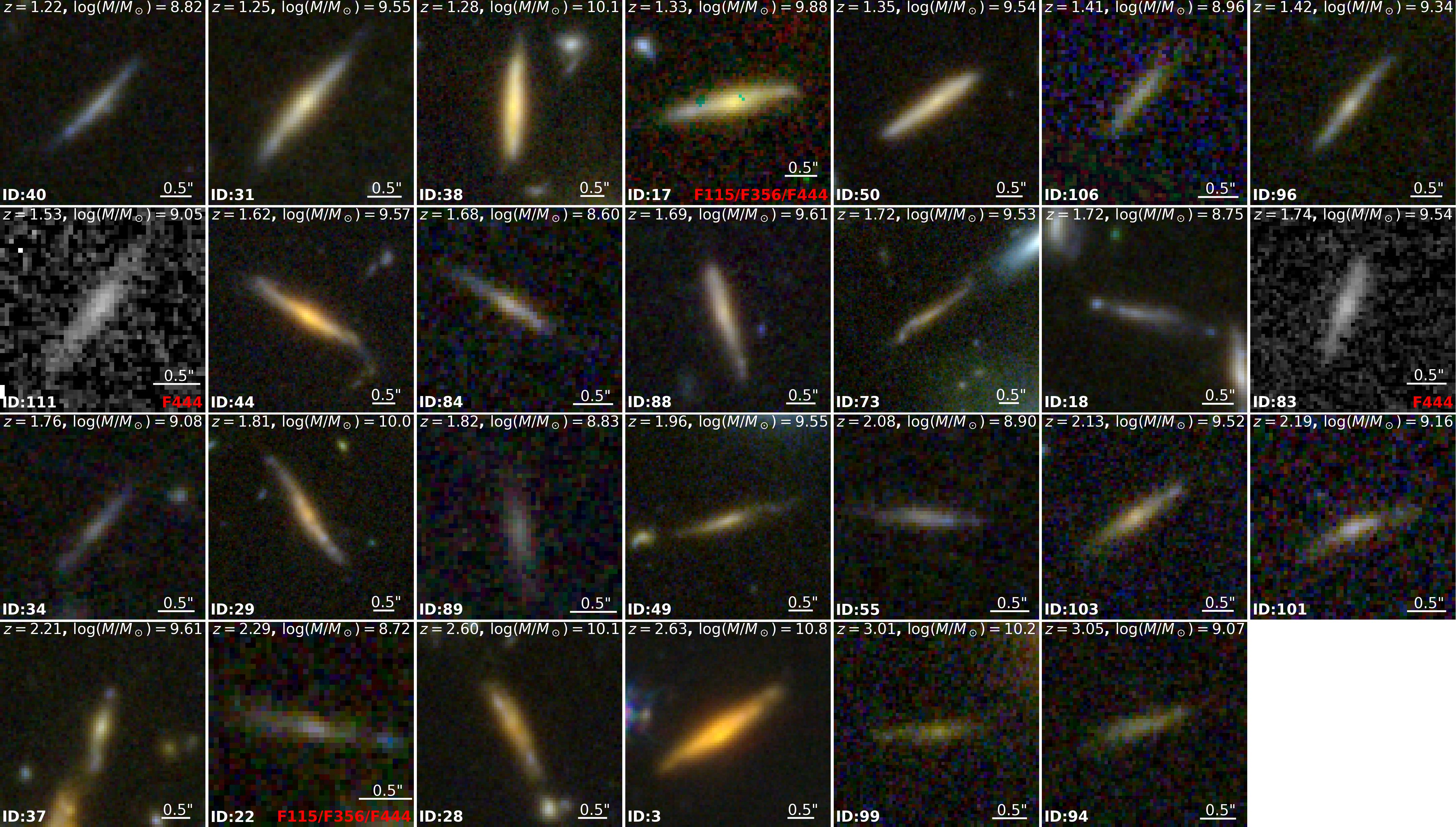}
    \caption{Continued from the Fig.\ref{fig:fig001}.}
    \label{fig:figb3}
\end{figure*}

\section{Fitting systematics and uncertainties}\label{sec:appendixc}
In this section we assessed systematic uncertainty on our structural measurements. 
\subsection{Minimum recoverable disc scale height}
\label{subsec:appendixc1}
\begin{figure}
    \centering
    \includegraphics[width=\linewidth]{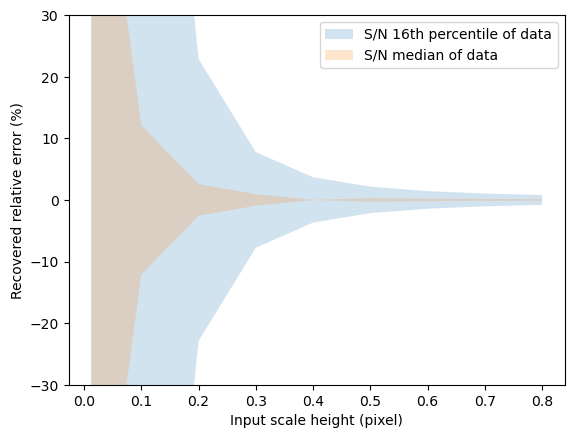}
    \caption{Relative error in scale height as a function of input scale height (in pixels). The shaded regions represent 1$\sigma$ uncertainty for 16th percentile of the signal-to-noise ratio (S/N)}
    \label{fig:figc1}
\end{figure}

The first consideration is how reliably we can recover the scale height relative to image pixel sizes. To demonstrate this, we generate mock simulated images with a typical disc scale radius of $\sim$10 pixels and a range of disc scale heights. We convolve the simulated disc with the F444W PSF, which provides the most conservative estimate of accuracy (as the F444W PSF is the largest compared to other filters F277W, F356W). We then refit the simulated image using an appropriate variance image, sky noise, read noise, and Poisson variance of the source based on the typical effective gain. Figure~\ref{fig:figc1} shows the relative error in the recovered scale height as a function of the input scale height for both the 16th percentile signal-to-noise (SN), where most of the sample (84\%) has a higher SN and thus better accuracy, and the median SN for the sample. This demonstrates that we can recover the scale height down to 0.2 pixels with $\sim$20\% accuracy for most of the data. We adopt this value as a conservative lower limit for galaxies that reach this boundary.

\subsection{Inclination deviation from a perfect edge-on}\label{subsec:appendixc2}
\begin{figure*}
    \centering
    \includegraphics[width=0.33\linewidth]{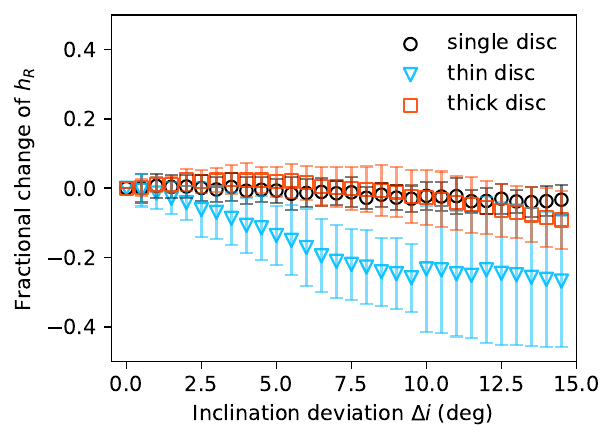}
    \includegraphics[width=0.33\linewidth]{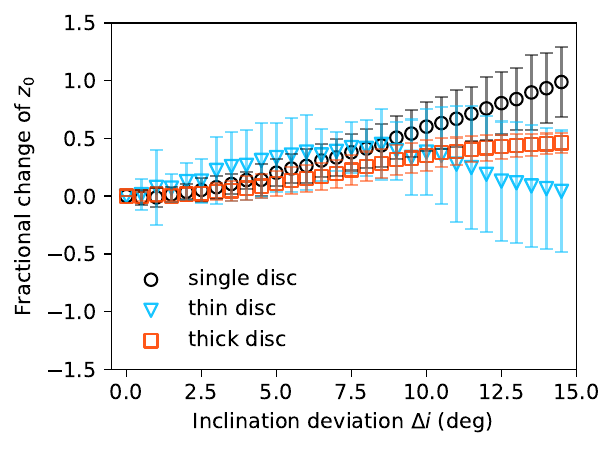}
    \includegraphics[width=0.33\linewidth]{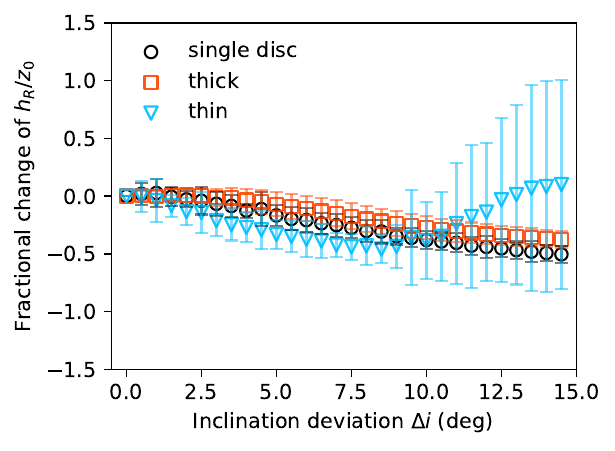}
    \caption{Fractional change of the derived parameters are shown as a function of inclination deviation from 90 deg $\Delta i$ (deg). The error bars denote the range encompassed by our sample of galaxies (1 $\sigma$)}
    \label{fig:figc2}
\end{figure*}

We quantify the bias introduced by deviations from a perfect edge-on orientation. Slight inclination deviations, of $<7$ deg, expected for our sample (64th percentile) has been shown to minimally affect structural measurements \citep{de_grijs_z-structure_1997}. However, this affect may vary depending on the image quality, the intrinsic properties of the discs, and the presence of a second disc or central bulge. To assess whether inclination effects affect our results we generate simulated images based on the measured structural parameters of our galaxy sample (111 best-fit models), varying the disc inclination over a range of $\Delta i = [0, 15$ deg] in 0.5 deg intervals. The simulated images were further convolved with the PSF and had Gaussian white noise added from the original variance map. We then refit these convolved simulated images using a model assuming a 90 degree inclination (as adopted in our study) and obtained the fractional biases in the structural parameters, shown in Fig.~\ref{fig:figc2}. 
We use the best fit model classified in Section \ref{sec:results} (e.g., a disc, a disc + bulge, two discs, and two discs + bulge), where single disc, thin disc, and thick disc are denoted in black circle, blue triangle, red square respectively. The error bar shows the standard deviation encompassing our sample of galaxies, representing galaxy to galaxy variation due to the intrinsic structure of disc and image quality for our sample. 

As expected the bias is largest for thin disc structures and smallest for thick disc structures. We found 30\%, 20\% and 10\% overestimation are expected for vertical scale height measurements if galaxies have $5$ deg deviation from the perfect edge-on for thin, single, and thick discs respectively. The same value of 30\% is found in \citet{de_grijs_z-structure_1997} for a similar intrinsic axial ratio q=0.11 of a thin disc.   
For galaxies with $q<0.3$ adopted in our sample (see Fig.~\ref{fig:fig003}), the assumption of 90 deg lead to median bias of 0.1\%, 12\%, 1\% underestimation for disc scale radii, 17\%, 27\%, 9\% overestimation for disc scale height, and 13\%, 27\%, 6\% underestimation for disc radii/height ratios (values are denoted for single disc, thin disc, and thick disc respectively). These variations contribute to the measured scatter in the reported values. 
 
\subsection{Systematics from nuisance disc sub-structures such as dust lane, disc truncation, bulges.}\label{subsec:appendixc3}
\begin{figure*}
    \centering
    \includegraphics[width=\linewidth]{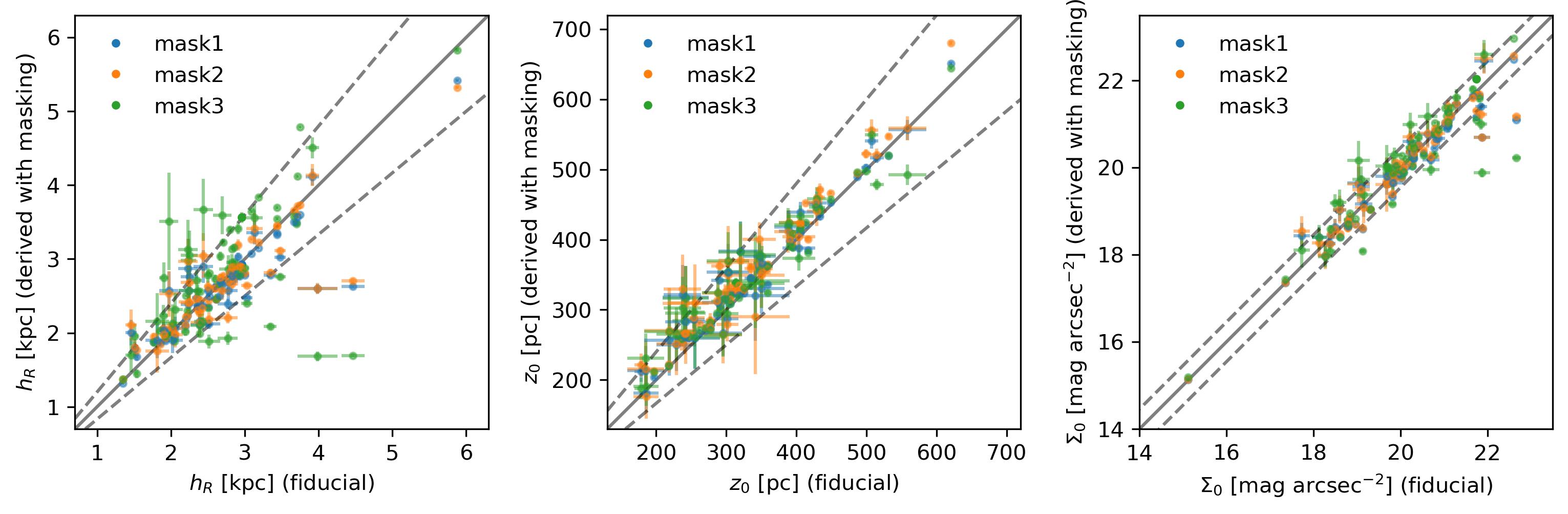}
    \caption{The measurements for galaxies classified as `Single disc' and `disc + bulge' are compared against fitting results with different masks: mask1 (`bulge mask'), mask2 (`bulge mask' + `midplane mask') and, mask3 (`bulge mask' + `disc truncation mask'). See the text for details. For visual aid, the solid line shows the 1:1 relation and the dashed lines show 20\% range (of linear quantities) from the 1:1 relation.}
    \label{fig:figc3}
\end{figure*}

\begin{figure*}
    \centering
    \includegraphics[width=\linewidth]{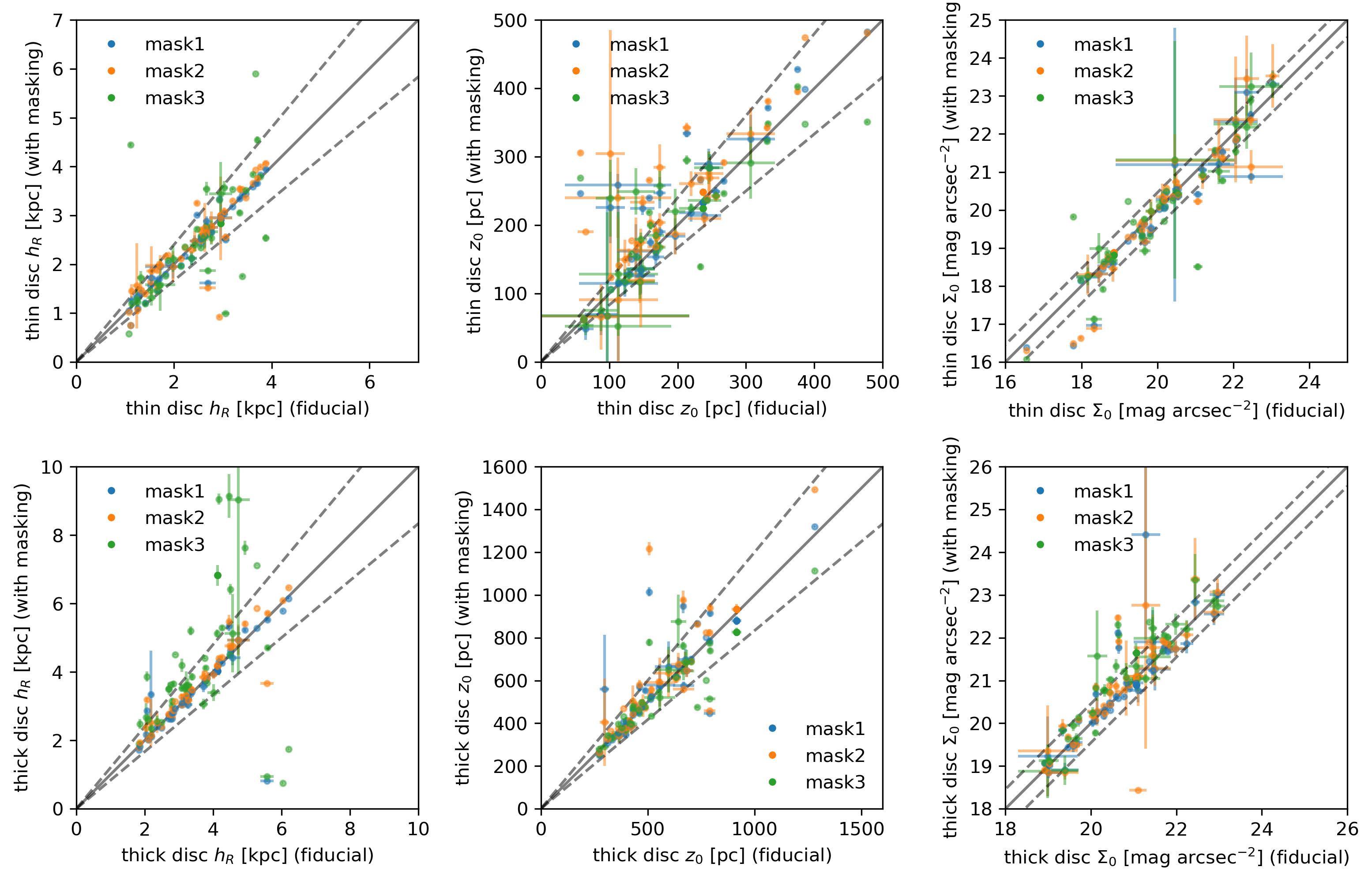}
    \caption{The measurements for galaxies classified as 'two discs' and 'two discs + bugle' are compared against fitting results with different masks (same as Fig.~\ref{fig:figc3}). For visual aid, solid line shows the one-to-one line and the dashed line shows +/- 20\% from the one-to-one line in linear scaled measurements.}
    \label{fig:figc4}
\end{figure*}

Figures~\ref{fig:figc3} and \ref{fig:figc4} summarize our assessment of systematic uncertainties in our modeling due to unaccounted disc substructures and potential dust extinction. To evaluate how much our fiducial results are affected by these substructures, we repeat the fitting procedure using different masks. A `bulge mask ($r<1.5$ kpc)', a circular mask with a 1.5 kpc radius centred on the galaxy, is used to exclude the central concentrated light. A `midplane mask ($z<z_0$)', masks the region where the latitude is less than the disc scale height $z_0$, is used to account for disc substructures and potential dust extinction. Finally, `disc truncation mask ($R>3h_R$)' masks the unmodeled outer trunctation of the disc to assess the potential impact of this omission. We use the combination of those masks: mask1 (`bulge mask'), mask2 (`bulge mask' + `midplane mask') and, mask3 (`bulge mask' + `disc truncation mask') and refit a two disc model for two disc galaxies and one disc for single disc galaxies without a bulge component, which is masked in all cases. 

Figures~\ref{fig:figc3} and \ref{fig:figc4} compare the refit results with masking against our fiducial fit results for galaxies classified as single- and double-disc galaxies, respectively. The impact of masking results is summarized by the table \ref{tab:fractional_error}. The scatter and mean values quantify the deviation introduced by masking, relative to the fiducial (unmasked) fit. The comparison shows that masking does not introduce a significant systematic shift in the refit results, as the mean deviation is smaller than the introduced scatter. This suggests that the reduced amount of available data to constrain the structural parameters has a greater effect than the structural differences in the masked regions. As expected, the midplane mask primarily affects disc heights, while the truncation mask has a greater impact on disc radii. This validates the fiducial fit and confirms that the potential midplane disc structure and disc truncations do not affect our measurements significantly. The moderate mean deviation for height measurements using midplane mask is consistent with the dust effect $\sim11\%$ estimated in similar bands \citep{bizyaev_structural_2009}.

\begin{table*}
    \centering
    \caption{Fractional scatter and mean deviation of measurements obtained using different masks relative to the fiducial (unmasked) fit.}
    \label{tab:fractional_error}
    \begin{tabular}{lcccc}
        \hline
        Parameter & Disc Component & Mask 1 (Bulge) & Mask 2 (Bulge + Midplane) & Mask 3 (Bulge + Truncation) \\
        \hline
        & & Scatter / Mean dev. (\%) & Scatter / Mean dev. (\%) & Scatter / Mean dev. (\%) \\
        \hline
        \( z_{\text{single}} \) & Single Disc & 8.21 / -2.59 & 8.78 / -6.04 & 8.82 / -3.09 \\
        \( z_{\text{thin}} \) & Thin Disc & 11.07 / -1.30 & 22.29 / -14.45 & 19.34 / 0.02 \\
        \( z_{\text{thick}} \) & Thick Disc & 6.79 / -0.94 & 8.90 / -3.40 & 7.37 / 1.34 \\
        \hline
        \( h_{\text{single}} \) & Single Disc & 12.16 / 1.62 & 12.47 / -0.49 & 21.92 / -6.70 \\
        \( h_{\text{thin}} \) & Thin Disc & 7.62 / 0.47 & 11.61 / -1.59 & 16.69 / 1.96 \\
        \( h_{\text{thick}} \) & Thick Disc & 3.01 / 0.79 & 4.54 / -2.89 & 40.26 / -14.31 \\
        \hline
    \end{tabular}
\end{table*}

Considering the discussions in this section and Appendix \ref{subsec:appendixc2}, the systematic uncertainty from the disc inclination dominates in our measurements. The thin disc is most affected by the inclination effect, introducing a median bias of approximately 12\% underestimation and 27\% overestimation in the measured disc size and height, with smaller biases for single and thick discs. Therefore, the systematic uncertainty is minor compared to the dynamic range of the measurements and seen trend (e.g., Fig. \ref{fig:fig005} and \ref{fig:fig006}). The visual exclusion of face-on discs with dust morphology further reduce the uncertainty than this estimate. These uncertainties contribute primarily to the scatter in the correlation, while the median trends are less affected.

\section{Distribution of galaxy and disc structural parameters}
\begin{figure*}
    \centering
    \includegraphics[width=1.\linewidth]{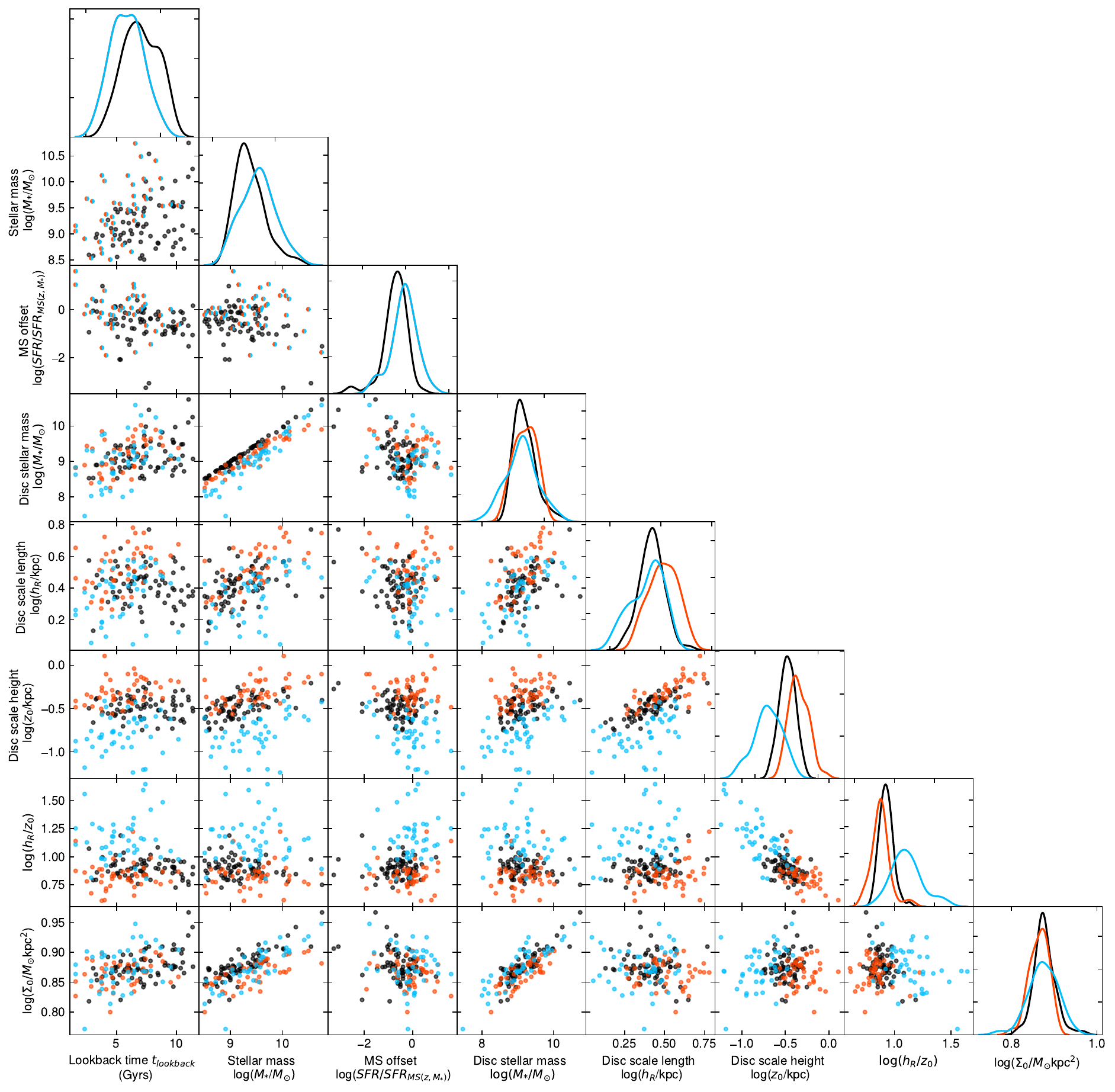}
    \caption{A corner plot of physical properties of host galaxies and individual discs, with single, thin, and thick discs represented by black, blue, and red points, respectively.}
    \label{fig:fige1}
\end{figure*}

\begin{figure*}
    \centering
    \includegraphics[width=1\linewidth]{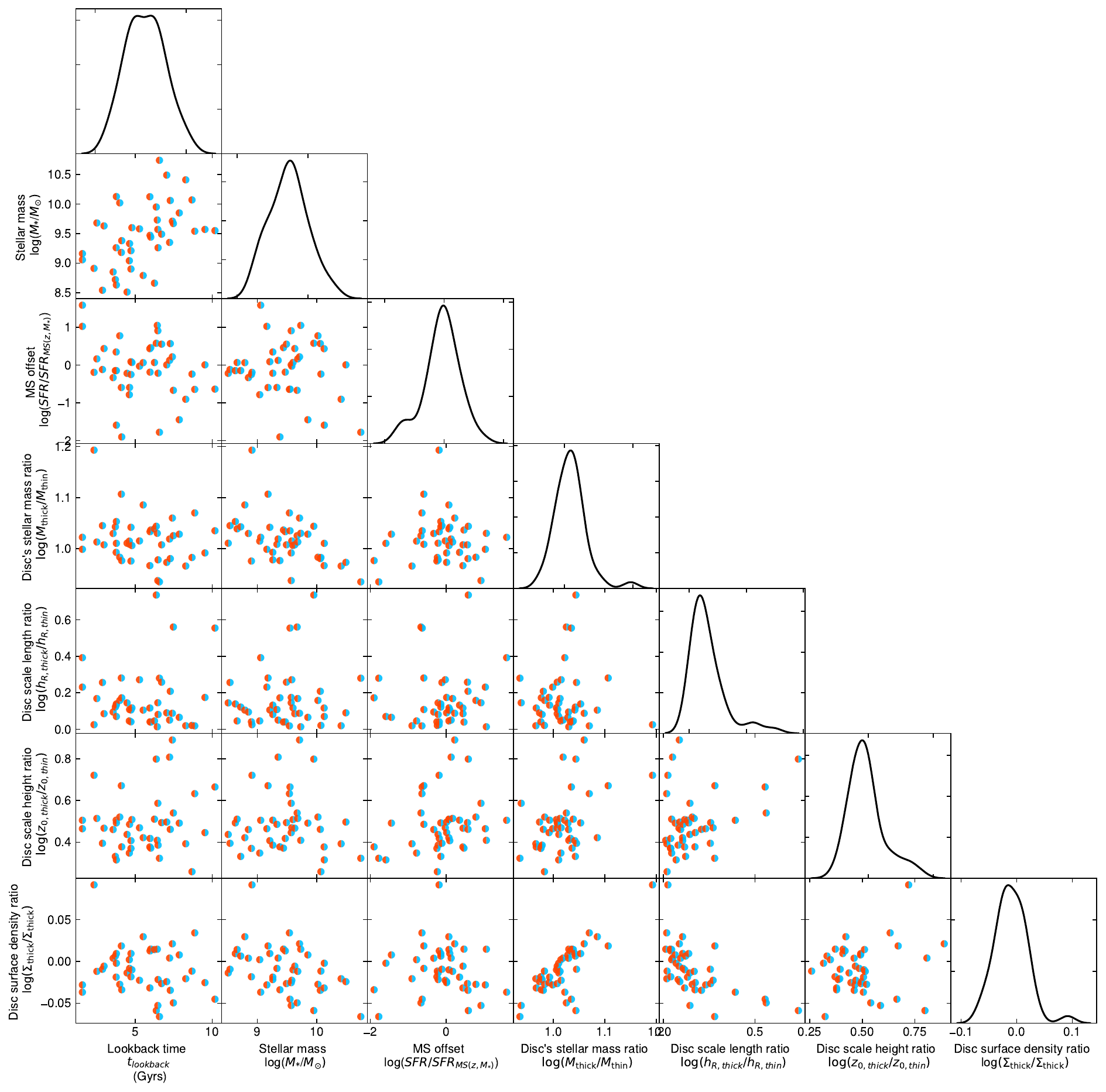}
    \caption{A corner plot of physical properties of two disc galaxies (blue-red points).}
    \label{fig:fige2}
\end{figure*}

Table.~\ref{tab:taba2} presents the properties of the edge-on galaxies in this study, including physical parameters extracted from 3D-HST \citep{skelton_3d-hst_2014, momcheva_3d-hst_2016} and measured disc structural parameters. Figure~\ref{fig:fige1} show the physical properties of host galaxies and individual discs (single, thin, and thick discs) with scatter plots illustrating the relationships between different parameters and diagonal panel displaying the histogram of each parameter. The offest from the galaxy main sequence is computed using the main sequence defined by \citet{popesso_main_2023}. Figure~\ref{fig:fige2} show similar corner plot showing the physical properties of two-disc galaxies, focusing on the inter-correlation of thin and thick discs within the same galaxies.  

\onecolumn
\begin{landscape}
\setlength\LTcapwidth{\linewidth}
\begin{longtable}{rrrcrrcccccccccc}
\caption{\textbf{Properties of edge-on galaxies in our sample:} galaxy ID, Right ascention (R.A.), Declination (Decl.), redshift ($z$), stellar mass (log($M_*/M_\odot$), and the best-fit model. The best-fit model is denoted as 0: a disc, 1: a disc + bulge, 2: two discs, 3 two discs + bulge (see details in Sec.~\ref{sec:3dmodel}). For each galaxy, it lists the radial scale length $h_{R}$, vertical scale height ($z_0$), midplane intensity $J_0$, disc stellar mass $M_{\mathrm{disc}}$ for either the single or thin disc (in the case of galaxies with two discs), followed by those for thick disc. R.A. and Decl. are sourced from the DJA photometric catalogue \citep{valentino_atlas_2023}. Redshift and stellar mass are extracted from HST-3D catalogue \citep{skelton_3d-hst_2014, momcheva_3d-hst_2016}. Individual disc masses are derived by assuming the same mass to light ratio for all components, based on the best-fit model and the total stellar mass (Sec.~\ref{sec:results}). Statistical uncertainties ($1\sigma$) are provided for individual parameters, and systematic uncertainties are discussed in Appendix~\ref{sec:appendixc}.}\label{tab:taba2}\\
    \hline
   ID & R.A. & Decl. & $z$ & log($M_{*}$) & Best & $h_{R, \mathrm{single/thin}}$ & $z_{0, \mathrm{single/thin}}$ & $J_{0, \mathrm{single/thin}}$ & log$M_{\mathrm{disc}, \mathrm{single/thin}}$ & $h_{R, \mathrm{thick}}$ & $z_{0, \mathrm{thick}}$ & $J_{0, \mathrm{thick}}$ & log$M_{\mathrm{disc}, \mathrm{thick}}$\\
    & (deg) & (deg) & & $M_\odot$ & model & (kpc) & (pc) & (mag arcsec$^{2}$) & $M_\odot$ & (kpc) & (pc) & (mag arcsec$^{2}$) & $M_\odot$\\
\hline
   1 &  53.0997 & -27.7863  & 0.731 &  9.57    & 3 & 2.96$^{\pm0.01}$ & 237$^{\pm3}$  & 19.25$^{\pm0.01}$ & 9.45$^{\pm0.05}$  & 4.13$^{\pm0.08}$ & 915$^{\pm23}$ & 22.56$^{\pm0.09}$ & 8.86$^{\pm0.06}$  \\
   2 &  53.1575 & -27.7765  & 0.84  &  8.85    & 1 & 2.96$^{\pm0.05}$ & 314$^{\pm7}$  & 22.71$^{\pm0.04}$ & 8.82$^{\pm0.98}$  & -                & -             & -                 & -                 \\
   3 &  53.0614 & -27.7728  & 2.632 & 10.75    & 0 & 2.22$^{\pm0.02}$ & 395$^{\pm4}$  & 19.60$^{\pm0.02}$ & -                 & -                & -             & -                 & -                 \\
   4 &  53.1368 & -27.7689  & 0.366 & 10.02    & 3 & 3.67$^{\pm0.00}$ & 386$^{\pm0}$  & 18.89$^{\pm0.00}$ & 9.78$^{\pm0.01}$  & 5.28$^{\pm0.00}$ & 1281$^{\pm1}$ & 21.00$^{\pm0.00}$ & 9.61$^{\pm0.01}$  \\
   5 &  53.1007 & -27.7668  & 0.893 &  9.43    & 1 & 2.71$^{\pm0.03}$ & 346$^{\pm5}$  & 21.01$^{\pm0.03}$ & 9.42$^{\pm0.22}$  & -                & -             & -                 & -                 \\
   6 &  53.1985 & -27.7442  & 1.155 & 10.41    & 3 & 3.62$^{\pm0.02}$ & 256$^{\pm6}$  & 19.91$^{\pm0.02}$ & 10.24$^{\pm0.08}$ & 3.78$^{\pm0.05}$ & 632$^{\pm23}$ & 21.81$^{\pm0.14}$ & 9.89$^{\pm0.09}$  \\
   7 & 189.22   &  62.153   & 0.457 &  8.9     & 3 & 2.55$^{\pm0.02}$ & 169$^{\pm5}$  & 20.95$^{\pm0.03}$ & 8.69$^{\pm0.19}$  & 2.80$^{\pm0.04}$ & 397$^{\pm14}$ & 22.51$^{\pm0.15}$ & 8.48$^{\pm0.18}$  \\
   8 & 189.217  &  62.1614  & 0.434 &  9.11    & 1 & 2.88$^{\pm0.01}$ & 348$^{\pm1}$  & 21.09$^{\pm0.01}$ & 9.05$^{\pm0.10}$  & -                & -             & -                 & -                 \\
   9 & 189.247  &  62.1581  & 0.866 &  8.59    & 0 & 1.53$^{\pm0.03}$ & 178$^{\pm10}$ & 22.23$^{\pm0.06}$ & -                 & -                & -             & -                 & -                 \\
  10 & 189.269  &  62.1677  & 0.226 &  8.6     & 1 & 3.75$^{\pm0.01}$ & 318$^{\pm1}$  & 21.56$^{\pm0.00}$ & 8.52$^{\pm0.05}$  & -                & -             & -                 & -                 \\
  11 & 189.275  &  62.1695  & 0.443 &  9.04    & 2 & 1.65$^{\pm0.03}$ & 123$^{\pm11}$ & 20.71$^{\pm0.07}$ & 8.67$^{\pm0.05}$  & 1.84$^{\pm0.03}$ & 274$^{\pm11}$ & 21.38$^{\pm0.19}$ & 8.80$^{\pm0.04}$  \\
  12 & 189.218  &  62.1722  & 0.457 &  9.21    & 3 & 3.34$^{\pm0.05}$ & 174$^{\pm8}$  & 21.27$^{\pm0.04}$ & 8.65$^{\pm0.09}$  & 4.26$^{\pm0.03}$ & 444$^{\pm6}$  & 21.65$^{\pm0.06}$ & 9.01$^{\pm0.08}$  \\
  13 & 189.295  &  62.1914  & 0.476 &  8.98    & 1 & 2.21$^{\pm0.01}$ & 261$^{\pm1}$  & 20.92$^{\pm0.01}$ & 8.96$^{\pm0.19}$  & -                & -             & -                 & -                 \\
  14 & 189.228  &  62.1909  & 0.254 &  9.63    & 3 & 2.22$^{\pm0.00}$ & 114$^{\pm1}$  & 18.56$^{\pm0.01}$ & 9.29$^{\pm0.01}$  & 2.71$^{\pm0.00}$ & 334$^{\pm1}$  & 19.76$^{\pm0.01}$ & 9.36$^{\pm0.01}$  \\
  15 & 189.048  &  62.2236  & 0.83  &  8.53    & 0 & 1.90$^{\pm0.06}$ & 253$^{\pm16}$ & 22.47$^{\pm0.08}$ & -                 & -                & -             & -                 & -                 \\
  16 & 189.155  &  62.2262  & 0.846 & 10.49    & 3 & 3.39$^{\pm0.02}$ & 233$^{\pm3}$  & 19.07$^{\pm0.02}$ & 10.30$^{\pm0.04}$ & 4.18$^{\pm0.04}$ & 731$^{\pm10}$ & 21.23$^{\pm0.05}$ & 10.02$^{\pm0.04}$ \\
  17 & 189.022  &  62.2328  & 1.332 &  9.88    & 1 & 2.83$^{\pm0.08}$ & 390$^{\pm15}$ & 21.07$^{\pm0.08}$ & 9.82$^{\pm0.21}$  & -                & -             & -                 & -                 \\
  18 & 189.354  &  62.2369  & 1.724 &  8.75    & 0 & 3.13$^{\pm0.11}$ & 388$^{\pm20}$ & 23.05$^{\pm0.08}$ & -                 & -                & -             & -                 & -                 \\
  19 & 189.096  &  62.242   & 0.191 &  8.91    & 2 & 2.36$^{\pm0.11}$ & 66$^{\pm12}$  & 22.97$^{\pm0.15}$ & 7.46$^{\pm0.10}$  & 2.50$^{\pm0.01}$ & 345$^{\pm1}$  & 21.24$^{\pm0.01}$ & 8.89$^{\pm0.00}$  \\
  20 & 189.281  &  62.2533  & 1.205 &  8.83    & 0 & 2.24$^{\pm0.10}$ & 185$^{\pm27}$ & 22.10$^{\pm0.16}$ & -                 & -                & -             & -                 & -                 \\
  21 & 189.263  &  62.2741  & 0.852 &  9.02    & 0 & 1.76$^{\pm0.02}$ & 237$^{\pm7}$  & 21.15$^{\pm0.04}$ & -                 & -                & -             & -                 & -                 \\
  22 & 189.177  &  62.2777  & 2.288 &  8.72    & 0 & 1.97$^{\pm0.13}$ & 219$^{\pm36}$ & 22.57$^{\pm0.19}$ & -                 & -                & -             & -                 & -                 \\
  23 & 189.134  &  62.2792  & 1.038 &  9.85    & 3 & 3.88$^{\pm0.07}$ & 148$^{\pm18}$ & 19.97$^{\pm0.06}$ & 9.38$^{\pm0.09}$  & 4.51$^{\pm0.04}$ & 460$^{\pm12}$ & 20.70$^{\pm0.10}$ & 9.64$^{\pm0.08}$  \\
  24 & 189.398  &  62.2865  & 0.64  & 10.1237  & 3 & 3.77$^{\pm0.01}$ & 332$^{\pm3}$  & 18.97$^{\pm0.01}$ & 9.95$^{\pm0.02}$  & 4.93$^{\pm0.04}$ & 791$^{\pm11}$ & 21.03$^{\pm0.06}$ & 9.62$^{\pm0.03}$  \\
  25 & 189.143  &  62.2891  & 0.446 &  9.33    & 3 & 2.55$^{\pm0.02}$ & 168$^{\pm4}$  & 19.95$^{\pm0.02}$ & 8.97$^{\pm0.06}$  & 3.27$^{\pm0.02}$ & 514$^{\pm6}$  & 21.24$^{\pm0.04}$ & 9.05$^{\pm0.05}$  \\
  26 & 189.392  &  62.293   & 0.531 &  8.9     & 1 & 2.39$^{\pm0.02}$ & 359$^{\pm5}$  & 21.32$^{\pm0.04}$ & 8.83$^{\pm0.19}$  & -                & -             & -                 & -                 \\
  27 & 189.199  &  62.2983  & 0.601 &  8.83    & 0 & 1.99$^{\pm0.02}$ & 292$^{\pm5}$  & 21.50$^{\pm0.03}$ & -                 & -                & -             & -                 & -                 \\
  28 & 189.332  &  62.3023  & 2.601 & 10.1     & 0 & 2.60$^{\pm0.04}$ & 432$^{\pm8}$  & 21.72$^{\pm0.03}$ & -                 & -                & -             & -                 & -                 \\
  29 & 189.135  &  62.3049  & 1.805 & 10.02    & 1 & 3.70$^{\pm0.04}$ & 500$^{\pm7}$  & 21.67$^{\pm0.04}$ & 9.97$^{\pm0.22}$  & -                & -             & -                 & -                 \\
  30 & 189.391  &  62.3072  & 0.545 &  9.48    & 1 & 3.09$^{\pm0.01}$ & 413$^{\pm2}$  & 20.49$^{\pm0.02}$ & 9.45$^{\pm0.06}$  & -                & -             & -                 & -                 \\
  31 & 189.26   &  62.3136  & 1.246 &  9.55    & 1 & 2.61$^{\pm0.07}$ & 303$^{\pm15}$ & 21.23$^{\pm0.09}$ & 9.47$^{\pm0.32}$  & -                & -             & -                 & -                 \\
  32 & 214.827  &  52.7416  & 0.903 &  8.51    & 0 & 2.23$^{\pm0.13}$ & 242$^{\pm33}$ & 23.05$^{\pm0.16}$ & -                 & -                & -             & -                 & -                 \\
  33 & 214.856  &  52.7623  & 0.244 &  8.54    & 2 & 1.13$^{\pm0.03}$ & 174$^{\pm9}$  & 21.73$^{\pm0.05}$ & 8.02$^{\pm0.04}$  & 2.04$^{\pm0.04}$ & 431$^{\pm11}$ & 22.45$^{\pm0.11}$ & 8.38$^{\pm0.02}$  \\
  34 & 214.833  &  52.747   & 1.764 &  9.08    & 0 & 2.44$^{\pm0.13}$ & 302$^{\pm30}$ & 23.12$^{\pm0.13}$ & -                 & -                & -             & -                 & -                 \\
  35 & 215.122  &  52.9561  & 0.65  &  9.44    & 3 & 3.06$^{\pm0.08}$ & 220$^{\pm18}$ & 21.13$^{\pm0.08}$ & 8.90$^{\pm0.11}$  & 3.35$^{\pm0.04}$ & 517$^{\pm12}$ & 21.35$^{\pm0.10}$ & 9.22$^{\pm0.10}$  \\
  36 & 215.074  &  52.9248  & 0.571 &  8.79    & 3 & 1.72$^{\pm0.30}$ & 113$^{\pm78}$ & 22.22$^{\pm0.42}$ & 8.00$^{\pm0.59}$  & 2.19$^{\pm0.07}$ & 298$^{\pm22}$ & 21.83$^{\pm0.33}$ & 8.68$^{\pm0.46}$  \\
  37 & 215.027  &  52.8952  & 2.211 &  9.61    & 1 & 3.35$^{\pm0.09}$ & 515$^{\pm10}$ & 22.92$^{\pm0.07}$ & 9.49$^{\pm0.61}$  & -                & -             & -                 & -                 \\
  38 & 214.876  &  52.7871  & 1.283 & 10.07    & 2 & 3.03$^{\pm0.07}$ & 307$^{\pm36}$ & 20.31$^{\pm0.21}$ & 9.85$^{\pm0.12}$  & 3.18$^{\pm0.12}$ & 556$^{\pm73}$ & 21.44$^{\pm0.68}$ & 9.67$^{\pm0.18}$  \\
  39 & 214.775  &  52.7504  & 0.733 &  9.39    & 1 & 3.19$^{\pm0.01}$ & 395$^{\pm2}$  & 20.54$^{\pm0.01}$ & 9.38$^{\pm0.09}$  & -                & -             & -                 & -                 \\
  40 & 214.92   &  52.8747  & 1.22  &  8.82    & 0 & 2.35$^{\pm0.06}$ & 302$^{\pm16}$ & 22.39$^{\pm0.07}$ & -                 & -                & -             & -                 & -                 \\
  41 & 214.814  &  52.805   & 0.778 &  9.49    & 2 & 2.47$^{\pm0.03}$ & 214$^{\pm6}$  & 20.12$^{\pm0.02}$ & 9.15$^{\pm0.02}$  & 4.47$^{\pm0.06}$ & 666$^{\pm11}$ & 21.83$^{\pm0.07}$ & 9.22$^{\pm0.02}$  \\
  42 & 214.895  &  52.8719  & 0.691 &  8.66    & 2 & 1.98$^{\pm0.12}$ & 113$^{\pm58}$ & 21.99$^{\pm0.20}$ & 8.17$^{\pm0.19}$  & 2.20$^{\pm0.08}$ & 367$^{\pm36}$ & 22.59$^{\pm0.36}$ & 8.49$^{\pm0.09}$  \\
  43 & 214.951  &  52.9364  & 0.713 &  9.95    & 3 & 1.11$^{\pm0.01}$ & <58            & 17.66$^{\pm0.02}$ & 9.30$^{\pm0.03}$  & 6.04$^{\pm0.02}$ & 362$^{\pm1}$  & 20.48$^{\pm0.01}$ & 9.71$^{\pm0.03}$  \\
  44 & 214.765  &  52.8184  & 1.616 &  9.57    & 3 & 2.69$^{\pm0.17}$ & 239$^{\pm24}$ & 20.63$^{\pm0.19}$ & 9.16$^{\pm0.17}$  & 4.02$^{\pm0.19}$ & 668$^{\pm50}$ & 22.37$^{\pm0.31}$ & 9.08$^{\pm0.18}$  \\
  45 & 214.845  &  52.9024  & 0.782 &  9.16    & 1 & 2.86$^{\pm0.02}$ & 335$^{\pm3}$  & 21.22$^{\pm0.02}$ & 9.13$^{\pm0.23}$  & -                & -             & -                 & -                 \\
  46 & 150.158  &   2.30622 & 0.886 &  9.35    & 2 & 2.42$^{\pm0.09}$ & <62            & 20.02$^{\pm0.10}$ & 8.96$^{\pm0.04}$  & 2.71$^{\pm0.09}$ & 399$^{\pm15}$ & 21.74$^{\pm0.11}$ & 9.13$^{\pm0.03}$  \\
  47 & 150.185  &   2.30793 & 0.339 &  9.26    & 3 & 1.52$^{\pm0.01}$ & 158$^{\pm2}$  & 19.37$^{\pm0.02}$ & 8.96$^{\pm0.05}$  & 2.06$^{\pm0.02}$ & 507$^{\pm6}$  & 21.13$^{\pm0.04}$ & 8.90$^{\pm0.05}$  \\
  48 & 150.168  &   2.31407 & 0.896 & 10.06    & 3 & 3.47$^{\pm0.02}$ & 243$^{\pm4}$  & 19.23$^{\pm0.01}$ & 9.84$^{\pm0.04}$  & 5.60$^{\pm0.06}$ & 704$^{\pm11}$ & 21.39$^{\pm0.07}$ & 9.65$^{\pm0.04}$  \\
  49 & 150.156  &   2.34599 & 1.961 &  9.55    & 2 & 1.32$^{\pm0.08}$ & 139$^{\pm27}$ & 21.19$^{\pm0.23}$ & 9.06$^{\pm0.12}$  & 4.74$^{\pm0.33}$ & 644$^{\pm36}$ & 23.45$^{\pm0.24}$ & 9.38$^{\pm0.06}$  \\
  50 & 150.143  &   2.39619 & 1.346 &  9.54    & 2 & 2.96$^{\pm0.23}$ & 97$^{\pm120}$ & 21.21$^{\pm0.80}$ & 8.83$^{\pm0.51}$  & 3.08$^{\pm0.07}$ & 417$^{\pm26}$ & 21.30$^{\pm0.22}$ & 9.45$^{\pm0.12}$  \\
  51 & 150.147  &   2.39797 & 0.574 &  9.23    & 1 & 1.86$^{\pm0.02}$ & 197$^{\pm3}$  & 20.58$^{\pm0.03}$ & 9.20$^{\pm0.26}$  & -                & -             & -                 & -                 \\
  52 & 150.175  &   2.43063 & 0.889 &  9.14    & 1 & 3.98$^{\pm0.27}$ & 558$^{\pm27}$ & 23.25$^{\pm0.18}$ & 8.95$^{\pm0.48}$  & -                & -             & -                 & -                 \\
  53 & 150.155  &   2.44345 & 0.34  & 10.1267  & 3 & 3.20$^{\pm0.01}$ & 268$^{\pm3}$  & 18.70$^{\pm0.01}$ & 9.75$^{\pm0.01}$  & 3.76$^{\pm0.01}$ & 553$^{\pm3}$  & 19.41$^{\pm0.03}$ & 9.85$^{\pm0.01}$  \\
  54 & 150.185  &   2.44561 & 0.679 &  9.07    & 0 & 2.66$^{\pm0.03}$ & 433$^{\pm5}$  & 21.86$^{\pm0.02}$ & -                 & -                & -             & -                 & -                 \\
  55 & 150.171  &   2.44687 & 2.085 &  8.9     & 0 & 2.06$^{\pm0.10}$ & 229$^{\pm28}$ & 22.41$^{\pm0.15}$ & -                 & -                & -             & -                 & -                 \\
  56 & 150.184  &   2.45766 & 0.378 &  9.18    & 2 & 1.24$^{\pm0.07}$ & 102$^{\pm21}$ & 21.05$^{\pm0.19}$ & 8.25$^{\pm0.11}$  & 2.37$^{\pm0.03}$ & 477$^{\pm8}$  & 21.24$^{\pm0.06}$ & 9.13$^{\pm0.02}$  \\
  57 & 150.15   &   2.45979 & 0.502 &  8.56    & 1 & 2.19$^{\pm0.05}$ & 307$^{\pm9}$  & 22.20$^{\pm0.07}$ & 8.52$^{\pm0.54}$  & -                & -             & -                 & -                 \\
  58 & 150.15   &   2.4813  & 0.683 &  9.44    & 1 & 2.40$^{\pm0.01}$ & 305$^{\pm2}$  & 19.61$^{\pm0.01}$ & 9.42$^{\pm0.06}$  & -                & -             & -                 & -                 \\
  59 & 150.058  &   2.19537 & 0.935 & 10.01    & 1 & 3.66$^{\pm0.02}$ & 531$^{\pm3}$  & 20.32$^{\pm0.01}$ & 9.98$^{\pm0.06}$  & -                & -             & -                 & -                 \\
  60 & 150.103  &   2.22138 & 0.651 &  9.21    & 1 & 2.51$^{\pm0.03}$ & 390$^{\pm5}$  & 21.17$^{\pm0.04}$ & 9.14$^{\pm0.17}$  & -                & -             & -                 & -                 \\
  61 & 150.086  &   2.22929 & 1.155 &  8.72    & 0 & 2.78$^{\pm0.12}$ & 350$^{\pm33}$ & 22.71$^{\pm0.11}$ & -                 & -                & -             & -                 & -                 \\
  62 & 150.057  &   2.23075 & 0.927 &  9.71    & 2 & 2.76$^{\pm0.14}$ & <63            & 19.32$^{\pm0.15}$ & 9.06$^{\pm0.06}$  & 3.38$^{\pm0.06}$ & 488$^{\pm10}$ & 20.41$^{\pm0.06}$ & 9.60$^{\pm0.02}$  \\
  63 & 150.062  &   2.26762 & 0.543 &  8.98    & 1 & 4.47$^{\pm0.16}$ & 290$^{\pm12}$ & 22.33$^{\pm0.09}$ & 8.71$^{\pm0.20}$  & -                & -             & -                 & -                 \\
  64 & 150.091  &   2.28517 & 0.751 & 10.7412  & 3 & 2.93$^{\pm0.02}$ & 375$^{\pm3}$  & 18.93$^{\pm0.01}$ & 10.59$^{\pm0.03}$ & 5.58$^{\pm0.19}$ & 790$^{\pm27}$ & 22.16$^{\pm0.20}$ & 9.90$^{\pm0.08}$  \\
  65 & 150.099  &   2.29026 & 0.362 & 10.71    & 3 & 1.08$^{\pm0.00}$ & 477$^{\pm1}$  & 17.76$^{\pm0.01}$ & 10.36$^{\pm0.01}$ & 6.21$^{\pm0.01}$ & 773$^{\pm1}$  & 20.47$^{\pm0.00}$ & 10.25$^{\pm0.01}$ \\
  66 & 150.089  &   2.29717 & 0.535 &  9.58    & 2 & 1.65$^{\pm0.02}$ & 160$^{\pm5}$  & 19.66$^{\pm0.02}$ & 9.20$^{\pm0.02}$  & 3.09$^{\pm0.03}$ & 471$^{\pm7}$  & 21.16$^{\pm0.06}$ & 9.34$^{\pm0.01}$  \\
  67 & 150.076  &   2.30481 & 0.123 &  9.05642 & 3 & 1.82$^{\pm0.00}$ & 102$^{\pm0}$  & 18.33$^{\pm0.00}$ & 8.63$^{\pm0.00}$  & 4.50$^{\pm0.01}$ & 327$^{\pm0}$  & 20.09$^{\pm0.01}$ & 8.82$^{\pm0.00}$  \\
  68 & 150.111  &   2.31308 & 0.482 &  9.48    & 1 & 3.71$^{\pm0.01}$ & 405$^{\pm2}$  & 21.12$^{\pm0.01}$ & 9.45$^{\pm0.09}$  & -                & -             & -                 & -                 \\
  69 & 150.08   &   2.31392 & 0.381 &  9.38    & 3 & 2.77$^{\pm0.02}$ & 331$^{\pm4}$  & 20.25$^{\pm0.01}$ & 9.13$^{\pm0.04}$  & 4.12$^{\pm0.05}$ & 789$^{\pm14}$ & 22.16$^{\pm0.07}$ & 8.92$^{\pm0.04}$  \\
  70 & 150.085  &   2.31729 & 0.944 &  9.19    & 1 & 3.00$^{\pm0.06}$ & 358$^{\pm8}$  & 21.35$^{\pm0.07}$ & 9.14$^{\pm0.25}$  & -                & -             & -                 & -                 \\
  71 & 150.062  &   2.31874 & 0.66  &  8.87    & 1 & 1.93$^{\pm0.05}$ & 298$^{\pm9}$  & 21.22$^{\pm0.09}$ & 8.82$^{\pm0.32}$  & -                & -             & -                 & -                 \\
  72 & 150.066  &   2.35129 & 0.34  &  8.63    & 2 & 1.53$^{\pm0.09}$ & 88$^{\pm24}$  & 21.27$^{\pm0.15}$ & 8.06$^{\pm0.12}$  & 2.11$^{\pm0.06}$ & 273$^{\pm16}$ & 21.78$^{\pm0.22}$ & 8.49$^{\pm0.05}$  \\
  73 & 150.087  &   2.36566 & 1.724 &  9.53    & 0 & 3.91$^{\pm0.09}$ & 429$^{\pm14}$ & 22.42$^{\pm0.05}$ & -                 & -                & -             & -                 & -                 \\
  74 & 150.087  &   2.37173 & 0.208 &  9.68    & 2 & 1.88$^{\pm0.00}$ & 140$^{\pm1}$  & 18.99$^{\pm0.01}$ & 9.31$^{\pm0.00}$  & 2.76$^{\pm0.01}$ & 458$^{\pm2}$  & 20.39$^{\pm0.01}$ & 9.44$^{\pm0.00}$  \\
  75 & 150.091  &   2.37952 & 0.454 &  8.78    & 1 & 2.37$^{\pm0.06}$ & 277$^{\pm7}$  & 21.71$^{\pm0.07}$ & 8.74$^{\pm0.31}$  & -                & -             & -                 & -                 \\
  76 & 150.091  &   2.4093  & 0.734 &  9.26    & 2 & 2.67$^{\pm0.07}$ & 246$^{\pm21}$ & 21.10$^{\pm0.07}$ & 9.05$^{\pm0.07}$  & 3.21$^{\pm0.16}$ & 598$^{\pm64}$ & 22.76$^{\pm0.43}$ & 8.85$^{\pm0.11}$  \\
  77 & 150.112  &   2.47552 & 0.794 &  9.19    & 0 & 2.91$^{\pm0.04}$ & 507$^{\pm9}$  & 21.66$^{\pm0.03}$ & -                 & -                & -             & -                 & -                 \\
  78 &  34.5301 &  -5.18858 & 0.294 &  9       & 1 & 3.03$^{\pm0.07}$ & 417$^{\pm5}$  & 21.46$^{\pm0.08}$ & 8.90$^{\pm0.06}$  & -                & -             & -                 & -                 \\
  79 &  34.3494 &  -5.18658 & 0.633 &  9.41    & 1 & 3.44$^{\pm0.02}$ & 449$^{\pm2}$  & 20.62$^{\pm0.01}$ & 9.38$^{\pm0.07}$  & -                & -             & -                 & -                 \\
  80 &  34.3995 &  -5.18517 & 0.725 &  9.73    & 3 & 3.70$^{\pm0.04}$ & 168$^{\pm12}$ & 19.42$^{\pm0.04}$ & 9.26$^{\pm0.05}$  & 3.82$^{\pm0.03}$ & 431$^{\pm8}$  & 19.78$^{\pm0.08}$ & 9.53$^{\pm0.04}$  \\
  81 &  34.5058 &  -5.18498 & 0.466 &  9.6     & 2 & 2.14$^{\pm0.02}$ & 128$^{\pm7}$  & 18.95$^{\pm0.03}$ & 9.27$^{\pm0.02}$  & 2.81$^{\pm0.03}$ & 414$^{\pm9}$  & 20.38$^{\pm0.09}$ & 9.33$^{\pm0.02}$  \\
  82 &  34.305  &  -5.1816  & 0.582 &  9.02    & 0 & 2.67$^{\pm0.03}$ & 246$^{\pm5}$  & 21.31$^{\pm0.03}$ & -                 & -                & -             & -                 & -                 \\
  83 &  34.4538 &  -5.18107 & 1.745 &  9.54    & 0 & 1.91$^{\pm0.06}$ & 241$^{\pm18}$ & 21.39$^{\pm0.09}$ & -                 & -                & -             & -                 & -                 \\
  84 &  34.4177 &  -5.17535 & 1.682 &  8.6     & 0 & 1.51$^{\pm0.05}$ & 185$^{\pm18}$ & 21.37$^{\pm0.11}$ & -                 & -                & -             & -                 & -                 \\
  85 &  34.436  &  -5.17003 & 0.285 &  8.93    & 1 & 2.53$^{\pm0.01}$ & 405$^{\pm1}$  & 20.95$^{\pm0.01}$ & 8.90$^{\pm0.06}$  & -                & -             & -                 & -                 \\
  86 &  34.254  &  -5.1685  & 0.907 &  9.73    & 0 & 2.85$^{\pm0.03}$ & 325$^{\pm5}$  & 20.93$^{\pm0.02}$ & -                 & -                & -             & -                 & -                 \\
  87 &  34.4198 &  -5.16491 & 0.485 &  9.09    & 1 & 2.80$^{\pm0.02}$ & 270$^{\pm2}$  & 20.95$^{\pm0.02}$ & 9.08$^{\pm0.16}$  & -                & -             & -                 & -                 \\
  88 &  34.3024 &  -5.16273 & 1.694 &  9.61    & 0 & 2.50$^{\pm0.05}$ & 351$^{\pm12}$ & 21.31$^{\pm0.05}$ & -                 & -                & -             & -                 & -                 \\
  89 &  34.2888 &  -5.14641 & 1.821 &  8.83    & 0 & 1.82$^{\pm0.16}$ & 341$^{\pm49}$ & 23.12$^{\pm0.20}$ & -                 & -                & -             & -                 & -                 \\
  90 &  34.2943 &  -5.14363 & 0.508 &  9.42    & 1 & 3.44$^{\pm0.01}$ & 487$^{\pm2}$  & 20.55$^{\pm0.01}$ & 9.38$^{\pm0.05}$  & -                & -             & -                 & -                 \\
  91 &  34.2498 &  -5.12784 & 1.032 &  8.96    & 1 & 2.52$^{\pm0.15}$ & 404$^{\pm24}$ & 22.54$^{\pm0.18}$ & 8.85$^{\pm0.71}$  & -                & -             & -                 & -                 \\
  92 &  34.2577 &  -5.27481 & 0.261 &  8.58    & 1 & 1.35$^{\pm0.01}$ & 218$^{\pm3}$  & 20.88$^{\pm0.04}$ & 8.53$^{\pm0.15}$  & -                & -             & -                 & -                 \\
  93 &  34.3184 &  -5.27302 & 0.698 &  9       & 1 & 2.77$^{\pm0.13}$ & 406$^{\pm19}$ & 22.82$^{\pm0.13}$ & 8.88$^{\pm0.63}$  & -                & -             & -                 & -                 \\
  94 &  34.3362 &  -5.27179 & 3.051 &  9.07    & 0 & 2.39$^{\pm0.12}$ & 295$^{\pm28}$ & 22.82$^{\pm0.13}$ & -                 & -                & -             & -                 & -                 \\
  95 &  34.4907 &  -5.26951 & 0.424 &  8.51    & 2 & 2.67$^{\pm0.07}$ & 247$^{\pm14}$ & 22.16$^{\pm0.05}$ & 8.16$^{\pm0.05}$  & 3.72$^{\pm0.12}$ & 676$^{\pm33}$ & 23.40$^{\pm0.18}$ & 8.25$^{\pm0.04}$  \\
  96 &  34.4726 &  -5.24115 & 1.422 &  9.34    & 1 & 2.70$^{\pm0.12}$ & 288$^{\pm23}$ & 22.03$^{\pm0.13}$ & 9.29$^{\pm0.73}$  & -                & -             & -                 & -                 \\
  97 &  34.2925 &  -5.23886 & 0.401 &  9.27    & 1 & 2.05$^{\pm0.04}$ & 277$^{\pm3}$  & 20.57$^{\pm0.07}$ & 9.22$^{\pm0.11}$  & -                & -             & -                 & -                 \\
  98 &  34.3119 &  -5.23517 & 1     & 10.54    & 1 & 5.88$^{\pm0.02}$ & 621$^{\pm2}$  & 20.28$^{\pm0.01}$ & 10.46$^{\pm0.03}$ & -                & -             & -                 & -                 \\
  99 &  34.3028 &  -5.23065 & 3.012 & 10.25    & 0 & 2.24$^{\pm0.13}$ & 320$^{\pm31}$ & 22.81$^{\pm0.13}$ & -                 & -                & -             & -                 & -                 \\
 100 &  34.2416 &  -5.22803 & 0.943 &  9.67    & 2 & 1.26$^{\pm0.05}$ & 197$^{\pm15}$ & 20.46$^{\pm0.10}$ & 9.24$^{\pm0.06}$  & 4.57$^{\pm0.24}$ & 682$^{\pm28}$ & 22.63$^{\pm0.18}$ & 9.47$^{\pm0.04}$  \\
 101 &  34.2535 &  -5.22677 & 2.195 &  9.16    & 0 & 1.46$^{\pm0.07}$ & 237$^{\pm27}$ & 21.56$^{\pm0.14}$ & -                 & -                & -             & -                 & -                 \\
 102 &  34.2875 &  -5.2261  & 1.033 &  9.68    & 1 & 3.48$^{\pm0.07}$ & 335$^{\pm9}$  & 21.10$^{\pm0.08}$ & 9.57$^{\pm0.16}$  & -                & -             & -                 & -                 \\
 103 &  34.4759 &  -5.22053 & 2.128 &  9.52    & 1 & 2.43$^{\pm0.11}$ & 347$^{\pm24}$ & 21.97$^{\pm0.11}$ & 9.48$^{\pm0.63}$  & -                & -             & -                 & -                 \\
 104 &  34.2433 &  -5.22055 & 0.345 &  8.97    & 1 & 2.83$^{\pm0.01}$ & 286$^{\pm1}$  & 19.86$^{\pm0.01}$ & 8.93$^{\pm0.03}$  & -                & -             & -                 & -                 \\
 105 &  34.2167 &  -5.2177  & 0.317 &  8.85    & 2 & 2.63$^{\pm0.13}$ & 146$^{\pm21}$ & 21.84$^{\pm0.11}$ & 8.41$^{\pm0.09}$  & 3.27$^{\pm0.13}$ & 422$^{\pm29}$ & 22.61$^{\pm0.23}$ & 8.66$^{\pm0.05}$  \\
 106 &  34.4552 &  -5.21709 & 1.408 &  8.96    & 0 & 2.02$^{\pm0.14}$ & 255$^{\pm36}$ & 22.14$^{\pm0.18}$ & -                 & -                & -             & -                 & -                 \\
 107 &  34.3351 &  -5.21707 & 0.642 &  9.47    & 3 & 2.59$^{\pm0.07}$ & 145$^{\pm22}$ & 19.90$^{\pm0.07}$ & 8.99$^{\pm0.11}$  & 2.85$^{\pm0.04}$ & 380$^{\pm13}$ & 20.30$^{\pm0.15}$ & 9.29$^{\pm0.08}$  \\
 108 &  34.3845 &  -5.21452 & 0.909 &  9.5     & 1 & 2.93$^{\pm0.07}$ & 428$^{\pm10}$ & 21.54$^{\pm0.06}$ & 9.44$^{\pm0.24}$  & -                & -             & -                 & -                 \\
 109 &  53.1022 & -27.912   & 0.123 &  9.16    & 3 & 1.70$^{\pm0.00}$ & 133$^{\pm0}$  & 18.67$^{\pm0.00}$ & 8.83$^{\pm0.01}$  & 2.89$^{\pm0.00}$ & 387$^{\pm0}$  & 20.44$^{\pm0.00}$ & 8.82$^{\pm0.01}$  \\
 110 &  53.0757 & -27.8904  & 0.333 &  8.72    & 3 & 1.41$^{\pm0.04}$ & 146$^{\pm9}$  & 21.25$^{\pm0.08}$ & 8.19$^{\pm0.18}$  & 1.87$^{\pm0.02}$ & 314$^{\pm7}$  & 21.51$^{\pm0.10}$ & 8.54$^{\pm0.16}$  \\
 111 &  53.2146 & -27.8741  & 1.533 &  9.05    & 0 & 2.25$^{\pm0.11}$ & 242$^{\pm29}$ & 21.46$^{\pm0.15}$ & -                 & -                & -             & -                 & -                 \\
\hline

\end{longtable}
\end{landscape}
\twocolumn

\section{M/L variation at observed bands F277W, F356W, and F444W}
\label{sec:appendixd}

In Sec. \ref{subsec:massratio}, We compare thin and thick discs mass ratios at redshifts $z = 0.1-2$ in observed bands (F227W, F356W, F444W) with a galaxy sample at $z = 0$ \citep[][at Spitzer 3.5 $\micron$ band $\approx$ F356W]{comeron_evidence_2014}. The derivation of the mass ratio depends on the mass-to-light ratio of thick and thin discs, $\Upsilon_{\mathrm{thick}}/\Upsilon_{\mathrm{thin}}$, for which we assumed a value of 1.2, as derived by \citet{comeron_thick_2011} based on a typical star formation history of the Milky Way, and adopted in \citet{comeron_evidence_2014}. 

\begin{figure}
    \centering
    \includegraphics[width=\linewidth]{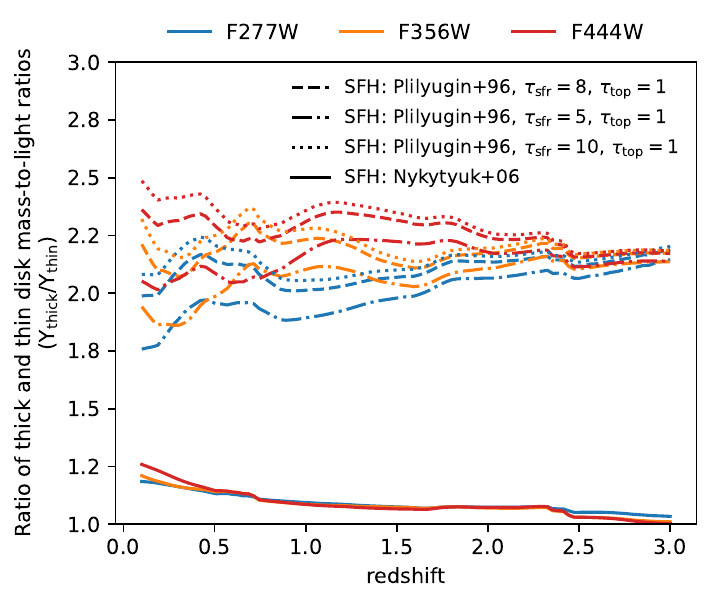}
    \caption{The evolution of the ratio of thick and thin disc mass-to-light ratios ($\Upsilon_{\mathrm{thick}}/\Upsilon_{\mathrm{thin}}$) as a function of redshift for the observing bands F227W, F356W, and F444W. Different lines represent various star formation histories (SFHs), including models from \citet[][see Sec. 3.3.1 in \citealt{comeron_thick_2011}]{pilyugin_chemical_1996} with varying $\tau_{\text{str}}$ and $\tau_{\text{top}}$, and an SFH from \citet[][see Sec. 3.3.2 in \citealt{comeron_thick_2011}]{nykytyuk_galactic_2006}. The $\Upsilon_{\mathrm{thick}}/\Upsilon_{\mathrm{thin}}$ ratio remains nearly constant across the explored redshift range, with variations of less than 0.25 among the different bands, supporting the use of $\Upsilon_{\mathrm{thick}}/\Upsilon_{\mathrm{thin}}$ at z = 0 and F356W for galaxies at redshifts 0.1–2.}
    \label{fig:figd1}
\end{figure}

A key question is whether this value can be used for galaxies with different redshift and observed at different bands although close to the band used in \citep{comeron_thick_2011}. To address this, we compute $\Upsilon_{\mathrm{thick}}/\Upsilon_{\mathrm{thin}}$ as a function of redshift at each observing band, with the same four star formation histories for thin and thick discs adopted in \citet{comeron_thick_2011}. For this computation, we use the python implementation of FSPS (Flexible stellar Population Synthesis) code \citep{conroy_fspsI_2009, conroy_fspsII_2010, johnson_pythonfspscode_2024}.
Figure~\ref{fig:figd1} shows that all $\Upsilon_{\mathrm{thick}}/\Upsilon_{\mathrm{thin}}$ remains nearly constant across the explored redshift range and exhibits similar values in all three observing bands (F227W, F356W, F444W) with difference being less than 0.25.  
This suggests that it is reasonable to adopt the $\Upsilon_{\mathrm{thick}}/\Upsilon_{\mathrm{thin}}$ value at $z=0$ and F356W also for our sample of galaxies at redshifts $z = 0.1-2$ in the observed bands. 

\begin{figure}
    \centering
    \includegraphics[width=\linewidth]{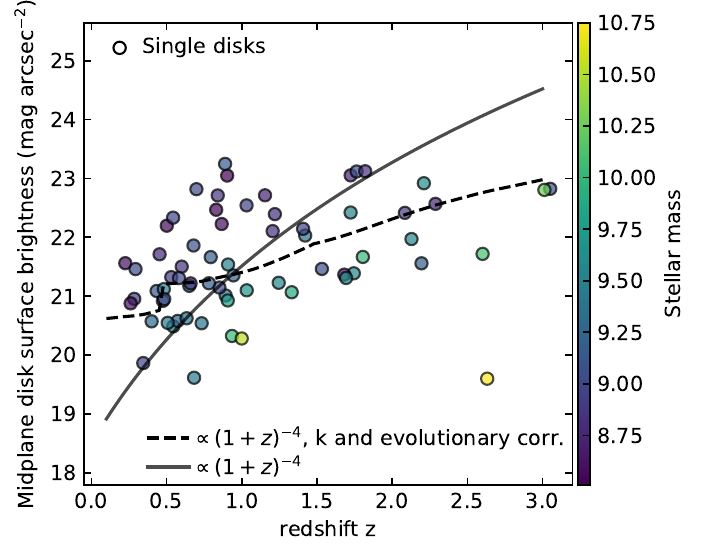}
    \caption{Midplane disc surface brightness of single-disc galaxies plotted against redshift, color-coded by stellar mass. The solid and dashed lines indicate the expected surface brightness dimming $(1+z)^{-4}$ and the combined effect of dimming, k-correction (rest-frame band shifting), and evolutionary correction (stellar population aging) derived from the star formation history used to compute Fig.~\ref{fig:figd1}. The observed trend shows that surface brightness dimming alone overpredicts the observed evolution, whereas including k-correction and evolutionary correction reproduces the data better. The errors are comparable to or smaller than the size of the markers.}
    \label{fig:figd2}
\end{figure}

Figure \ref{fig:figd2} shows the disc midplane intensities of single-disc galaxies plotted against their redshifts, with data points color-coded by galaxy mass. As expected, galaxies become fainter due to surface brightness dimming, and higher-mass galaxies are brighter at a fixed redshift. The expected surface brightness dimming trend, $(1+z)^{-4}$, appears to over-predict the evolution of our galaxies (solid line). Incorporating an additional k-correction (accounting for rest-frame band shifting), and evolutionary correction (accounting for stellar population aging), using the SFH used to derive Fig.~\ref{fig:figd1} (see Sec. 3.3.2 in \citealt{comeron_thick_2011}) more accuratly reproduces the observed trend (dashed line).  

\bsp	
\label{lastpage}
\end{document}